\documentclass[epj]{svjour}
\usepackage{graphicx}
\usepackage{amssymb}
\usepackage{psfrag}
\begin{document}

\title{Gap and out-gap breathers in a binary modulated
discrete nonlinear Schr\"odinger model}
\author{Andrey V. Gorbach \and Magnus Johansson
}                     
\authorrunning{A.V. Gorbach \and M. Johansson}

\titlerunning{Gap and out-gap breathers in a binary modulated
DNLS model}

%
%
\institute{Department of Physics and Measurement Technology (IFM),\\
Link\"{o}ping University, S-581 83 Link\"{o}ping, Sweden,
\email{mjn@ifm.liu.se}}
\date{Received: date / Revised version: date}
%
\abstract{
We consider a modulated discrete nonlinear Schr\"odinger (DNLS) model with 
alternating
 on-site potential, having a linear spectrum 
with two branches separated by a 'forbidden' gap. 
Nonlinear localized
time-periodic solutions with frequencies in
the gap and near the gap -- discrete gap and out-gap
breathers (DGBs and DOGBs) -- are investigated. Their linear stability 
is studied varying the system parameters from the continuous to
the anti-continuous limit, and different types of oscillatory and real 
instabilities are revealed. It is shown, that generally DGBs in infinite 
 modulated
DNLS chains with hard (soft) nonlinearity do not possess any oscillatory 
instabilities 
for breather frequencies in the lower (upper) half of the gap. 
Regimes of
'exchange of stability' between symmetric and antisymmetric DGBs are observed,
where an increased breather  mobility is expected. 
The transformation
from DGBs to DOGBs when the breather frequency enters the linear 
spectrum is studied, and
the general bifurcation picture for DOGBs
with tails of different wave numbers is described.
Close to the anti-continuous limit, the localized linear
eigenmodes and their corresponding eigenfrequencies are calculated analytically
for several gap/out-gap breather configurations, yielding explicit
proof of their linear stability or instability close to this limit.
\PACS{
      {63.20.Pw}{Localized modes}   \and
      {63.20.Ry}{Anharmonic lattice modes} \and
      {42.65}{Nonlinear waveguides}
     } 
} 
\maketitle
\section{Introduction}
\label{intro}

The existence of spatially localized nonlinear excitations in 
discrete models -- 'discrete brea\-thers' (DBs) -- has attracted
much attention, as they can play a significant role in
condensed matter physics, biophysics, nonlinear optics, etc.  
(for reviews see e.g. \cite{Aubry,Flach}). In
particular, being localized in space and typically stable,
DBs can contribute to energy and information transfer processes.
For example, breathers can trap
energy during long periods of time and cause non-exponential
thermal relaxation in nonlinear lattices \cite{Entrap}. By contrast,
under certain conditions \cite{Chen,Cretegny_PhD} DBs can be
rather mobile and become good energy carriers.

The presence of intrinsic structure of a medium can influence the
DB properties. In particular, new types of DBs --
\emph{discrete gap breathers} (DGBs) -- appear with
frequencies inside the forbidden gaps in the linear waves spectrum.
DGBs are the discrete analogues of gap solitons, first
discovered in a nonlinear optical medium with modulated refractive index
\cite{Chen&Mills}. Their specific features 
are due to the existence of two neighboring bands
of the linear dispersion curve, with opposite signs of dispersion
close to a gap.

A classical example of a system with two bands in the linear spectrum
is a diatomic one-dimensional lattice.
DGBs in diatomic lattices with
nonlinear interatomic potentials [Fermi-Pasta-Ulam (FPU) models] 
were studied analytically as well as numerically
within the rotating-wave approximation, neglecting 
higher harmonics generation 
\cite{Chub,Kiselev1,Aoki,Kiselev2,Franchini1}.
Later the existence of DGBs in diatomic FPU
lattices was proved rigorously \cite{Livi,James}. 
Existence and linear
stability properties of DGBs in diatomic FPU chains were also studied
numerically for some particular values of system parameters
\cite{Cretegny,Zolotaryuk,Maniadis}. We recently \cite{we} performed 
a detailed analysis of DGBs in a diatomic lattice with {\em nonlinear on-site
potential} -- {\em Klein-Gordon (KG) model} --
in the complete regime of
continuation from the anti-continuous (AC) to the continuous limit, and 
in particular we described the dynamics resulting from
several types of instability mechanisms 
(oscillatory as well as non-oscillatory). 

Very recently, much attention has been attracted to modulated structures 
described by equations of 
{\em discrete
nonlinear Schr\"odinger} (DNLS) type, in particular within nonlinear optics
for the description of
electromagnetic waves in arrays of weakly coupled optical waveguides 
\cite{DNLSopt}. For such a system with even and odd  
waveguides of different widths, discrete gap \cite{DGBopt,newKivshar}
as well as multigap \cite{SKPRL} solitons were found. A related model 
with two coupled DNLS-like equations describing a 
diffraction-managed waveguide array was also studied recently, and found 
to support different types of discrete gap solitons \cite{Kevrekidis}.
Much recent attention 
has also been given to  DNLS models in describing Bose-Einstein 
condensates (BECs) trapped in optical periodic potentials 
\cite{Trombeltoni,Cataliotti}. For a recent review 
on properties and applications of the DNLS equations, see \cite{EJ}. 

As the DNLS equations also approximate the small-amplitude dynamics of weakly 
coupled KG-chains \cite{Peyrard,Morgante}, the DNLS discrete gap solitons
should share many 
features of DGBs in KG (and also FPU) lattices (henceforth they will 
therefore also be referred to as DGBs). In particular, in the continuum 
limit these models 
are described by the same general class of coupled equations with 
exact gap soliton solutions analyzed e.g.\ in \cite{KF92,mePRE}. 

The aim of the present study is to give a
detailed and thorough analysis of DGB properties 
in the modulated DNLS model. Our motivation is twofold. First, we wish 
to provide, for the benefit e.g. of experimentalists
working in related areas (nonlinear optics, BECs), charts of 
different types of instabilities etc., and information about the 
parameter regimes where they are expected to be seen. 
Particularly interesting is the possible observation of moving DGBs. 
Generally DBs can not freely propagate
along the system due to the existence of the effective periodic 
Peierls-Nabarro (PN)
potential (caused by the discreteness of the system). 
However, in the diatomic KG
model we found \cite{we} parameter regimes with increased DGB mobility
due to 'exchange of stability'
between symmetric and antisymmetric DGB configurations, and thus
it is of interest to look for similar effects in the modulated DNLS models, 
proposing that the increased DGB mobility 
should be directly observable in experiments. 

Second, it was not carefully analyzed before, 
within a fully discrete model, what happens with 
DGBs when their frequencies approach gap boundaries.
In continuous models, gap solitons 
delocalize and vanish at one gap boundary, 
but bifurcate into a new type of excitations --
out-gap solitons -- at the other boundary \cite{Peyraurd}. Out-gap solitons
have frequencies inside the linear spectrum, and their 
structure can be viewed as a superposition of two fields. One field has a 
'dark' soliton shape with non-zero 
amplitudes at the infinities and a localized decrease 
in the center. The other field has either a 
'bright' soliton form with tails exponentially decaying to zero, 
or forms a 'bright soliton on a pedestal' ('anti-dark soliton'), 
similar to a 'dark' soliton but with a localized {\em increase} of the 
amplitude in the center. 
Although out-gap solitons can radiate energy
through linear waves, they may still represent long-lived 
localized excitations persistent to  
perturbations (e.g. \cite{Bose_outgap}).

Thus, analogously it is interesting to look for 
\textit{discrete out-gap breathers} (DOGBs) in discrete models, and 
to study their stability and the bifurcations from 
DGBs to DOGBs. 
This 
was not analyzed in earlier studies of DGBs in KG or FPU models, 
partly  due to technical problems: since DGBs delocalize
when approaching gap boundaries, numerical computations become time 
consuming when increasing the 
system size to avoid boundary effects. However, 
DNLS models are more conveniently analyzed numerically, 
since time-periodic DGBs become 'stationary' solutions with 
purely harmonic time-dependence,
that can be removed by transforming into a rotating frame.

Another important question concerns the possible existence of truly localized
excitations with frequencies inside the linear wave spectrum -- analogues
of 'embedded' solitons (e.g. \cite{embed}) -- and the nature of tails in 
such 
hypothetical localized DOGBs. This issue is also most conveniently 
analyzed with the modulated DNLS model.

Our study of the spatially binary modulated DNLS model and the properties 
of its DGB and DOGB solutions will be 
structured as follows. In Sec.~\ref{sec_model} the
model is described, as well as the  
properties of its linear spectrum. In Sec.~\ref{sec_solutions} we discuss the
linear stability analysis of stationary solutions, and in 
Sec.~\ref{sec_res} we review the  numerical 
construction of breather solutions from the anti-continuous limit, 
discussing  some particular features of the 
continuation in coupling and frequency of DGBs and DOGBs.  In 
Sec.~\ref{sec_numer} we present numerical and analytical 
results for linear stability properties of different types of 
DGBs and DOGBs. Sec.~\ref{subsec_DGB} describes the continuation in coupling 
of DGBs, and we compare the stability results 
with those obtained for the diatomic  KG
chain \cite{we}. In Sec.~\ref{subsec_freq}
we study the  transition from DGBs into DOGBs 
while changing the breather frequency, describing the 
bifurcations which may occur to DGBs when their
frequency approach the gap boundary, and furthermore penetrate the 
linear spectrum as DOGBs. In 
Sec.~\ref{subsec_freq2} bifurcations of DOGBs
with different types of tails are described. 
We study the subsequent bifurcations of 
DOGBs into 'on-top' DBs (DOTBs) with frequencies above the 
linear spectrum, and obtain a general bifurcation picture of 
DOGBs with tails of different wave numbers. We discuss the linear stability 
properties of the most important
DOGB and DOTB solutions  in
Sec.~\ref{subsec_DOGB}, and
in Sec.~\ref{sec_conclude} conclusions 
are made. Some details of the analytical investigation of linear 
stability properties of DGBs and DOGBs at small  coupling 
are deferred to Appendix~\ref{append}.

\section{The Model}
\label{sec_model}

The standard DNLS equation with cubic nonlinearity reads:
\begin{equation}
\label{DNLS}
i\frac{d\psi_n}{dt}=\lambda\psi_n-C\left(\psi_{n-1}+\psi_{n+1}-2\psi_n\right)+
\gamma\left|\psi_n\right|^2\psi_n.
\end{equation}
When modelling an array of weakly coupled optical 
wave\-guides  \cite{DNLSopt,DGBopt}, 
$\psi_n$ 
is the normalized amplitude of the electric field in the EM wave, 
propagating along the array. The coefficient 
$\lambda$ then characterizes the linear propagation constant in the 
waveguide, $C$ the coupling coefficient, $\gamma$  the 
effective nonlinear coefficient, and the time-like variable $t$ 
measures distance along the array. Here we consider
positive $\gamma$, corresponding to 'hard' nonlinearity.
The case of 'soft' nonlinearity ($\gamma<0$) can easily be recovered by 
a simple transformation, see footnote \ref{foot} below. 

To introduce a gap parameter, one can modulate any of the
constants $\lambda, C, \gamma$ (or two or three of them 
simultaneously).  We choose to modulate 
the coefficient $\lambda$, corresponding in the optical model to 
a variation of the width of 
even and odd waveguides (see e.g. \cite{DGBopt}). Thus, we put:
\begin{equation}
\label{lambda}
\lambda\rightarrow\lambda_n=1+(-1)^n\delta^2,
\end{equation}
where the parameter $\delta$ defines the width $\Delta$ 
of the gap in the linear waves 
spectrum: $\Delta=2\delta^2$ (see below).

\begin{figure}
\rotatebox{270}{
\resizebox{0.7\columnwidth}{!}{%
  \includegraphics{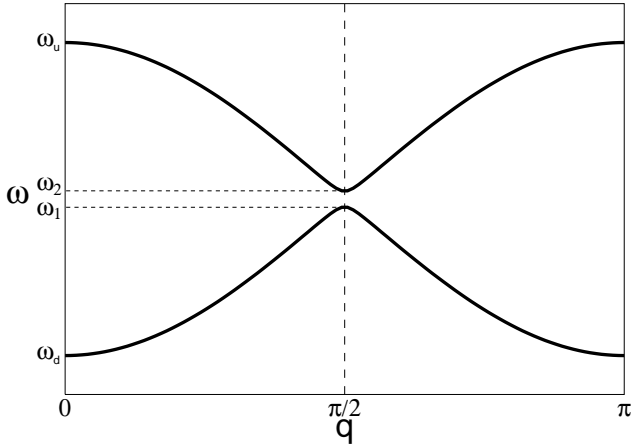}
}
}
\caption{The dispersion relation of linear waves (\ref{dispersion}).}
\label{fig:spectr}
\end{figure}

The dispersion relation for solutions:
\begin{eqnarray}
\label{linear-sol}
&&\psi_n=\left\{
\begin{array}{l}
A\exp(iqn-i\omega t), \qquad n=2k, \\
B\exp(iqn-i\omega t), \qquad n=2k+1,
\end{array}
\right. \\
\nonumber
&&k=0,1,2,...
\end{eqnarray}
of the 
linearized system (\ref{DNLS}) with 
$\lambda$ in the form (\ref{lambda}) reads:
\begin{equation}
\label{dispersion}
\omega_{o,a}=1+2C\pm\sqrt{\delta^4+4C^2\cos^2(q)},
\end{equation}
where indices $o$ and $a$ stand for the upper ('optic-like') and lower 
('acoustic-like') branches of the spectrum\footnote{Note however that 
also for the "acoustic-like" branch the dispersion is quadratic and not 
linear for small $q$.}.
It possesses a gap at the wave number $q=\pi/2$ (see 
Fig.~\ref{fig:spectr}). The gap boundaries 
are defined as:
\begin{equation}
\label{gap}
\omega_{2,1}=1+2C\pm\delta^2.
\end{equation} 
The ratio of the amplitudes $A$ and $B$ in a linear wave 
(\ref{linear-sol}) is defined by its frequency $\omega$:
\begin{equation}
\label{ratio}
\frac{A}{B}=\sqrt{\left|\frac{\omega-\omega_1}{\omega-\omega_2}\right|}.
\end{equation}
In linear waves with frequencies belonging to the upper branch of the 
spectrum ($\omega>\omega_2$), 
amplitudes in even waveguides ($A$) are always higher than 
in odd waveguides ($B$), while for frequencies 
in the lower branch ($\omega<\omega_1$) the 
odd waveguides are dominant. Therefore, the sub-field of
even (odd) waveguide amplitudes can be referred to as 
the upper (lower) band sub-field, respectively. With an analogy to 
a diatomic chain, even and odd waveguides correspond to light and heavy 
atoms, and 
will therefore be referred to as 'light' and 'heavy' sites, 
respectively.

Another type of modulated DNLS model is obtained by multiplying the
left-hand side of equation (\ref{DNLS}) with $m_n$, defined 
analogously to $\lambda_n$ in (\ref{lambda}). This model, 
corresponding to simultaneous modulation of all constants in 
(\ref{DNLS}), can be considered as a simplified model of
a diatomic lattice ($m_n$ playing the role of atom masses), 
neglecting higher harmonics generation. 
As shown in \cite{Franchini2} for the diatomic FPU 
model, the effect of higher harmonics on DB properties is often 
rather negligible, although additional instabilities and bifurcations 
generally may 
appear for large values of the coupling and/or large-amplitude oscillations 
(e.g. \cite{Morgante}).
Investigating numerically breather properties in the 
two modulated DNLS models we have found qualitatively similar 
results, and therefore we focus our discussion only on model 
(\ref{DNLS}) with coefficients $\lambda$ as defined in (\ref{lambda}). 

We denote the system size (the number of waveguides) with
$N$ ($n=1,2,...,N$) and impose 
periodic boundary conditions $\psi_{N+1}\equiv\psi_1$, 
$\psi_{0}\equiv\psi_N$.
The DNLS equations (\ref{DNLS}) are then the Hamiltonian equations with the 
Hamiltonian:
\begin{eqnarray}
\label{energy}
&&H(\{i\psi_n\},\{\psi_n^*\})=\\
\nonumber
&&\qquad\sum_{n=1}^{N}\left(C\left|\psi_{n+1}-\psi_n\right|^2 
+\lambda\left|\psi_n\right|^2+\frac{\gamma}{2}\left|\psi_n\right|^4\right).
\end{eqnarray}
%
In what follows we put  $\gamma=+1$ in (\ref{DNLS}), 
(\ref{energy}) (without loss of generality, 
since varying $\gamma$ is equivalent to rescaling $\psi_n$).

\section{Stationary solutions and linear stability}
\label{sec_solutions}

Stationary solutions of equations (\ref{DNLS}) 
have the form:
\begin{equation}
\label{sol}
\psi_n(t)=\phi_n e^{-i\omega_b t},
\end{equation}
with time-independent amplitudes $\phi_n$, where $\omega_b$ is the excitation 
frequency. DGBs have frequencies 
$\omega_b$ lying inside the gap (\ref{gap}), so that 
$\omega_1<\omega_b<\omega_2$. For DOGBs one has 
$\omega_b>\omega_2$, while 'on-top' breathers (DOTBs)
have frequencies above the linear spectrum, $\omega_b>\omega_{u}$, 
where
\begin{equation}
\label{top_freq}
\omega_{u}=1+2C+\sqrt{\delta^4+4C^2}.
\end{equation}

To analyze the linear stability of a particular solution 
$\{\phi_n^{(0)}\}$, 
we add a small perturbation $\{\epsilon_n(t)\}$ to it:
\begin{equation}
\label{perturb}
\psi_n(t)=\left(\phi_n^{(0)}+\epsilon_n(t)\right) e^{-i\omega_b t},
\end{equation}
and linearize the equations (\ref{DNLS}):
\begin{eqnarray}
\label{eq_per}
i\frac{d\epsilon_n}{dt}&=&\left(\lambda_n+2C-\omega_b\right)\epsilon_n -
C\left(\epsilon_{n-1}+\epsilon_{n+1}\right)+\\
\nonumber
&&+2\gamma\left|\phi_n^{(0)}
\right|^2\epsilon_n+\gamma\left(\phi_n^{(0)}\right)^2\epsilon_n^*.
\end{eqnarray}
Expanding $\epsilon_n(t)$ in real and imaginary
parts, $\epsilon_n(t)=\alpha_n(t)+i\beta_n(t)$,
equations
(\ref{eq_per}) can, for real solutions $\{\phi_n^{(0)}\}$, be written in 
matrix form \cite{CE85}:
\begin{equation}
\label{eq_per_matr}
\left(
\begin{array}{c}
\{\dot{\alpha}_n\}\\
\{\dot{\beta}_n\}
\end{array}
\right)=
\left(
\begin{array}{cc}
0 & \mathcal{L}_0\\
-\mathcal{L}_1 & 0
\end{array}
\right)
\left(
\begin{array}{c}
\{\alpha_n\}\\
\{\beta_n\}
\end{array}
\right),
\end{equation}
with $\mathcal{L}_0$ and $\mathcal{L}_1$ defined as follows:
\begin{equation}
\label{eq_per_real}
\left\{
\begin{array}{rcl}
\mathcal{L}_0 \beta_n & \equiv &
\left[\lambda_n+2C-\omega_b+\gamma\left(\phi_n^{(0)}\right)^2\right]\beta_n
-\\
&&- C \left(\beta_{n-1}+\beta_{n+1}\right),\\
\mathcal{L}_1 \alpha_n & \equiv &
\left[\lambda_n+2C-\omega_b+3\gamma\left(\phi_n^{(0)}\right)^2\right]
\alpha_n -\\
&&- C \left(\alpha_{n-1}+\alpha_{n+1}\right).\\
\end{array}
\right.
\end{equation}
The solution $\{\phi_n^{(0)}\}$ is linearly stable 
if and only if the perturbation $\{\epsilon_n(t)\}$ remains
bounded in time, i.e. if and only if all eigenvalues $i\omega_e$
of the matrix in (\ref{eq_per_matr}) are imaginary.

Note that writing the general solution of (\ref{eq_per}) as
\begin{equation}
\label{perturb_sol}
\epsilon_n(t)=\frac12\left(a_n+b_n\right)e^{-i\omega_e t}+
  \frac12\left(a_n^*-b_n^*\right)e^{i\omega_e^* t}
\end{equation}
leads to 
the following set of algebraic equations (e.g. 
\cite{Morgante}):
\begin{equation}
\label{eq_per_real2}
\left\{
\begin{array}{rcl}
\mathcal{L}_0 b_n & = & \omega_e a_n,\\
\mathcal{L}_1 a_n & = & \omega_e b_n,\\
\end{array}
\right.
\end{equation}
with $\mathcal{L}_0$ and $\mathcal{L}_1$ defined above (\ref{eq_per_real}).
Consequently, the condition of the linear stability of a solution
$\{\phi_n^{(0)}\}$ is that
all the eigenfrequencies $\omega_e$, obtained from (\ref{eq_per_real2}),
are real.

If $\omega_e$ is a real eigenfrequency of (\ref{eq_per_real2}) and
$(a_n,b_n)$ is the corresponding real solution, 
then $i\omega_e$ is an eigenvalue of
(\ref{eq_per_matr}) with eigenvector 
$(\{\alpha_n\equiv a_n\},\{\beta_n\equiv ib_n\})$.
Therefore with each real eigenfrequency $\omega_e>0$ one can associate a
Krein signature $\kappa(\omega_e)$ (e.g. \cite{Bri,Morgante} and
references therein):
\begin{equation}
\label{krein}
\kappa(\omega_e)=\textnormal{sign} \sum_n a_n b_n.
\end{equation}
For DGBs with frequencies inside the gap,
eigenmodes with positive (negative) $\kappa$
correspond to excitations inside even (odd) waveguides in the
AC limit $C\rightarrow 0$. Physically, the Krein
signature is the sign of the Hamiltonian energy carried by the
corresponding eigenmode (e.g. \cite{Bri,Skryabin}).

\section{Construction of breather solutions}
\label{sec_res}

The main idea for the numerical methods used to obtain breather solutions, 
based on
the proof of their existence \cite{Aubry,Marin&Aubry}, 
is to perform numerical continuation
of an exact breather solution, known for some
particular parameter values, to other, arbitrary, values of
these parameters. For stationary solutions (\ref{sol}), Eq. (\ref{DNLS})
with modulated coefficient $\lambda$ (\ref{lambda})
reduces to a set of algebraic equations 
($\gamma=1$)\footnote{\label{foot}Note that 
the case of soft nonlinearity $\gamma=-1$ is also covered through the 
transformations $\phi_n' = (-1)^n\phi_n$, $\omega_b'=2+4C-\omega_b$, 
$\delta'^2=-\delta^2$. Essentially, this reverses the roles of upper/lower 
bands and even/odd sites.}:
\begin{eqnarray}
\label{DNLS-algebra}
&&\left[1+(-1)^n\delta^2-\omega_b\right]\phi_n-\\
\nonumber
&&\qquad-C\left(\phi_{n-1}+\phi_{n+1}-2\phi_n\right)+
\left|\phi_n\right|^2\phi_n=0,
\end{eqnarray}
and solutions can be found by Newton schemes
(e.g. \cite{Eilbeck}).

Usually breather solutions at $C=0$ (AC limit) are taken as initial 
solutions and continued to 
non-zero values of $C$. At $C=0$ one has an array of uncoupled 
waveguides. Hence, in the AC limit, for real solutions 
each amplitude $\phi_n$ should be taken 
from the set $\phi_n\in\{0,\pm A_n\}$, with:
\begin{equation}
\label{ampl}
A_n=\sqrt{\omega_b-\lambda_n}.
\end{equation}
Consequently, to each breather solution one can associate a 
corresponding coding sequence, where we use symbol notation 
with up- and down- arrows for $+A_n$ and $-A_n$, respectively. 
To make the intrinsic structure of the solutions more clear, we
use different sets of symbols for odd and even amplitudes 
(lower and upper band sub-fields): $\{\mathbb{O} \Uparrow \Downarrow \}$
and $\{0 \uparrow \downarrow\}$, respectively.

Note, however, that these coding sequences are not unique for a given 
solution. Indeed, one can construct a DGB with some particular coding 
sequence and then continue it in frequency up to values above 
the upper gap boundary $\omega_b>\omega_2$. Above the gap this 
solution will have another coding sequence if continued to 
$C=0$, though 
the continuations in frequency and coupling  can be done smoothly without 
any true 
bifurcations to other solutions. Similar effects were observed for 
second-harmonic penetration into a phonon band for 
breathers
in a monoatomic Klein-Gordon lattice \cite{phantom}. 
In fact, one should distinguish two types of coding sequences: those 
corresponding to the AC limit inside and above the gap of the linear
spectrum 
('gap' and 'out-gap' coding sequences, respectively)\footnote{The existence 
of two types of coding sequences is
connected to the existence of two types of AC 
limits. In general, the AC limit $\omega_b/C\rightarrow 
\infty$ can be reached either by taking $C=0$ or 
$\omega_b\rightarrow\infty$. These are not 
equivalent for modulated systems.}. 
We 
will use the superscripts $G$ or $O$ in coding sequences to indicate 
whether they are 'gap' or 'out-gap', respectively.

Restricting to  solutions with spatial 
symmetry or anti\-symmetry 
with respect to the center, $N_c = \mathrm{Int} (N/2) + 1$, 
the coding 
sequences can be simplified by omitting the codes for 
sites $n=1,2..N_c-1$, and using subscripts $S$ or $A$  to mark 
whether a breather has \emph{symmetric} or \emph{antisymmetric} spatial 
configuration respectively. 
As follows from (\ref{ampl}) and (\ref{lambda}),
in the case of hard nonlinearity ($\gamma>0$ in (\ref{DNLS})), 
the simplest symmetric
DGBs are
centered on a 'heavy' (odd) site with non-zero amplitude, while the simplest
antisymmetric DGBs are centered on a 'light' (even) site
with zero
amplitude and with non-zero amplitudes of opposite signs on two neighboring
'heavy' sites.

Note 
that the structure of the linear spectrum, and in
particular the gap boundaries $\omega_{1,2}$ (\ref{gap}),
depends on the coupling $C$ (although the gap width
$\Delta=2\delta^2$ is independent on $C$). Thus, if one
constructs a DGB with frequency $\omega_b$ inside the gap at
some particular value of $C$ and tries to continue it versus  $C$ at fixed
frequency, it can reach one of the
gap boundaries in the process of continuation. Consequently, a DGB
can disappear (on the lower boundary) or transform into a DOGB
(on the upper boundary) while continued in coupling. The same
effect can occur for DOGBs, as the frequency may
 reach that of the linear wave corresponding to its tail. 
To avoid this problem we fix not the breather
frequency, but the {\em frequency detuning}: 
\begin{equation}
\label{dw}
\Delta\omega\equiv\omega_b-\omega_o(q,C),
\end{equation}
for the continuation in coupling. 
Here $\omega_o(q,C)$ is the
frequency of the linear wave (\ref{dispersion}) with the same wave number
$q$ as that of the brea\-ther tail, belonging to the upper branch of the
spectrum. All stationary DGB solutions have tails with wave number
$q=\pi/2$, so that the frequency detuning (\ref{dw}) is
equal to $\Delta\omega=\omega_b-\omega_2(C)<0$ for such breathers.

Hence, the frequency detuning $\Delta\omega$ (\ref{dw}) is a more convenient
parameter than the frequency  $\omega_b$, and
it can be considered as
the only dynamical parameter of stationary DGBs and DOGBs.
Note also that although Eq.\ (\ref{DNLS-algebra}) for stationary 
solutions
has three parameters ($\omega_b$, $\delta^2$, $C$), only two
 are independent, since 
varying the third parameter is equivalent to rescaling $\psi_n$.
For instance, one can choose as independent parameters: $\delta^2/C$ --
the 'modulation' parameter, and $\omega_b/C$ -- the 'discreteness'
parameter.
In
what follows we fix the 'gap' parameter $\delta^2$ and investigate the
properties of stationary solutions while varying the frequency
detuning $\Delta\omega$ (\ref{dw}) and the coupling $C$.

\begin{figure*}
\rotatebox{270}{
\resizebox{0.73\columnwidth}{!}{%
  \includegraphics{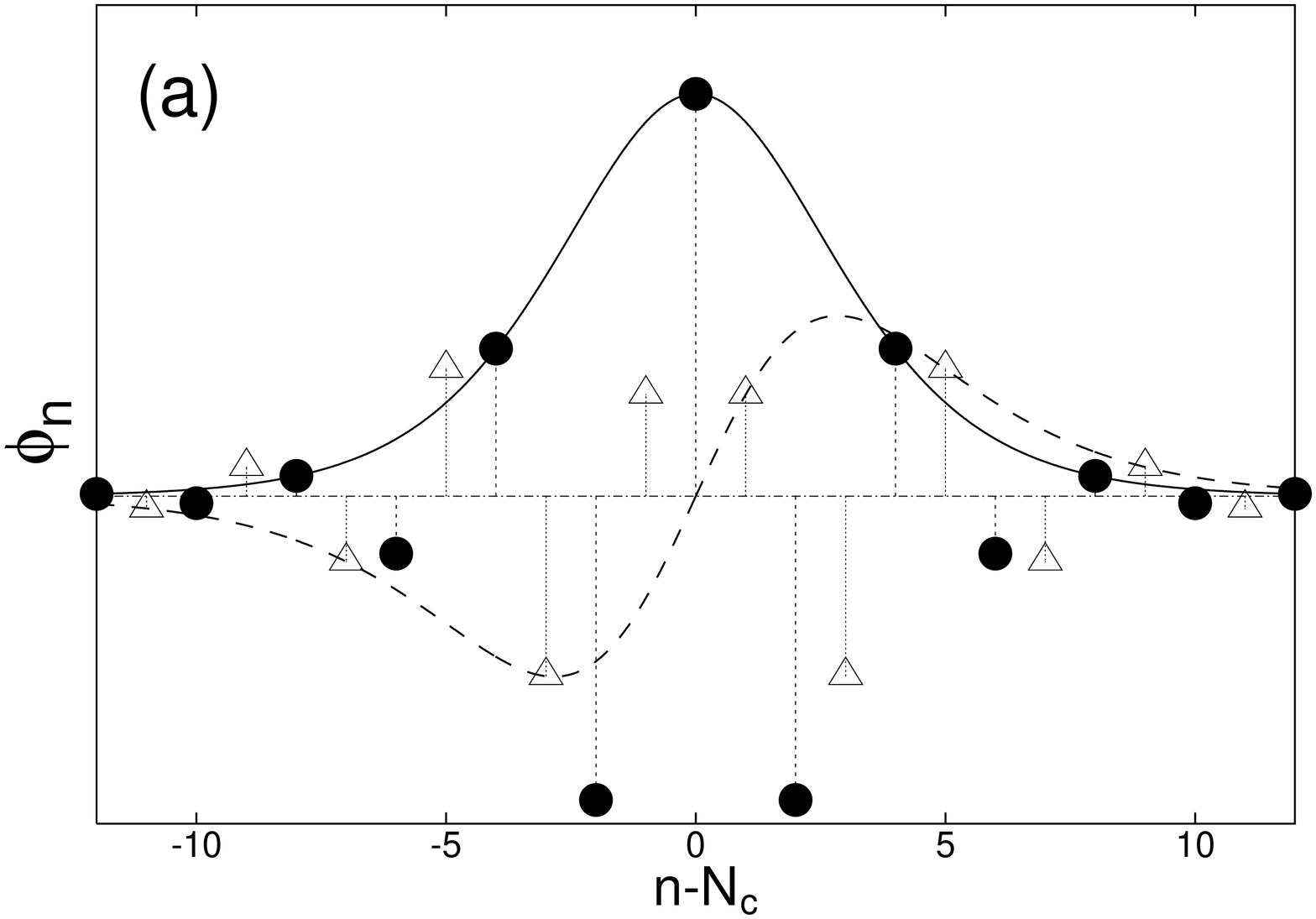}
}
}
\rotatebox{270}{
\resizebox{0.73\columnwidth}{!}{%
  \includegraphics{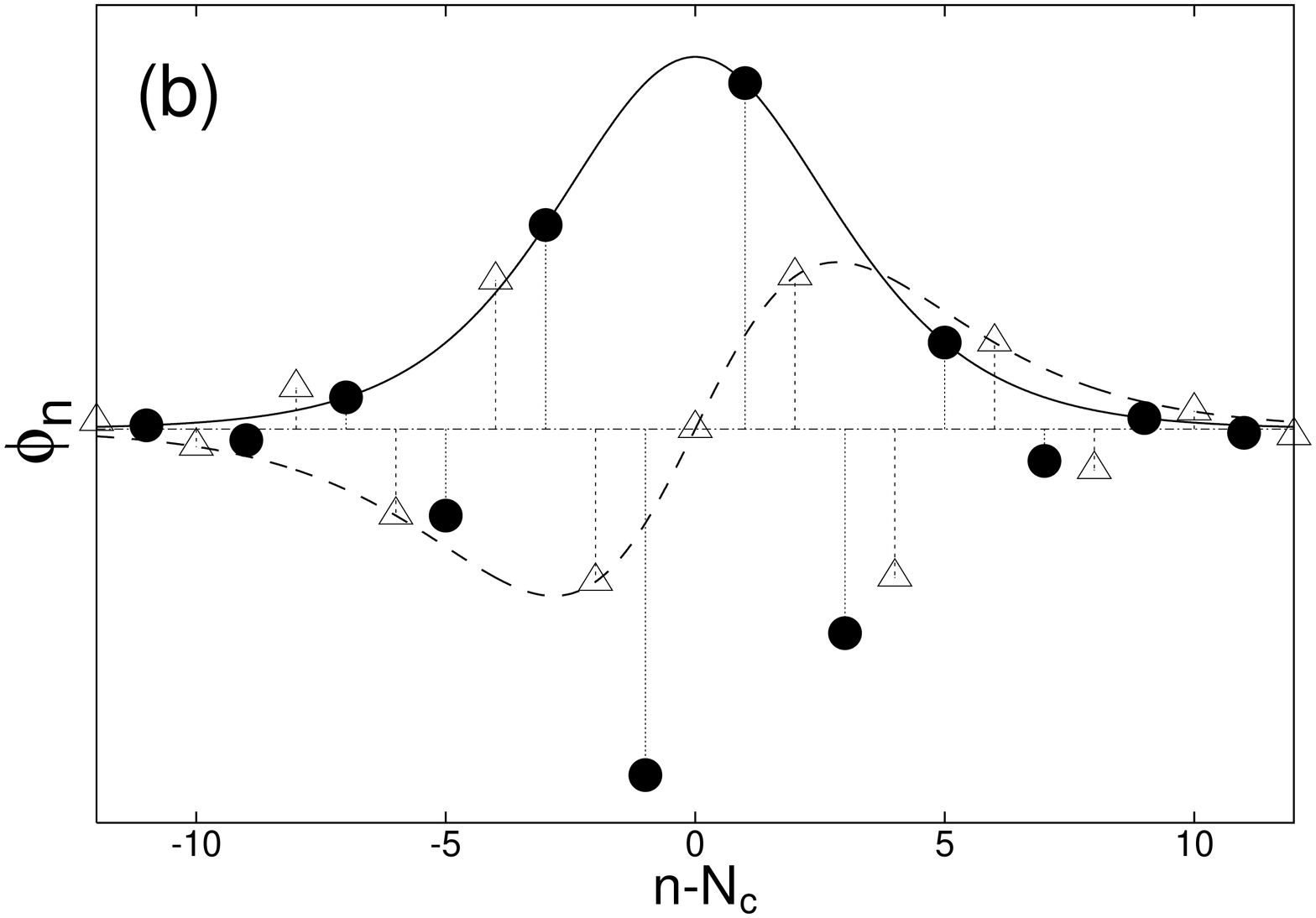}
}
}
\caption{Symmetric (a) and antisymmetric (b) DGBs
at non-zero coupling. Circles (triangles) denote 'heavy' 
('light')
amplitudes.}
\label{fig:DGB}
\end{figure*}
\section{Numerical results}
\label{sec_numer}

We here present the results of numerical continuation and
stability analysis of DGB and DOGB solutions. 

\subsection{Stability of gap breathers}
\label{subsec_DGB}
At frequencies inside the gap, $\omega_1<\omega_b<\omega_2$,
only 'heavy' (odd) amplitudes $\phi_n$ can be non-zero 
in the AC limit (\ref{ampl}). 
Thus, there are two basic configurations of DGB solutions 
with coding sequences 
$\{\Uparrow(0\mathbb{O})\}_S^G$ and $\{0\Uparrow(0\mathbb{O})\}_A^G$ 
(codes in 
parenthesis are repeated), representing a symmetric DGB centered on a 
'heavy' 
site 
with non-zero amplitude, and 
an antisymmetric DGB centered on a 'light' site with zero amplitude, 
respectively (Fig.~\ref{fig:DGB}). 
Since DGBs
have exponentially decaying tails of wave number $q=\pi/2$, 
neighboring even amplitudes, as well as neighboring odd amplitudes, 
have opposite signs. In addition, there is a phase shift of
'light' (even)  
amplitudes in the center (Fig.~\ref{fig:DGB}), known 
from gap solitons in the continuum
limit (e.g. \cite{mePRE}).

\begin{figure*}
\rotatebox{270}{
\resizebox{0.7\columnwidth}{!}{%
  \includegraphics{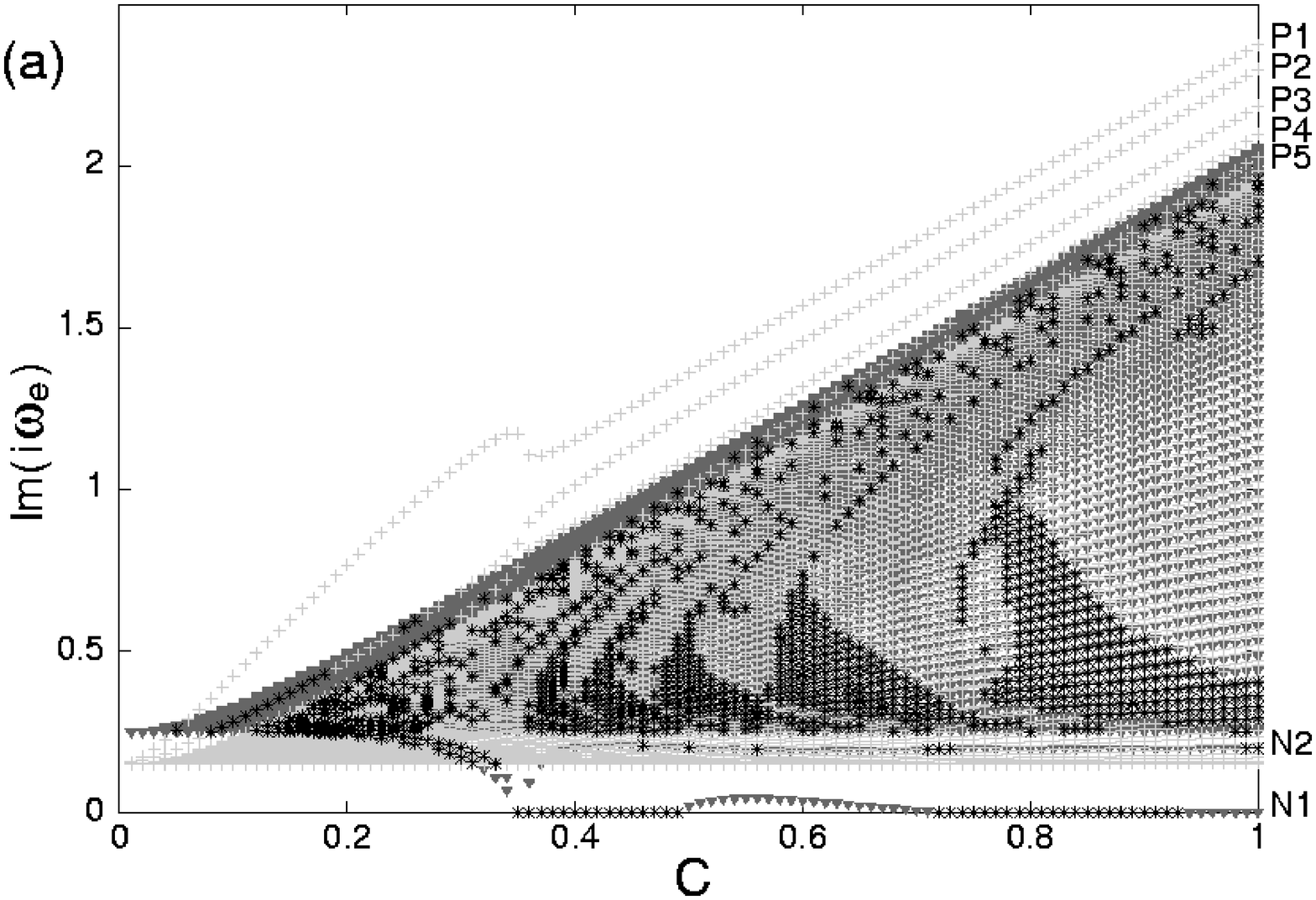}
}
\resizebox{0.7\columnwidth}{!}{%
  \includegraphics{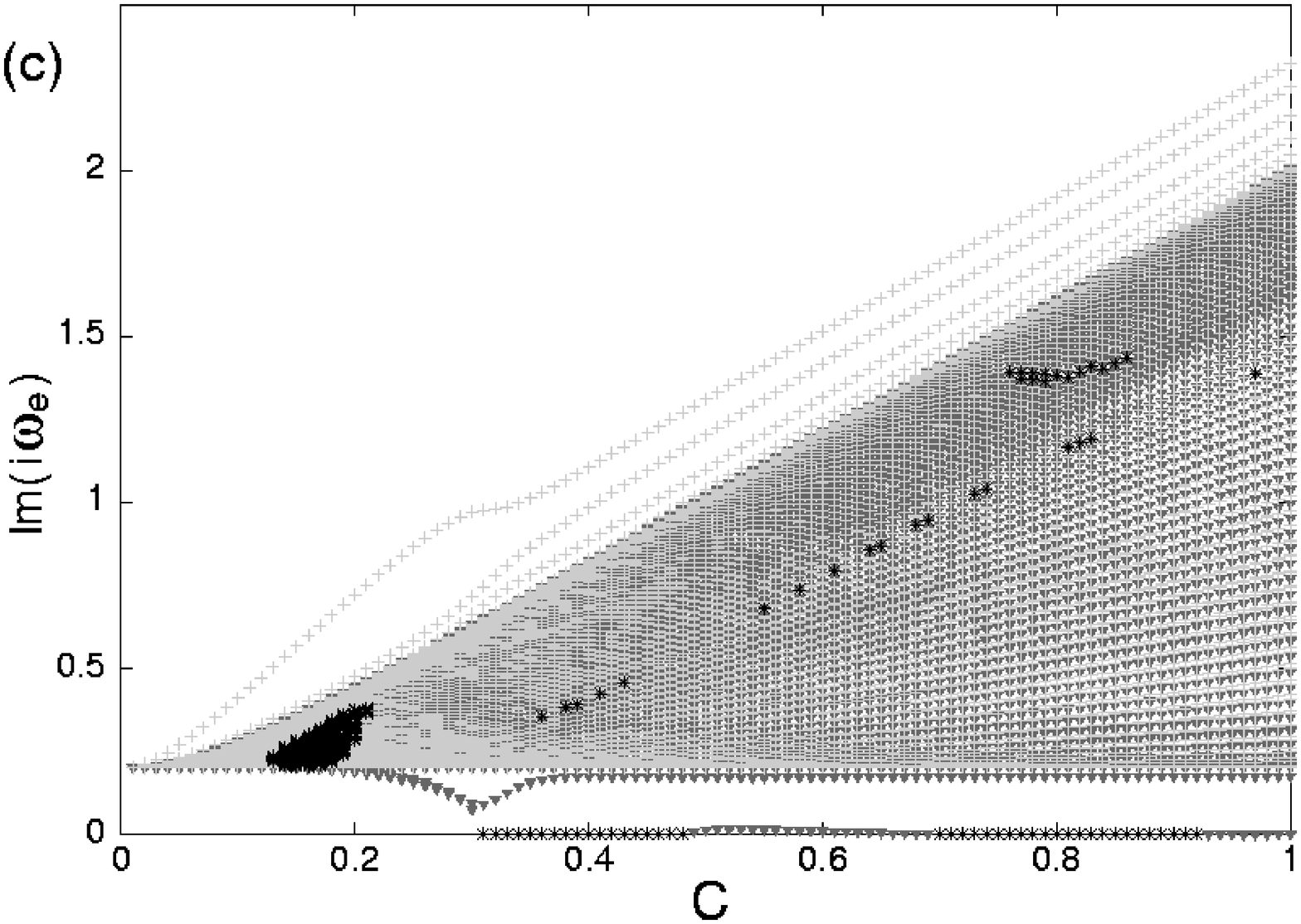}
}
\resizebox{0.7\columnwidth}{!}{%
  \includegraphics{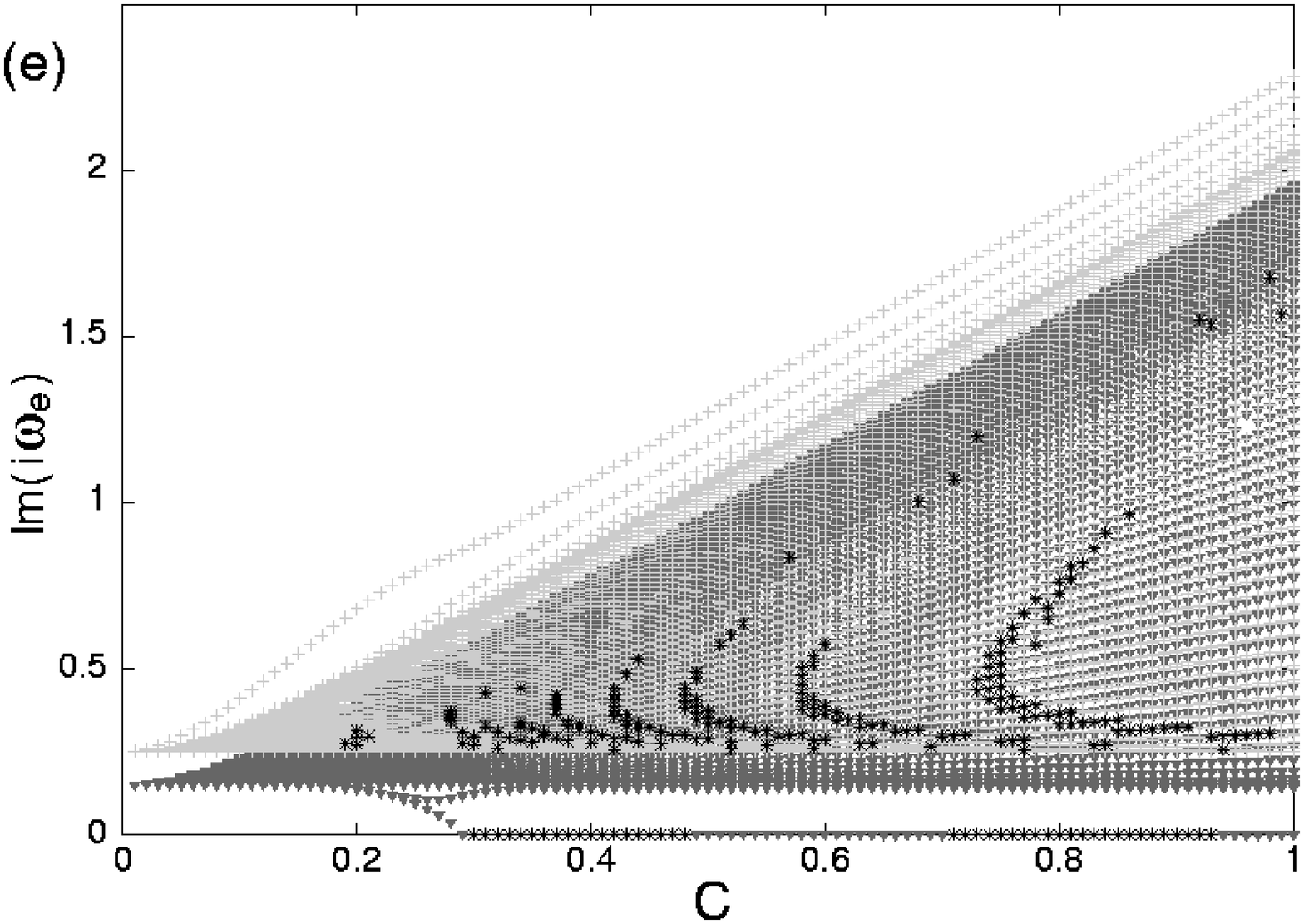}
}
}
\rotatebox{270}{
\resizebox{0.7\columnwidth}{!}{%
  \includegraphics{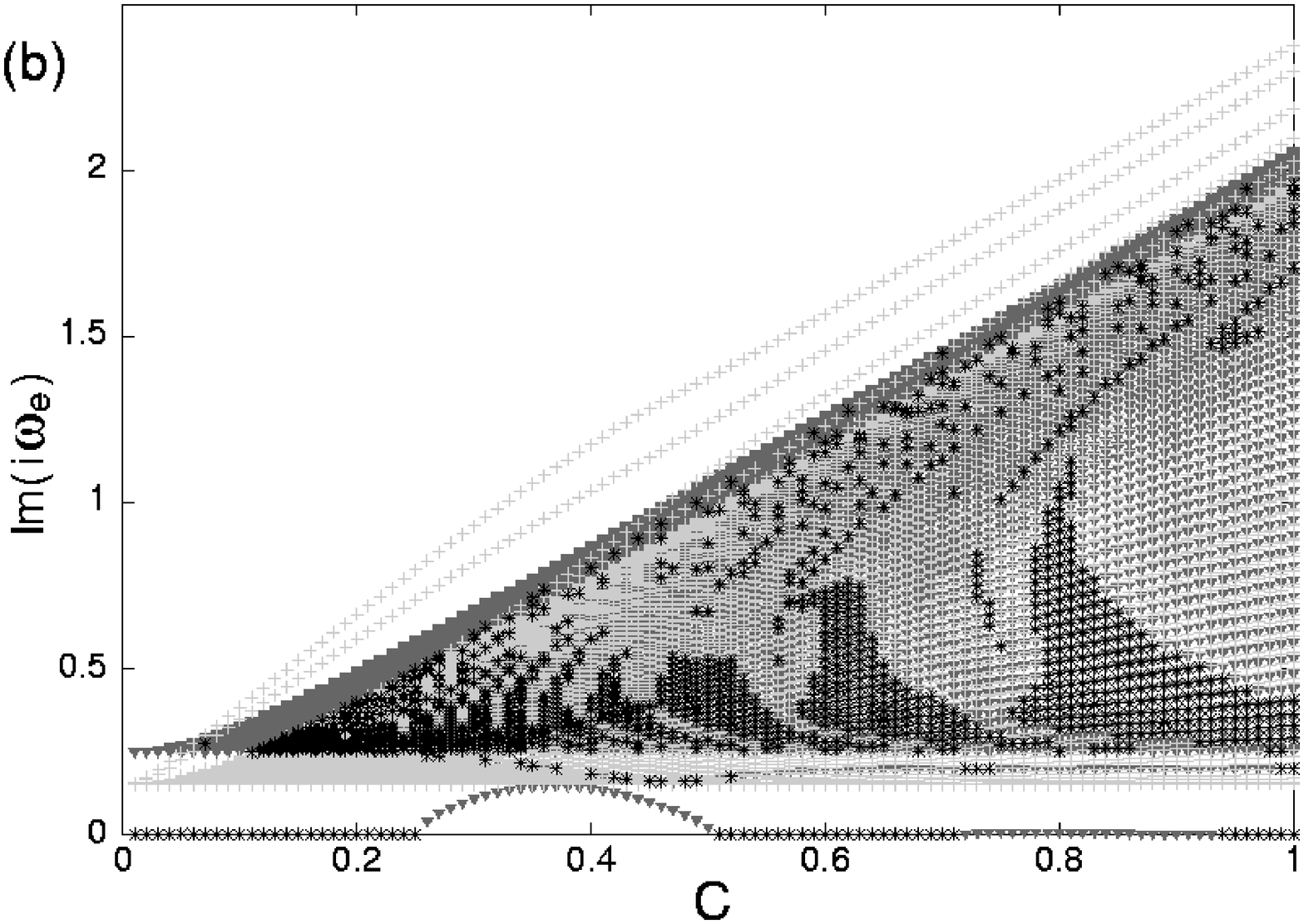}
}
\resizebox{0.7\columnwidth}{!}{%
  \includegraphics{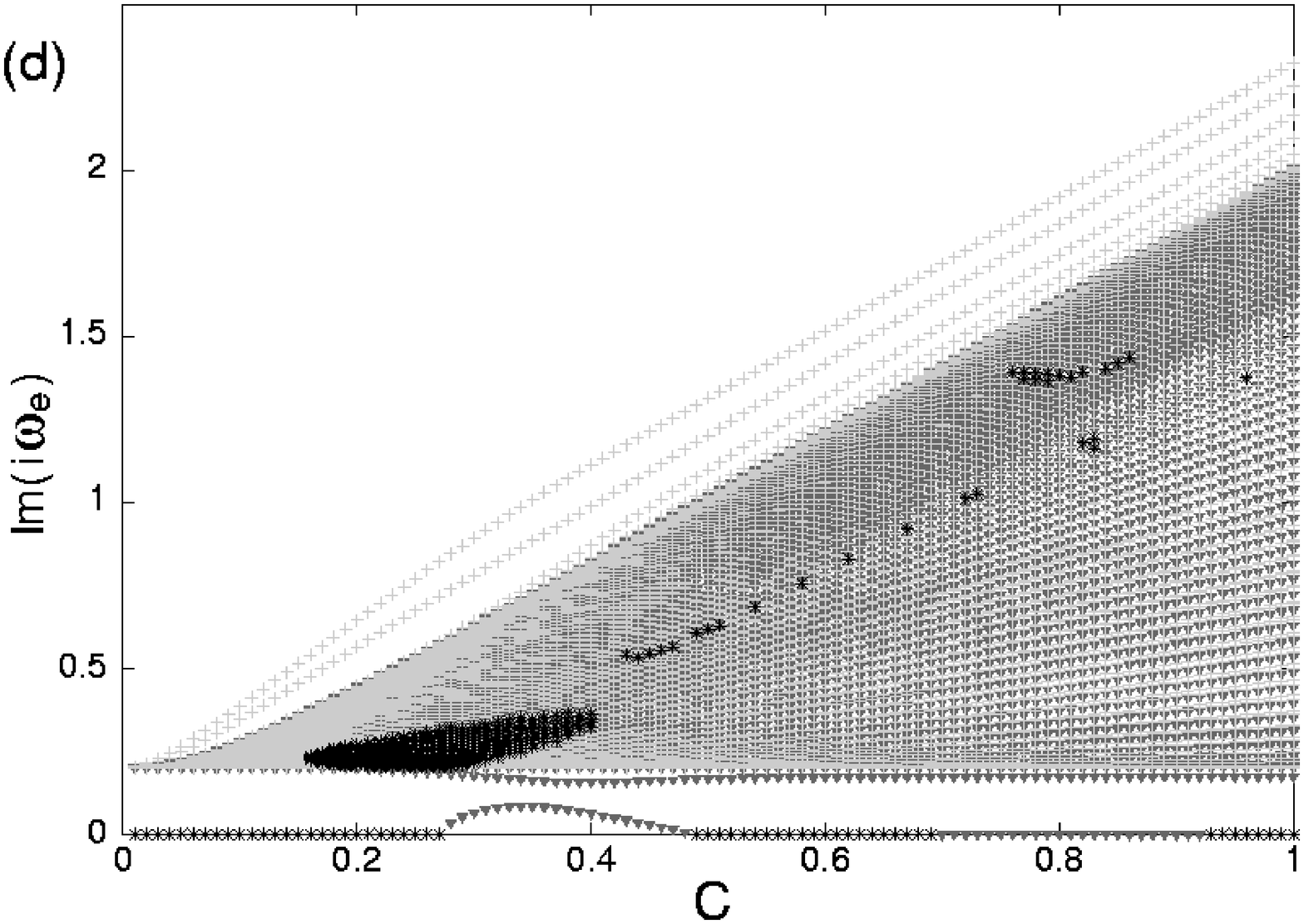}
}
\resizebox{0.7\columnwidth}{!}{%
  \includegraphics{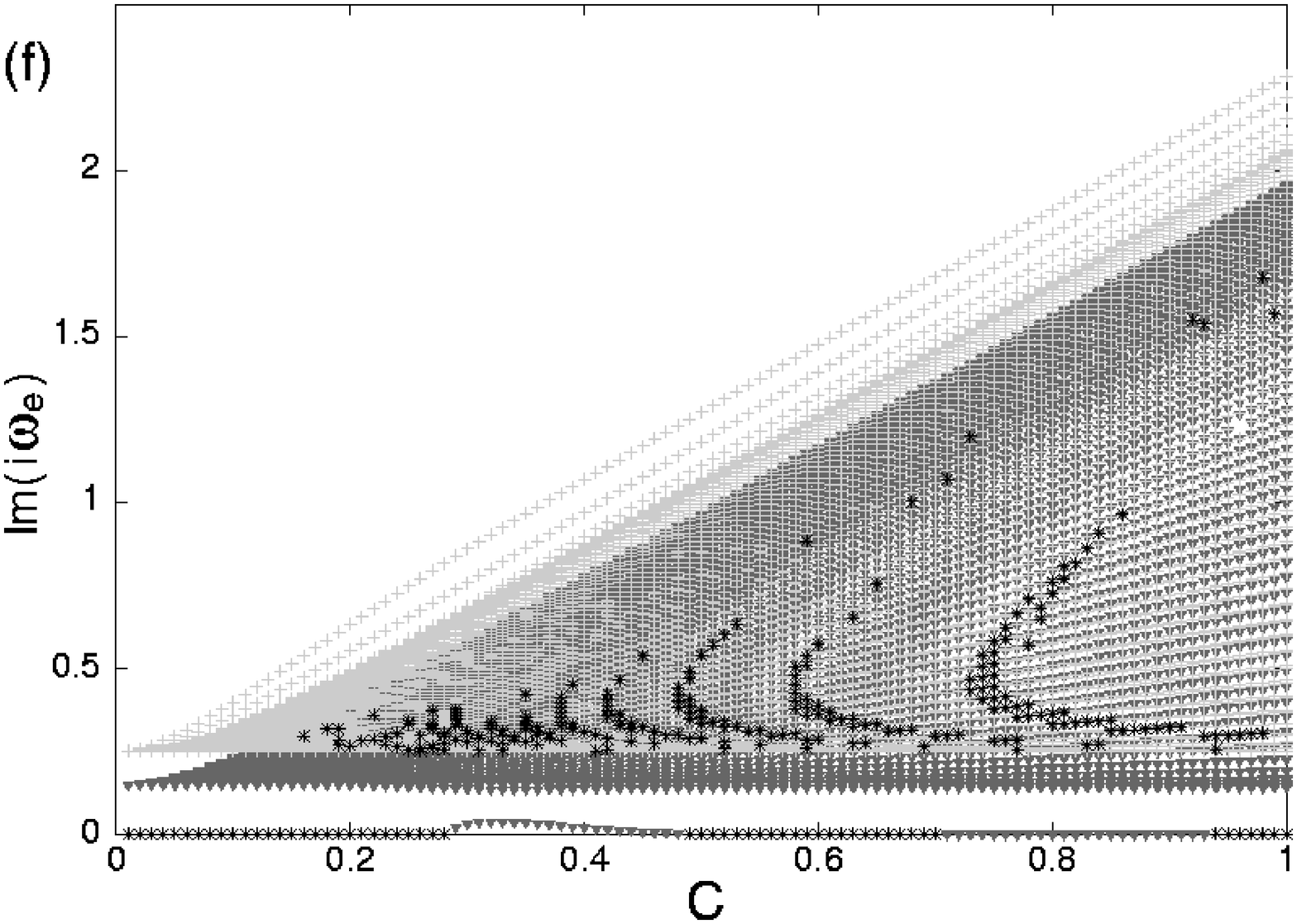}
}
}
\caption{Imaginary parts of eigenvalues $i\omega_e$ of (\ref{eq_per_matr}) 
for symmetric (a),(c),(e) and antisymmetric (b),(d),(f) breathers 
at different values of the frequency detuning (\ref{dw}): 
(a),(b) $\Delta\omega=-0.75\cdot\delta^2$ (upper half of the gap); 
(c),(d) $\Delta\omega=-\delta^2$ (middle of the gap);
(e),(f) $\Delta\omega=-1.25\cdot\delta^2$ (lower half of the gap).
Light-grey crosses correspond to eigenvalues with positive Krein 
signature, dark-grey triangles to eigenvalues with negative 
Krein signature, and black stars  to unstable eigenvalues.
In all cases 
the system size $N=242$ 
and the gap parameter $\delta^2=0.2$.}
\label{fig:DGB_eigen}
\end{figure*}

In Fig.~\ref{fig:DGB_eigen} we show the results of stability analysis of 
symmetric and antisymmetric DGBs for the continuation in 
coupling at three different frequency detunings (\ref{dw}): 
in the upper half of the gap,  in the gap center, and in 
the lower half of the gap, respectively.
The stability eigenvalues can be divided into two groups, corresponding to 
spatially localized and extended
eigenvectors, respectively. 

The 'extended' eigenvalues are independent on the brea\-ther spatial 
configuration and would be obtained also for the system without 
breather. They correspond to extended excitations of the 
upper and lower band sub-fields and form two bands $i\omega_+$ and
$i\omega_-$
with positive and negative Krein signatures, respectively:
\begin{eqnarray}
\label{ev_ext}
i\omega_+=i\left(\omega_o(q)-\omega_b\right),\\
\nonumber
i\omega_-=i\left(\omega_b-\omega_a(q)\right).
\end{eqnarray}
Here $\omega_o(q)$ and $\omega_a(q)$ are the frequencies of linear 
waves (\ref{linear-sol}), (\ref{dispersion})
with wavenumbers $q$ (defined by the system size), 
corresponding to the upper and lower
bands,
respectively. In the limit of an infinite system the extended 
eigenvalues will have continuous spectrum.
The boundaries of these bands
are determined by the DGB frequency 
$\omega_b$ and the boundaries of the lower and upper bands of the 
linear spectrum $\omega_{1,2},\omega_{u,d}$ (see 
Fig.~\ref{fig:spectr}).
Defining the deviation $\alpha\equiv\Delta\omega+\delta^2$ 
of a DGB frequency from 
the gap center: $\omega_b=1+2C+\alpha$ 
($-\delta^2\le\alpha\le\delta^2$), yields
\begin{eqnarray}
\label{bands_limits2}
\delta^2-\alpha\le\omega_+\le\sqrt{\delta^4+4C^2}-\alpha, \\
\nonumber
\delta^2+\alpha\le\omega_-\le\sqrt{\delta^4+4C^2}+\alpha.
\end{eqnarray}

Thus, having the same shape, the two bands of extended eigenvalues are 
shifted by $2\alpha$ with respect to each other. While increasing the 
coupling constant $C$, these 
bands broaden and at a certain value $C_0$,
\begin{equation}
\label{c0}
C_0=\sqrt{\delta^2\left|\alpha\right|+\alpha^2},
\end{equation}
they overlap. (In the gap center $\alpha=0$, the two bands coincide 
as shown in 
Fig.~\ref{fig:DGB_eigen}(c),(d).) The overlapping of the 
bands will lead to collisions of extended eigenvalues with opposite 
Krein signatures with generation of instabilities (see 
Fig.~\ref{fig:DGB_eigen}). For larger systems  more 
collisions will occur, as the bands become more 'dense', but
the instabilities produced will become weaker and should 
completely disappear in the limit of an infinite system
\cite{size_effect}\footnote{This is true generically for localized 
breathers, while for non-localized solutions (e.g.\ breathers with extended 
tails) such instabilities survive also in the 
infinite-size limit (e.g.\ \cite{Morgante}).}.

The localized eigenvalues play a more important role, as 
their collisions with each other or with extended 
eigenvalues can produce strong instabilities independent on the 
system size. The corresponding eigenvectors are localized around the 
breather when the eigenvalues lie outside the extended bands. 
There are two types of localized eigenvalues, bifurcating from the 
top of the extended band with 
$\kappa=+1$
($P1,P2,P3,...$ in Fig.~\ref{fig:DGB_eigen}(a)), and 
from the bottom of the band with 
$\kappa=-1$ ($N1$, $N2$), respectively. In addition, 
there is also the 'phase' mode
with  $i\omega_e\equiv 0$, whose eigenvector 
is the same as the breather solution, and associated
with the phase rotation of the breather.

The eigenvector corresponding to the eigenvalue $N1$ 
has symmetry opposite to the DGB. 
Colliding with its complex conjugate a real 
instability is produced, connected to the 'exchange of stability' between 
symmetric and antisymmetric DGBs 
\footnote{Here 'stability' only means w.r.t. the PN-mode, since
Krein instabilities may exist also in this regime (see 
Fig.~\ref{fig:DGB_eigen}(a),(b)).} 
[Peierls-Nabarro (PN) mode, see below].
The other eigenvalue 
bifurcating from the negative Krein 
signature band, $N2$, has an eigenvector with same symmetry as 
the breather.
For DGB frequencies in 
the upper half of the gap, both $N1$ and $N2$ eigenvalues 
penetrate the positive Krein signature band, yielding oscillatory 
instability 
(Fig.~\ref{fig:DGB_eigen}(a), and $N2$ in (b)). In particular regimes 
the $N2$
 eigenmode can collide with its complex conjugate producing 
real instabilities, connected to
transitions from 'discrete-like' to 'continuous-like' 
DGBs \cite{we} (see Sec.~\ref{subsec_freq}.)

The eigenvalues $P1,P2,P3,...$ bifurcate one by one
when increasing 
the coupling, corresponding alternately to symmetric or antisymmetric 
eigenvectors. The number of these localized modes increases as, with 
increasing coupling, the central part of the DGB will 
occupy a larger number of sites. These eigenvalues can 
also produce oscillatory instabilities in the upper half of the gap, 
penetrating the extended 
band with opposite Krein signature (see 
Fig.~\ref{fig:DGB_eigen}(a),(b)).
Unlike other eigenvalues,  $P1$ and $P2$ 
(and for the antisymmetric DGB also $P3$) can 
bifurcate from the extended band immediately when $C$ becomes 
non-zero.
Analysis shows, that 
for each of these eigenvalues there is a critical value of 
the breather frequency detuning $\Delta\omega_{cr}$ (in the lower half 
of the gap for $P1$ and $P2$), specific for symmetric 
and antisymmetric DGBs, above which they
stay outside the
band (i.e., the eigenmodes are localized) 
at arbitrarily small non-zero coupling (see 
Appendix~\ref{append} for details). 
The deviation of these eigenvalues from the top 
of the extended band is 
proportional to $C^2$ at small values of $C$ (see 
Eqs.(\ref{dw_p1_s_g}),(\ref{dw_p2_s_g})).
By contrast, below  $\Delta\omega_{cr}$ the eigenvalue is 
inside the band at small $C$.

Similar structure of extended and localized eigenvalues was reported 
for the diatomic KG model \cite{we}. However, an
important difference  is that in the 
modulated DNLS model, DGBs with 
frequencies in the lower half of the gap \emph{do not 
possess any oscillatory instability} associated with  localized 
eigenvalues.
Indeed, for frequencies in the lower half of the gap 
($\alpha\le 0$ in (\ref{bands_limits2})), the bottom of the extended 
band with $\kappa=-1$ is below the bottom of the  
$\kappa=+1$ band for all values of $C$
(see Fig.~\ref{fig:DGB_eigen}(c)-(f)), so  
the $N2$ eigenvalue does not penetrate the extended band. For similar 
reason the localized eigenvalues $P1,P2,P3...$ do not penetrate the 
extended band, as the top of the $\kappa=+1$  band is above the top of the 
other  band for all $C$. 
By contrast, in the diatomic KG model DGBs possess oscillatory 
instabilities 
associated with localized eigenmodes of $P$-type 
both in the upper and  lower 
halves of the gap for larger values of the coupling, 
as the  bands have different shapes \cite{we}. 

The two types of oscillatory instabilities of DGB 
solutions were recently observed also in \cite{newKivshar}, where 
the stability of DGBs was analyzed for a 
continuation in 
frequency at a particular value of coupling within a similar 
model of a coupled waveguide array. The $P1,P2,P3,...$ instabilities can be 
associated with \emph{external resonances}, as the frequencies of the 
corresponding localized modes are above the linear spectrum. By 
contrast, the $N2$ instability can be associated with \emph{internal 
resonances}, as the frequency of this mode is inside the gap 
\cite{newKivshar}. 

The real instability $N1$ was also observed in 
\cite{newKivshar} for the antisymmetric DGB solution, which was shown 
to be unstable at all frequencies inside the gap for the particular 
used
value of coupling. However, we emphasize that in 
fact {\em both symmetric and antisymmetric DGBs can possess this 
type of instability} (see Fig.~\ref{fig:DGB_eigen}) at different 
values of coupling. This \emph{exchange of stability} 
between symmetric and antisymmetric modes is very important for
breather mobility issues \cite{Cretegny_PhD,we}. Generally, 
for the antisymmetric DGB at small $C>0$ there are two localized 
modes associated with antisymmetric
and symmetric coupled oscillations in the two 'heavy'
sites with non-zero codes. The former is the 'phase' mode
with $i\omega_e\equiv 0$, while the latter is the $N1$
 mode 
which becomes unstable immediately when $C>0$
(see Fig.~\ref{fig:DGB_eigen}(b),(d),(f) and Appendix~\ref{append}).
Thus, antisymmetric
DGBs always possess a real instability at small $C$, with a
growth rate proportional to $C$ that can be 
obtained analytically (Appendix~\ref{append},Table~\ref{tab:1}, 
eigenmode $(N1,A,G)$).

\subsection{Transition from DGB to DOGB solutions. Discrete- and 
continuous-like breathers.}
\label{subsec_freq}

To study in details the transformation of DGBs into 
DOGBs, we 
perform the continuation in the frequency detuning (\ref{dw}) of 
the breather solutions,
at fixed coupling. 
We focus our attention on symmetric breather 
configurations.

\begin{figure}
\rotatebox{270}{
\resizebox{0.75\columnwidth}{!}{%
  \includegraphics{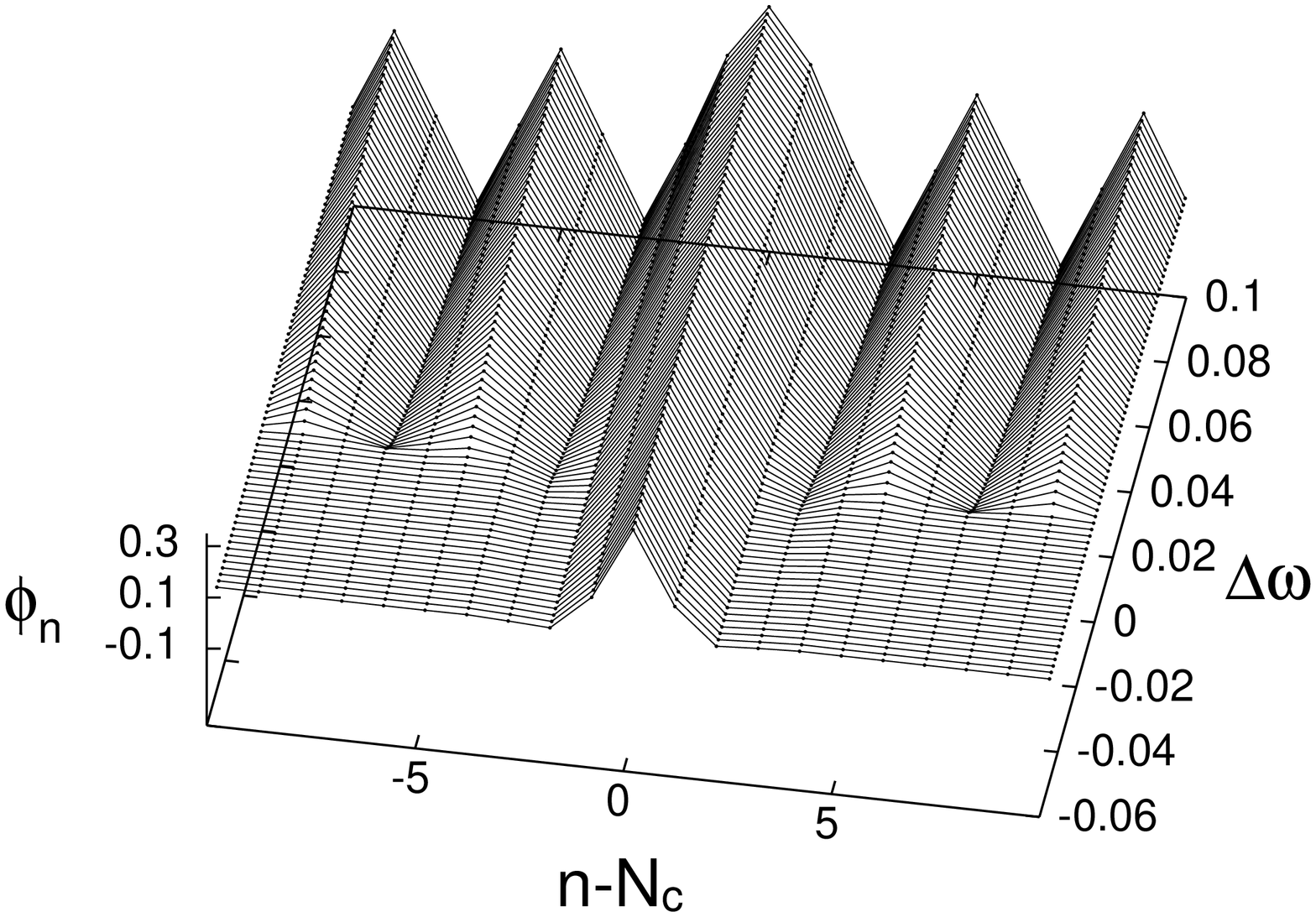}
}
}
\caption{Amplitudes 
 of the central part 
of a breather 
for the continuation in 
frequency detuning $\Delta\omega$ from DGB 
($\Delta\omega=-0.6\cdot\delta^2$) to DOGB
($\Delta\omega=\delta^2$). 
$N=50$, 
$\delta^2=0.1$ and $C=0.02$.
}
\label{fig:DGB-DOGB}
\end{figure}

In Fig.~\ref{fig:DGB-DOGB} the profile of a symmetric breather 
is shown 
for the continuation in frequency detuning, at fixed coupling, from a value 
inside the gap to a value above the linear spectrum. 
At the upper gap boundary ($\Delta\omega=0$) 
the DGB transforms into a 
DOGB with non-zero tails. The structure of this DOGB is 
qualitatively the same as that of a DGB (see Fig.~\ref{fig:DGB}(a)) but 
with non-zero asymptotes of the upper band sub-field ('light' amplitudes)
$\left|\phi^{(u)}\right|\sim\sqrt{\Delta\omega}$, 
while the lower band sub-field ('heavy' 
amplitudes) still exponentially decays to zero.
The wave number of the tails in this 
DOGB is $q=\pi/2$ as for a DGB, and the 
corresponding coding sequence is 
$\{\Uparrow(\uparrow\mathbb{O}\downarrow\mathbb{O})\}_S^O$. In what 
follows we will call this configuration a 'normal' DOGB.

\begin{figure}
\rotatebox{270}{
\resizebox{0.7\columnwidth}{!}{%
  \includegraphics{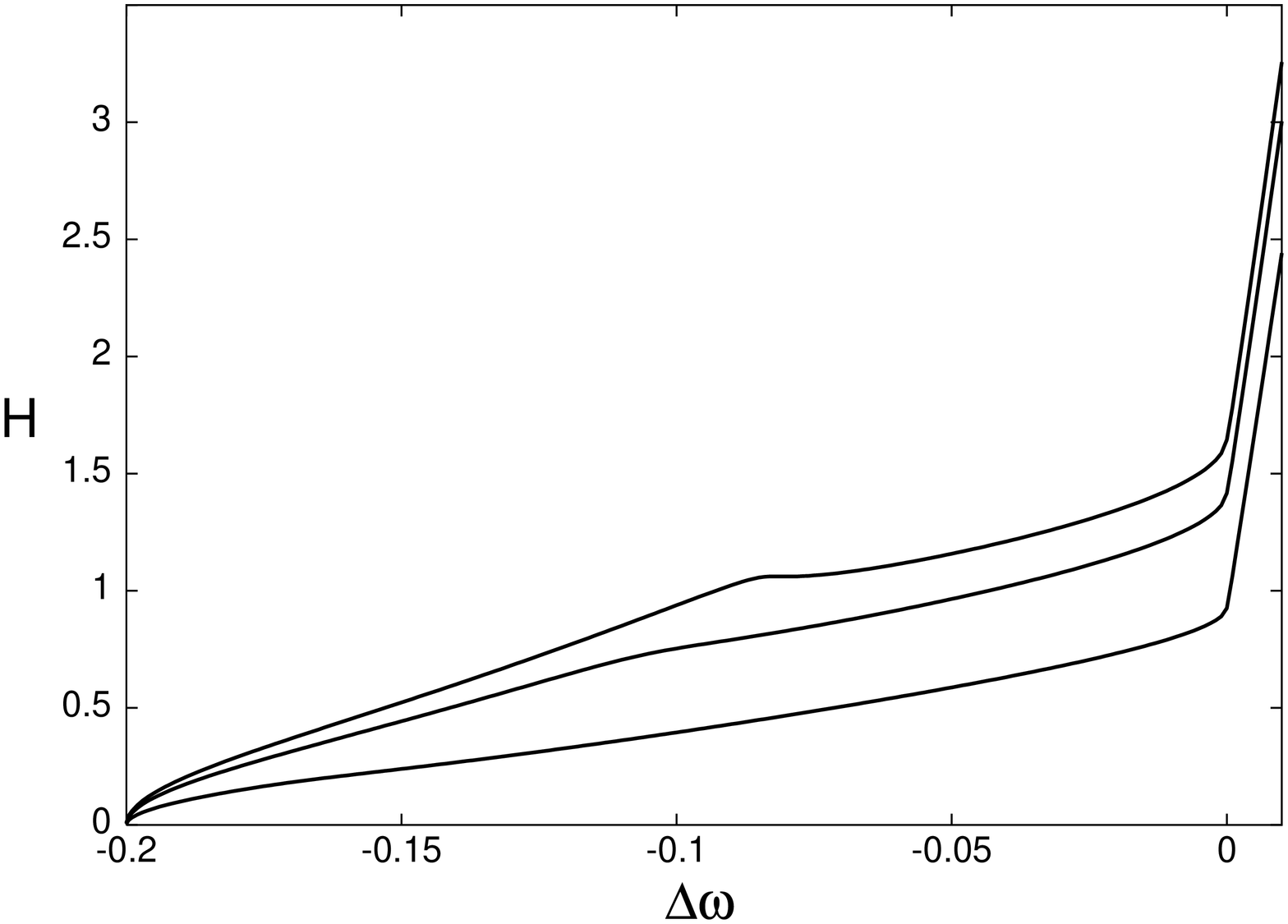}
}
}
\caption{The dependencies of the Hamiltonian energy $H$ (\ref{energy}) of a 
symmetric breather 
on its frequency detuning $\Delta\omega$ (\ref{dw}) for different 
values of the coupling (from top to bottom): $C=0.17$, $C=0.15$ and $C=0.1$.
In all cases $\delta^2=0.1$ and $N=242$.
}
\label{fig:enDGB-DOGB}
\end{figure}

In Fig.~\ref{fig:enDGB-DOGB} we show
the dependencies of  Hamiltonian $H$ 
(\ref{energy}) on  frequency detuning $\Delta\omega$ 
(\ref{dw}), for 
continuation of symmetric breathers from $\Delta\omega=-2\delta^2$ 
(bottom of the gap) to $\Delta\omega=0.1\delta^2$ (inside the upper 
band) at different fixed values of 
the coupling. 
At the upper gap boundary $\Delta\omega=0$, a DGB 
$\{\Uparrow(0\mathbb{O})\}_S^G$ transforms into a DOGB 
$\{\Uparrow(\uparrow\mathbb{O}\downarrow\mathbb{O})\}_S^O$, the energy of 
which rapidly grows with
frequency due to the tails. 

Increasing the coupling, the dependence $H(\Delta\omega)$ 
can become non-monotonous (e.g.\ $C=0.17$ in 
Fig.~\ref{fig:enDGB-DOGB}), which is connected to an additional 
transition from discrete- to continuous-like breather solution 
\cite{we}. Close to the AC limit 
($C\ll 1$ or $\omega_b\gg\omega_1$),
the lower-band sub-field of the breather 
is well localized, and the two peaks of the upper-band 
sub-field (to the left and to the 
right of the breather center, see Fig.~\ref{fig:DGB}) are 
situated on the 'light' sites nearest to the breather center. We call
this configuration 'discrete-like' breather. By contrast, 
close to the continuous limit the lower-band sub-field 
extends over many lattice sites, and the peaks of the upper band 
sub-field are located far from the breather center - 
'continuous-like' breather. Consequently, the continuation of a 
breather from the anti-continuous to the continuous limit is 
always accompanied with an infinite number of tranformations, 
where each transformation corresponds to a 'jump' of upper band 
sub-field peaks from positions $n=N_c\pm n_0$ to positions 
$n=N_c\pm(n_0+2)$ with $n_0=1,3,5,...$ for symmetric breathers and
$n_0=2,4,6,...$ for antisymmetric breathers.

\begin{figure*}
\rotatebox{270}{
\resizebox{0.95\columnwidth}{!}{%
  \includegraphics{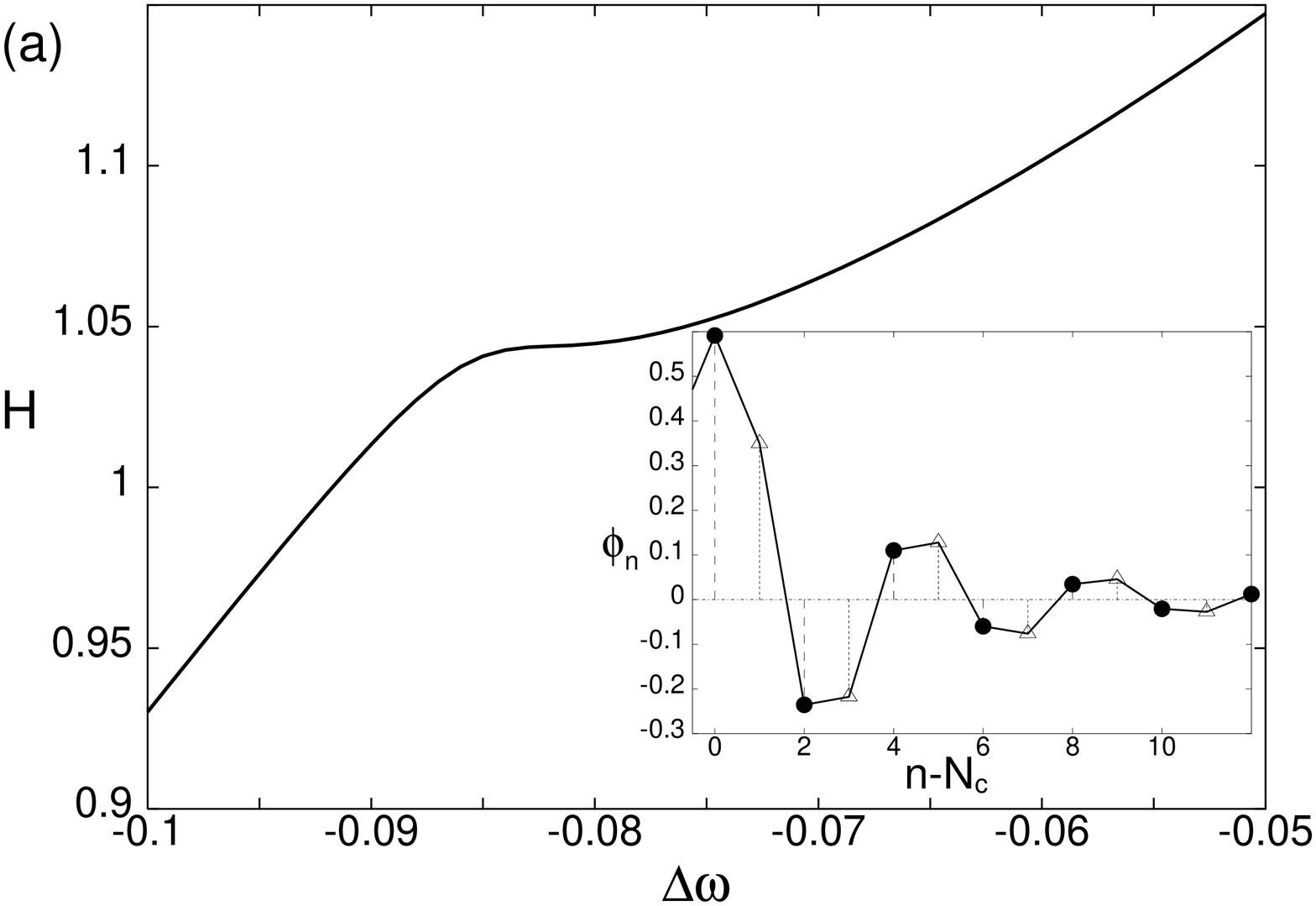}
  \includegraphics{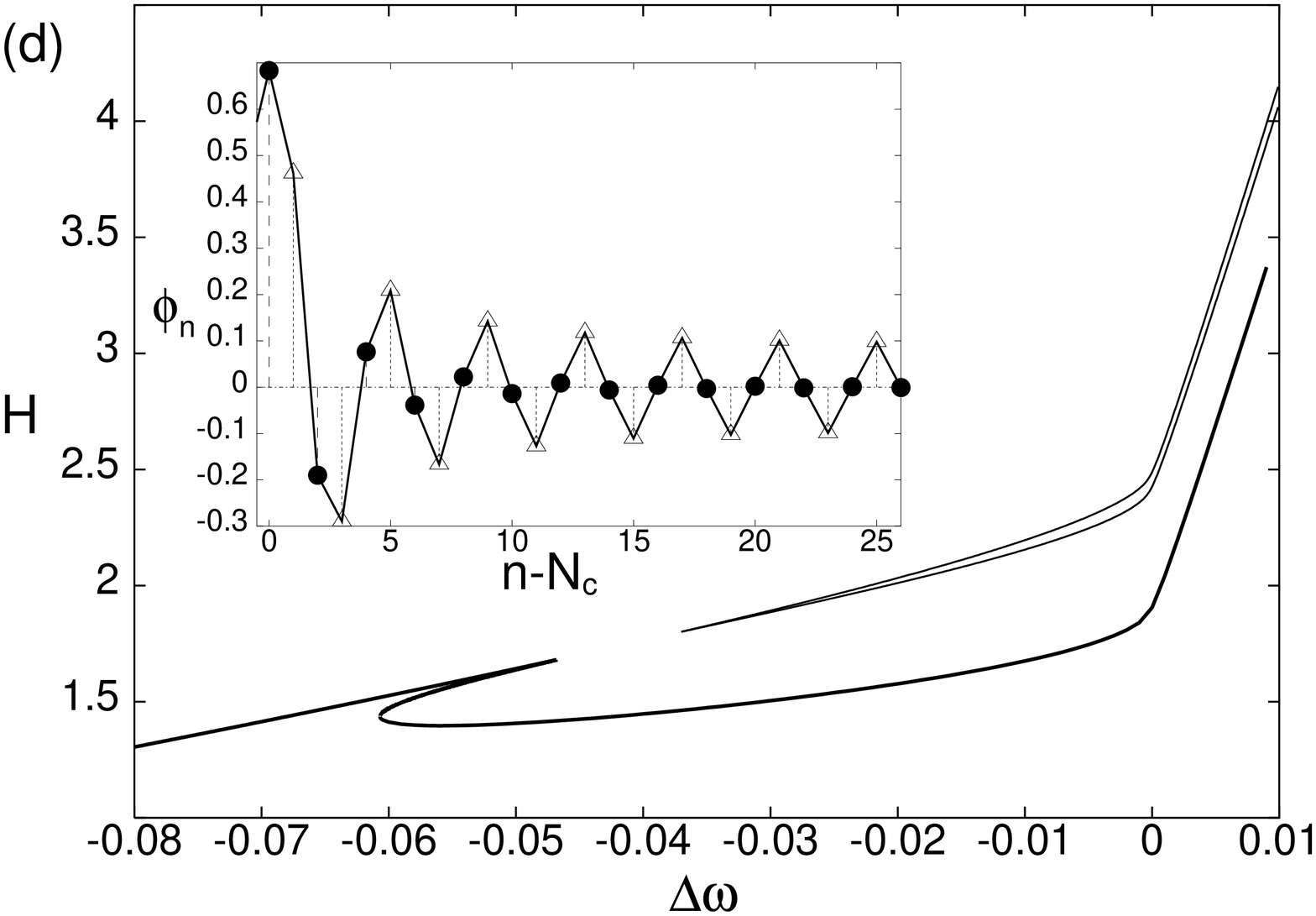}
}
}
\rotatebox{270}{
\resizebox{0.95\columnwidth}{!}{%
  \includegraphics{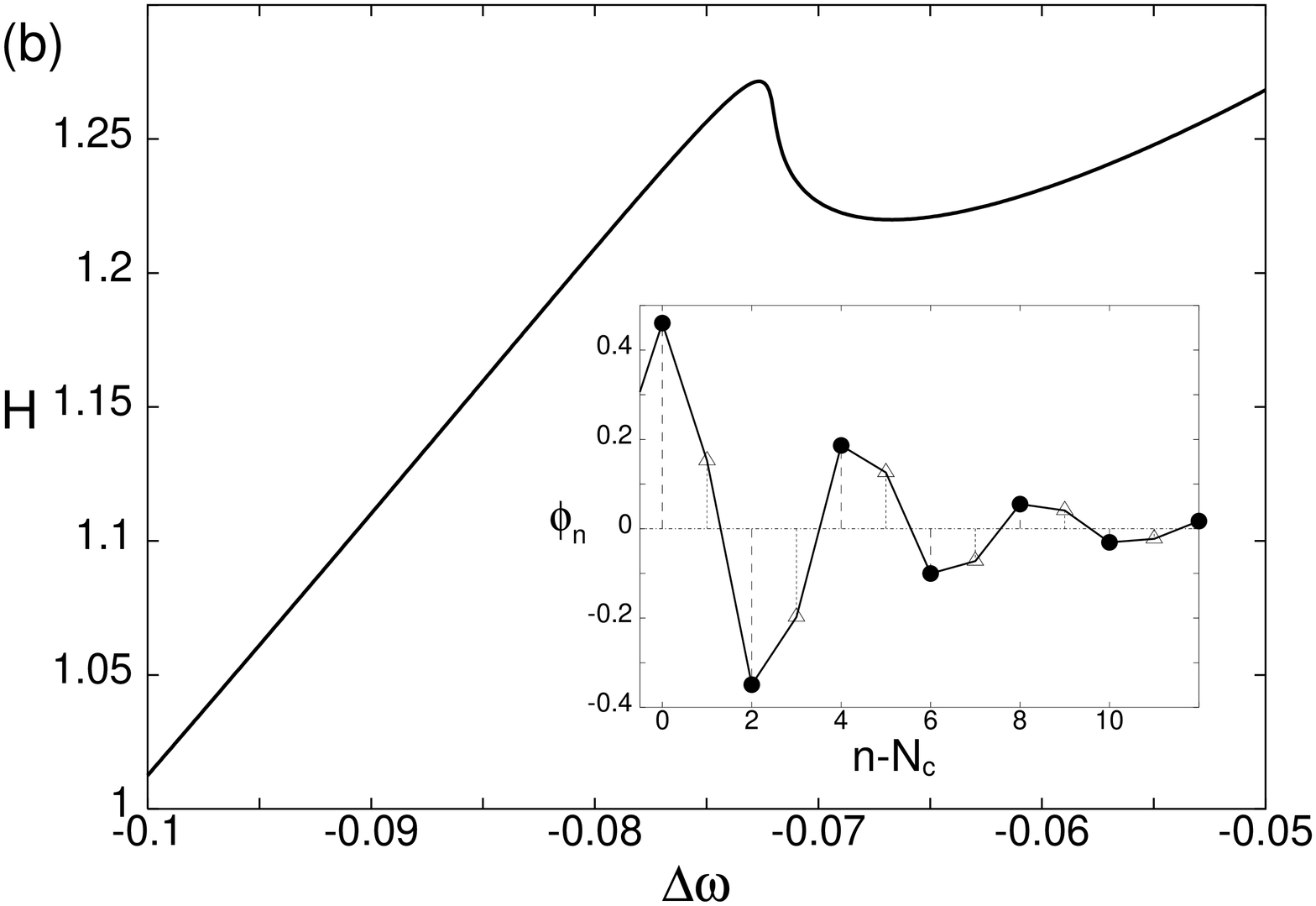}
  \includegraphics{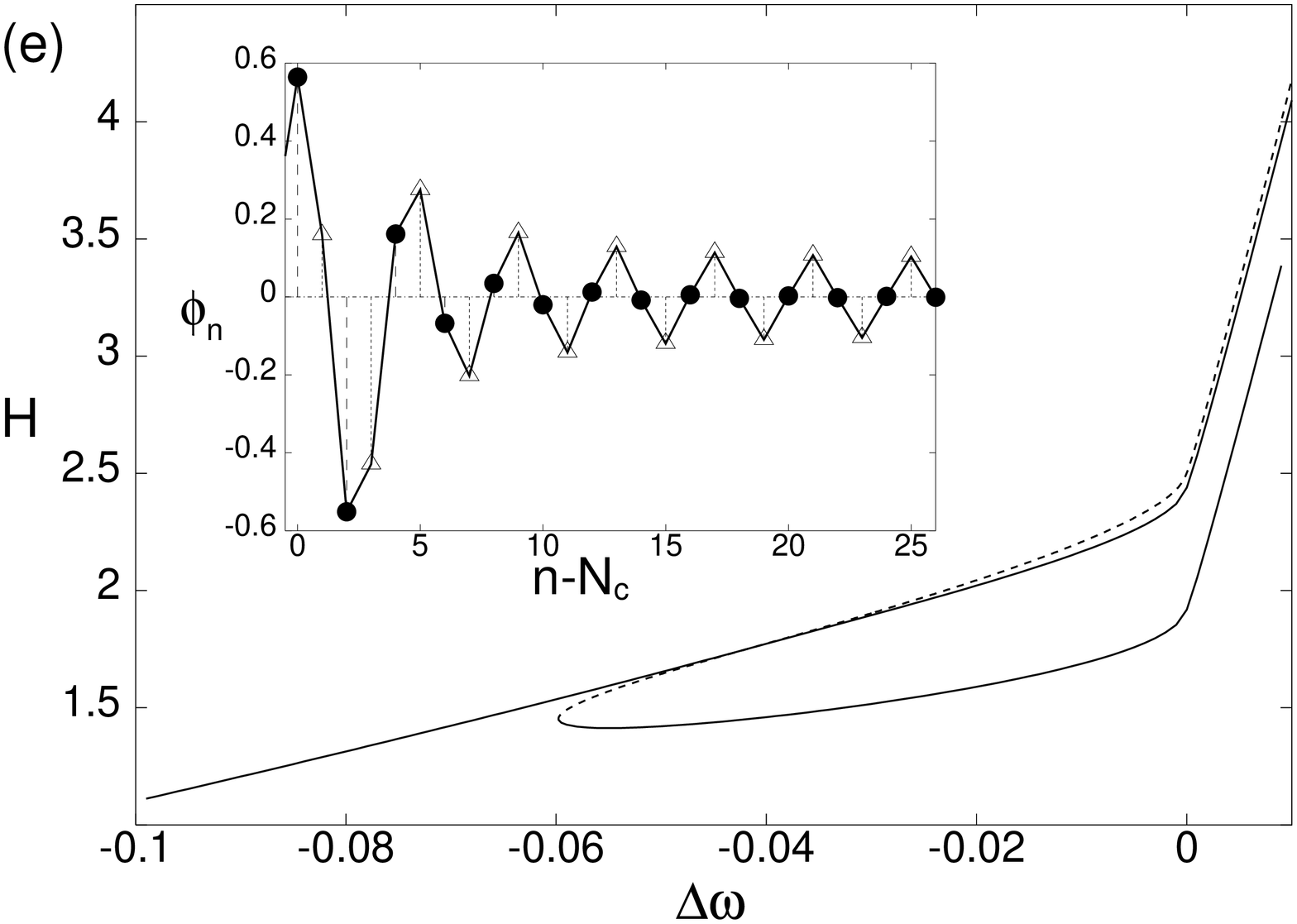}
}
}
\rotatebox{270}{
\resizebox{0.95\columnwidth}{!}{%
  \includegraphics{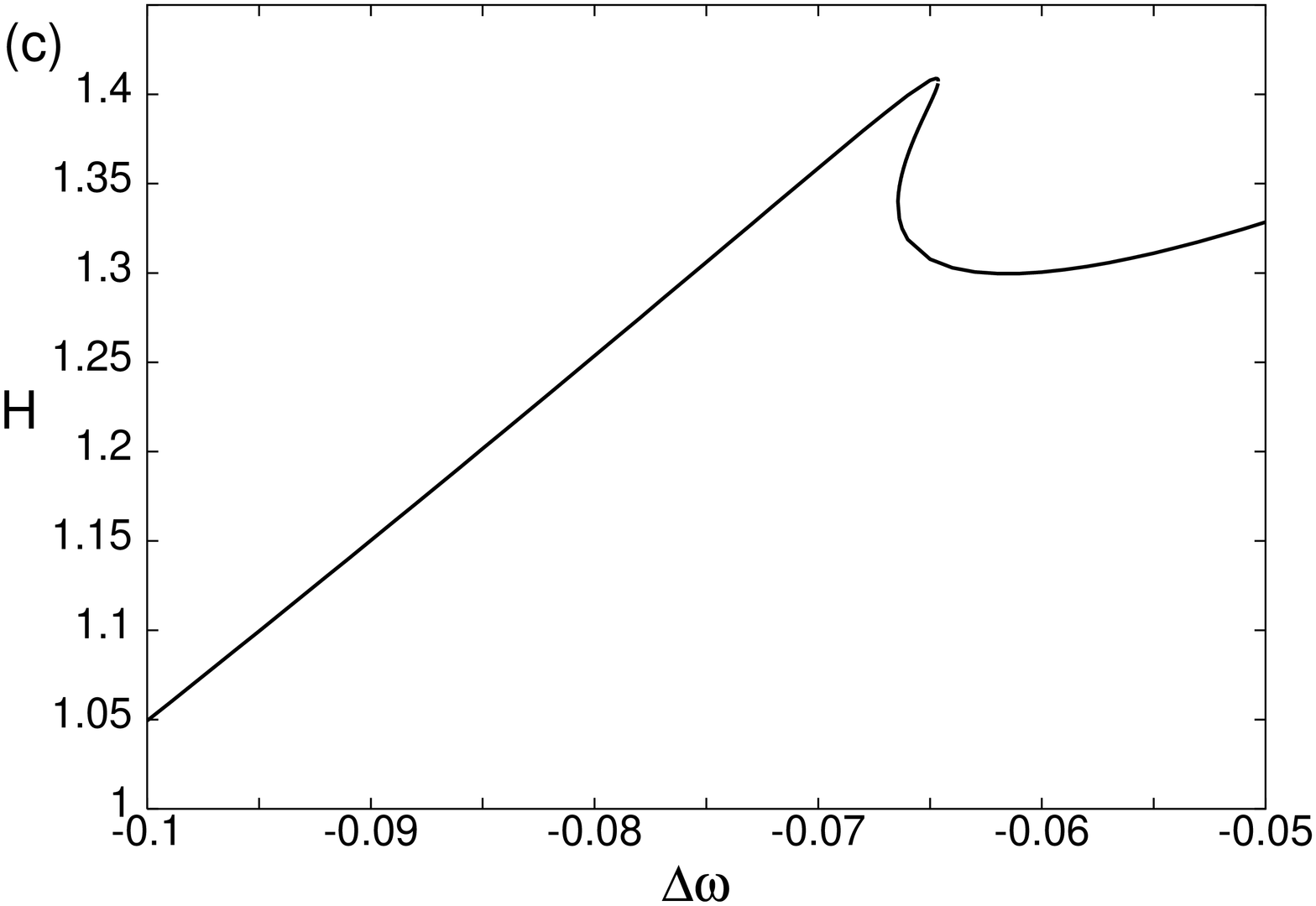}
  \includegraphics{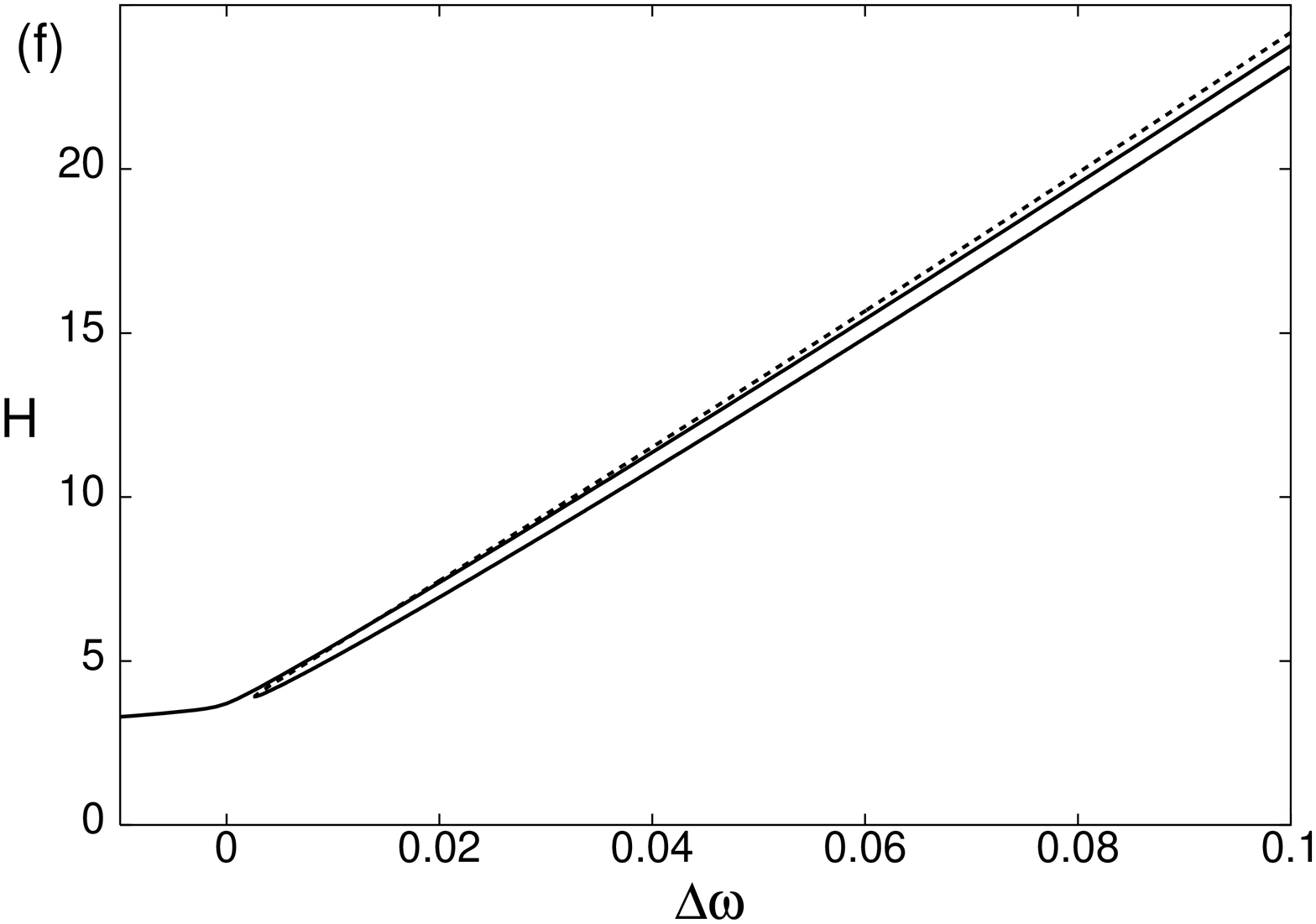}
}
}
\caption{Transitions from 'discrete-like' (larger $\Delta\omega$) 
to 'continuous-like' (smaller $\Delta\omega$) 
breathers. The dependencies of breather Hamiltonian 
energies $H$ (\ref{energy}) on 
frequency detunings $\Delta\omega$ (\ref{dw}) at different fixed 
values of the coupling: (a) $C=0.169$; (b) $C=0.18$; (c) $C=0.185$;
(d) $C=0.191$; (e) $C=0.192$; (f) $C=0.28$. 
The insets in (a),(b),(d) and (e) show the structure of 'discrete-like' 
DGB ($C=0.169, \Delta\omega=-0.05$), 
'continuous-like' 
DGB ($C=0.18, \Delta\omega=-0.1$), 'discrete-like' 
DOGB ($C=0.191, \Delta\omega=0.01$, lower curve) and 'continuous-like' 
DOGB ($C=0.192, \Delta\omega=0.01$, middle curve), respectively. 
In all cases, $\delta^2=0.1$ and $N=242$.
}
\label{fig:en_Dis-Con}
\end{figure*}

In 
Fig.~\ref{fig:en_Dis-Con} different regimes of transitions from 
'discrete-like' breathers with peaks of the upper band sub-field 
at $n=N_c\pm 1$ (see insets in Fig.~\ref{fig:en_Dis-Con}(a),(d))
to 'con\-ti\-nu\-ous-like' breathers 
with peaks of the upper band sub-field at $n=N_c\pm 3$ 
(insets in Fig.~\ref{fig:en_Dis-Con}(b),(e))
are demonstrated in the 
plane $(H,\Delta\omega)$ for continuation in frequency 
at different fixed values of coupling. For 
small enough $C$, a 'discrete-like' brea\-ther smoothly transforms into a 
'continuous-like' one while decreasing the frequency towards the lower gap
boundary ($\Delta\omega\rightarrow -2\delta^2$). At a certain 
critical coupling $C_1$ the derivative $\partial H/\partial
\omega_b$ turns to zero at the transition point (see 
Fig.~\ref{fig:en_Dis-Con}(a)). Further increase of the coupling will 
lead to the appearence of a part of the $H(\Delta\omega)$ curve with a
negative slope in the transition region (see 
Fig.~\ref{fig:en_Dis-Con}(b)). As a consequence, discrete-like and 
continuous-like breathers will possess real instabilities of the 
Vakhitov-Kolokolov type \cite{Vakh} near the transition point. These 
instabilities are produced by the collisions of the $N2$ localized 
eigenmode with its complex conjugate (see 
Sec.~\ref{subsec_DGB}).

At the second critical value $C_2>C_1$ the slope of the 'unstable' part of 
the curve $H(\Delta\omega)$ will become vertical, and 
for $C>C_2$ it bends over and the dependence  $H(\Delta\omega)$ becomes
multi-valued in the transition region
(Fig.~\ref{fig:en_Dis-Con}(c) and lower curve in 
Fig.~\ref{fig:en_Dis-Con}(d)).
In this regime the transition between discrete-like and continuous-like 
breathers  is possible only via an 
intermediate solution. 

For large enough coupling, multi-breather solutions get
involved in the transition from discrete-like to 
continuous-like breathers. 
In the upper right part of Fig.~\ref{fig:en_Dis-Con}(d) 
the bifurcation loop for the breathers 
$\{\Uparrow 0 \Downarrow (\downarrow \mathbb{O} \uparrow 
\mathbb{O})\}_S^O$ (lower part of the loop) and 
$\{\Uparrow \downarrow \Downarrow (\downarrow \mathbb{O} \uparrow 
\mathbb{O})\}_S^O$ (upper part) is shown. Being constructed 
at frequencies above the linear spectrum, these DOGBs were 
continued in frequency down to the values inside the gap\footnote{Note 
that these two solutions cannot 
be continued in frequency inside the gap at small values of coupling. 
Therefore, it is impossible to associate any
'gap' coding sequences to them.}, where  
they bifurcate with 
each other close to the bifurcation point of the continuous-like DGB with 
the 
intermediate breather. 
When the coupling reaches a third 
critical value $C_3$, these two bifurcation points coincide 
(Fig.~\ref{fig:en_Dis-Con}(e)), and the four bifurcating 
solutions are identical in that point. 
For $C>C_3$ 
the bifurcation picture drastically changes. The curves for a
continuous-like DGB $\{\Uparrow (0\mathbb{O})\}_S^G$ and for a 
DOGB $\{\Uparrow 0 \Downarrow (\downarrow \mathbb{O} \uparrow 
\mathbb{O})\}_S^O$ coincide, so that a single-breather solution
DGB $\{\Uparrow (0\mathbb{O})\}_S^G$ can be smoothly continued up to 
frequencies inside the linear spectrum (upper solid line in
Fig.~\ref{fig:en_Dis-Con}(e)). 
At the same time the curves for the intermediate breather, 
connecting the continuous-like and discrete-like DGBs, and for the 
DOGB $\{\Uparrow \downarrow \Downarrow (\downarrow \mathbb{O} \uparrow 
\mathbb{O})\}_S^O$ also coincide (dashed line in 
Fig.~\ref{fig:en_Dis-Con}(e)). Therefore, the discrete-like DGB 
$\{\Uparrow (0\mathbb{O})\}_S^G$ connected to the 
DOGB $\{\Uparrow(\uparrow\mathbb{O}\downarrow\mathbb{O})\}^O_S$
now bifurcates with the DOGB
$\{\Uparrow \downarrow \Downarrow (\downarrow \mathbb{O} \uparrow 
\mathbb{O})\}_S^O$ at frequencies inside the gap 
(lower solid line in Fig.~\ref{fig:en_Dis-Con}(e)).

Further increase of $C$ will lead to the movement of the bifurcation 
point of the discrete-like DGB $\{\Uparrow (0\mathbb{O})\}_S^G$ and 
the DOGB
$\{\Uparrow \downarrow \Downarrow (\downarrow \mathbb{O} \uparrow 
\mathbb{O})\}_S^O$ towards the upper boundary of the gap. Finally, 
above the forth critical value of the coupling $C\ge C_4$, the 
discrete-like DOGB $\{\Uparrow (\uparrow \mathbb{O} 
\downarrow\mathbb{O})\}_S^O$, as well as the DOGB 
$\{\Uparrow \downarrow \Downarrow (\downarrow \mathbb{O} \uparrow 
\mathbb{O})\}_S^O$ cannot be continued in frequency inside the gap, 
since they bifurcate with each other above the upper boundary of the gap
 $\Delta\omega=0$ 
(see Fig.~\ref{fig:en_Dis-Con}(f)). Consequently, for such large
coupling only continuous-like DGBs 
$\{\Uparrow (0\mathbb{O})\}_S^G$ can exist in the gap.

\begin{figure*}
\rotatebox{270}{
\resizebox{0.73\columnwidth}{!}{%
  \includegraphics{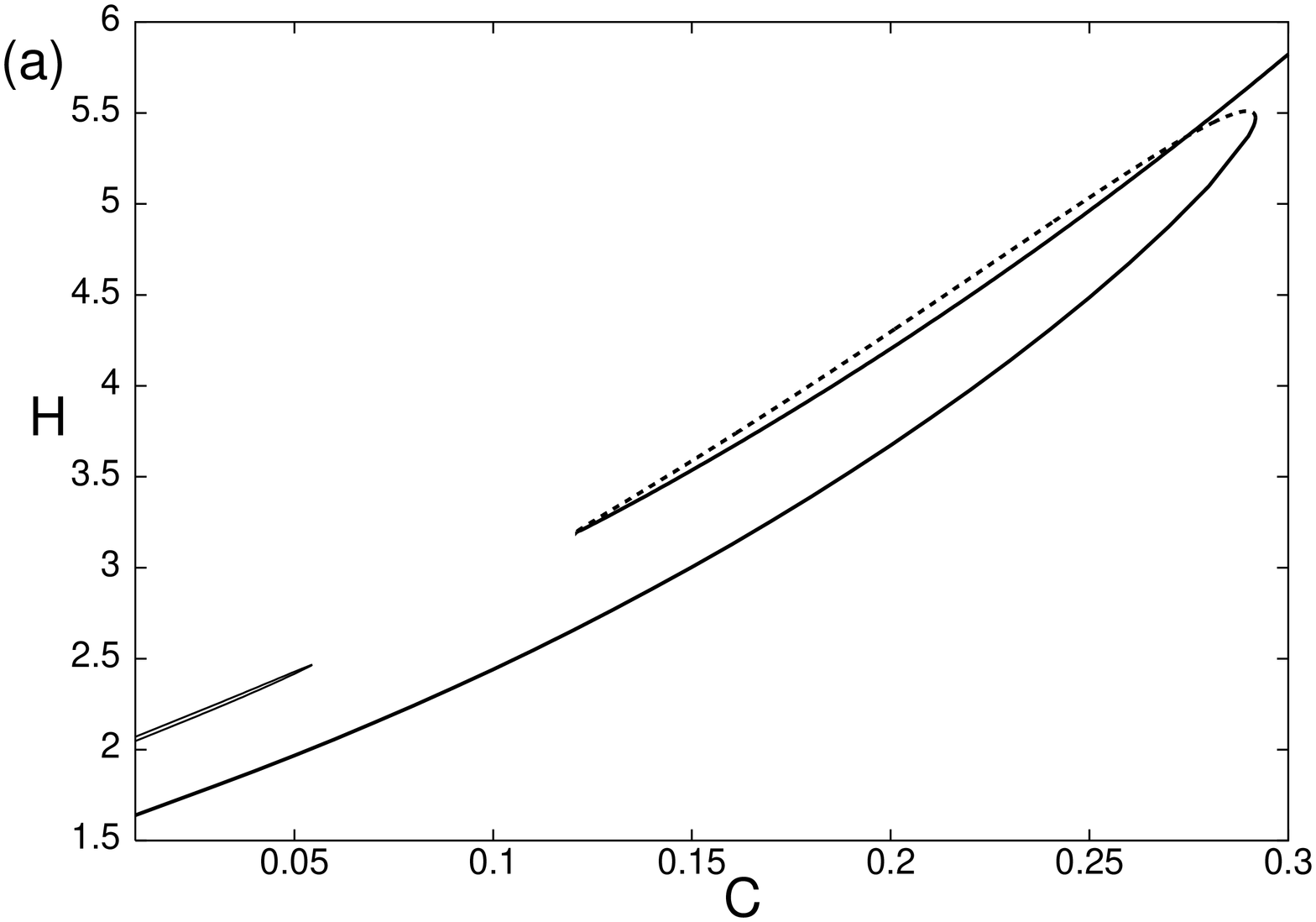}
}
}
\rotatebox{270}{
\resizebox{0.73\columnwidth}{!}{%
  \includegraphics{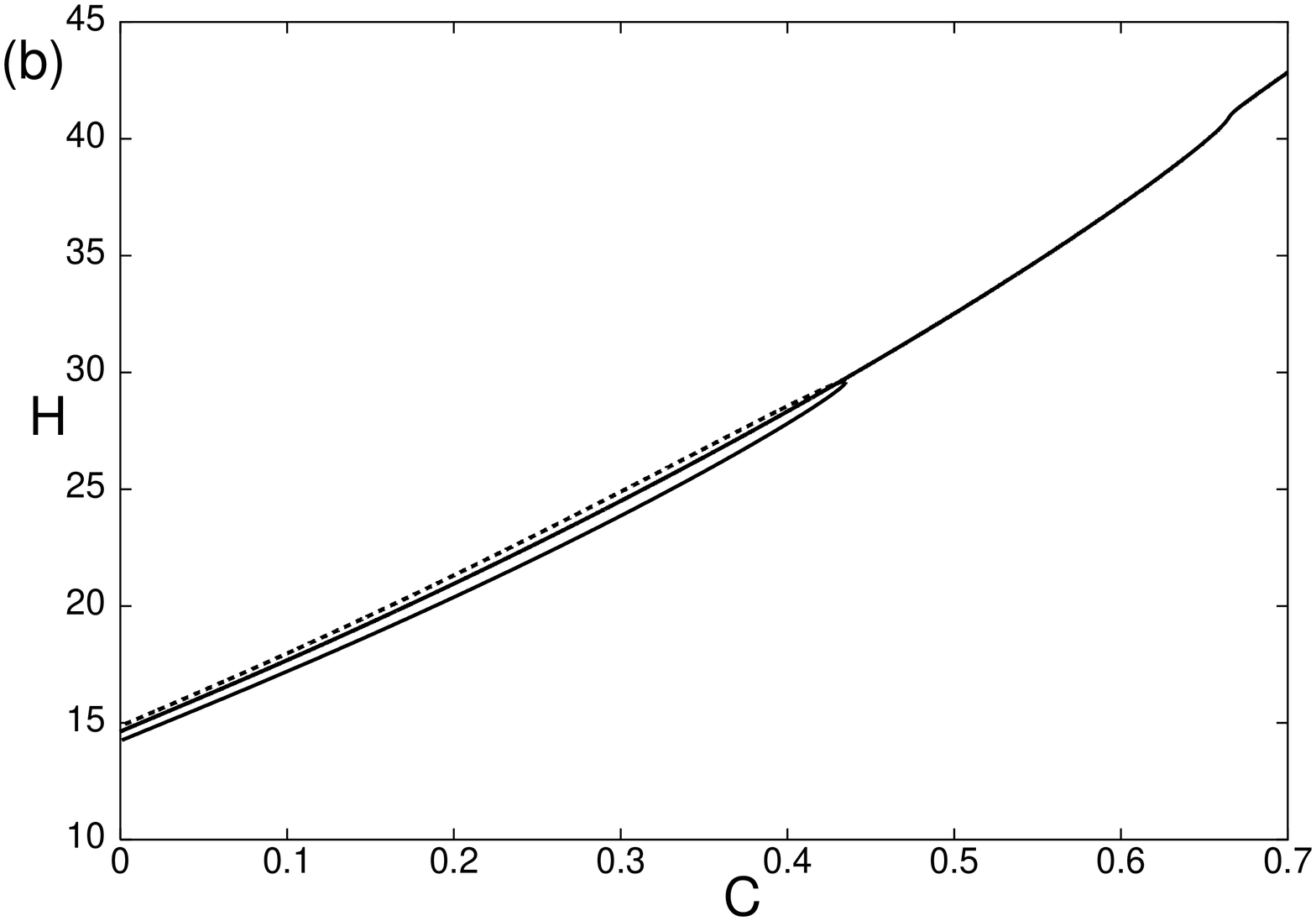}
}
}
\caption{Transitions from 'discrete-like' to 'continuous-like' 
out-gap breathers (see text for explanations). 
The dependencies of breather Hamiltonian 
energies $H$ (\ref{energy}) on 
the coupling $C$ at different fixed 
values of the frequency detuning $\Delta\omega$: 
(a) $\Delta\omega=0.1\cdot\delta^2$ and (b) $\Delta\omega=\delta^2$. 
The values of the system parameters are the same as in 
Fig.~\ref{fig:en_Dis-Con}.
}
\label{fig:en_Dis-Con2}
\end{figure*}

Similar transitions from discrete- to continuous-like  breather 
solutions also occur for the continuation in coupling of DGBs and DOGBs at 
a fixed value of the frequency detuning $\Delta\omega$ (\ref{dw}). 
In the lower half of the gap 
($-2\delta^2\le\Delta\omega\le -\delta^2$) the transitions are 
smooth. In the upper half of the gap the energy 
curve $H(C)$ bends over in a similar way as in Fig.~\ref{fig:en_Dis-Con}(c), 
and the discrete-like breather 
transforms into a continuous-like one 
via an intermediate solution. Above the upper gap 
boundary ($\Delta\omega>0$) the DOGBs 
$\{\Uparrow 0 \Downarrow (\downarrow \mathbb{O} \uparrow 
\mathbb{O})\}_S^O$ and 
$\{\Uparrow \downarrow \Downarrow (\downarrow \mathbb{O} \uparrow 
\mathbb{O})\}_S^O$ get involved in the transition 
between discrete- and continuous-like solutions (Fig.~\ref{fig:en_Dis-Con2}). 
In the lower left part of Fig.~\ref{fig:en_Dis-Con2}(a) the 
bifurcation loop for these two solutions is shown, while the upper 
right part shows the loop for transition from the 
continuous-like (upper solid 
line) into the discrete-like (lower solid line) DOGBs 
$\{\Uparrow(\uparrow\mathbb{O}\downarrow\mathbb{O})\}^O_S$ via the 
intermediate out-gap breather (dashed line).
At a certain critical  $\Delta\omega$
the bifurcation point of the DOGBs
$\{\Uparrow 0 \Downarrow (\downarrow \mathbb{O} \uparrow 
\mathbb{O})\}_S^O$ and 
$\{\Uparrow \downarrow \Downarrow (\downarrow \mathbb{O} \uparrow 
\mathbb{O})\}_S^O$
coincide with that of the continuous-like and
intermediate out-gap breathers. 
At higher frequencies the continuous-like DOGB 
can be continued smo\-othly in coupling down to $C=0$, 
with the coding sequence 
$\{\Uparrow 0 \Downarrow (\downarrow \mathbb{O} \uparrow 
\mathbb{O})\}_S^O$ (upper solid line in Fig.~\ref{fig:en_Dis-Con2}(b)),
while the discrete-like DOGB 
$\{\Uparrow(\uparrow\mathbb{O}\downarrow\mathbb{O})\}_S^O$ bifurcates 
with the DOGB 
$\{\Uparrow \downarrow \Downarrow (\downarrow \mathbb{O} \uparrow 
\mathbb{O})\}_S^O$ (lower solid and dashed lines 
in Fig.~\ref{fig:en_Dis-Con2}(b), respectively). An analogous 
scenario will then be
repeated at higher frequencies, for the transition from 
a solution with peaks of the upper band sub-field at 
$n=N_c\pm 3$ to one with 
 upper band sub-field peaks at $n=N_c\pm 5$. In 
Fig.~\ref{fig:en_Dis-Con2}(b) one can see a bending of the $H(C)$ curve 
for the continuous-like solution at $C\approx 0.68$, associated with 
this subsequent transition.

A similar switching of the bifurcation scenario with change of 
coupling was also observed for 'phantom' brea\-thers associated with 
second-harmonic
resonances in mono\-atomic KG chains \cite{phantom}.

\subsection{Bifurcations of DOGBs with different tails}
\label{subsec_freq2}

\begin{figure*}
  \psfrag{\{+\(o0\)\}sg}{$\{\Uparrow(0\mathbb{O})\}_S^G$}
  \psfrag{\{+\(+0+0\)\}so}{$\{\Uparrow(\uparrow\mathbb{O}\downarrow\mathbb{O})\}_S^O$}
  \psfrag{\{++\(0o\)\}so}{$\{\Uparrow\uparrow(0\mathbb{O})\}_S^O$}
  \psfrag{\{++O+\(0o\)\}so}{$\{\Uparrow\uparrow\mathbb{O}\uparrow(\mathbb{O}0)\}_S^O$}
  \psfrag{\{+\(o0\)\}so}{$\{\Uparrow(0\mathbb{O})\}_S^O$}
  \psfrag{\{+-\(0o\)\}so}{$\{\Uparrow\downarrow(\mathbb{O}0)\}_S^O$}
  \psfrag{\{\(0+0-\)\}so}{$\{(\mathbb{O}\uparrow\mathbb{O}\downarrow)\}_S^O$}
  \includegraphics[scale=0.34,angle=270]{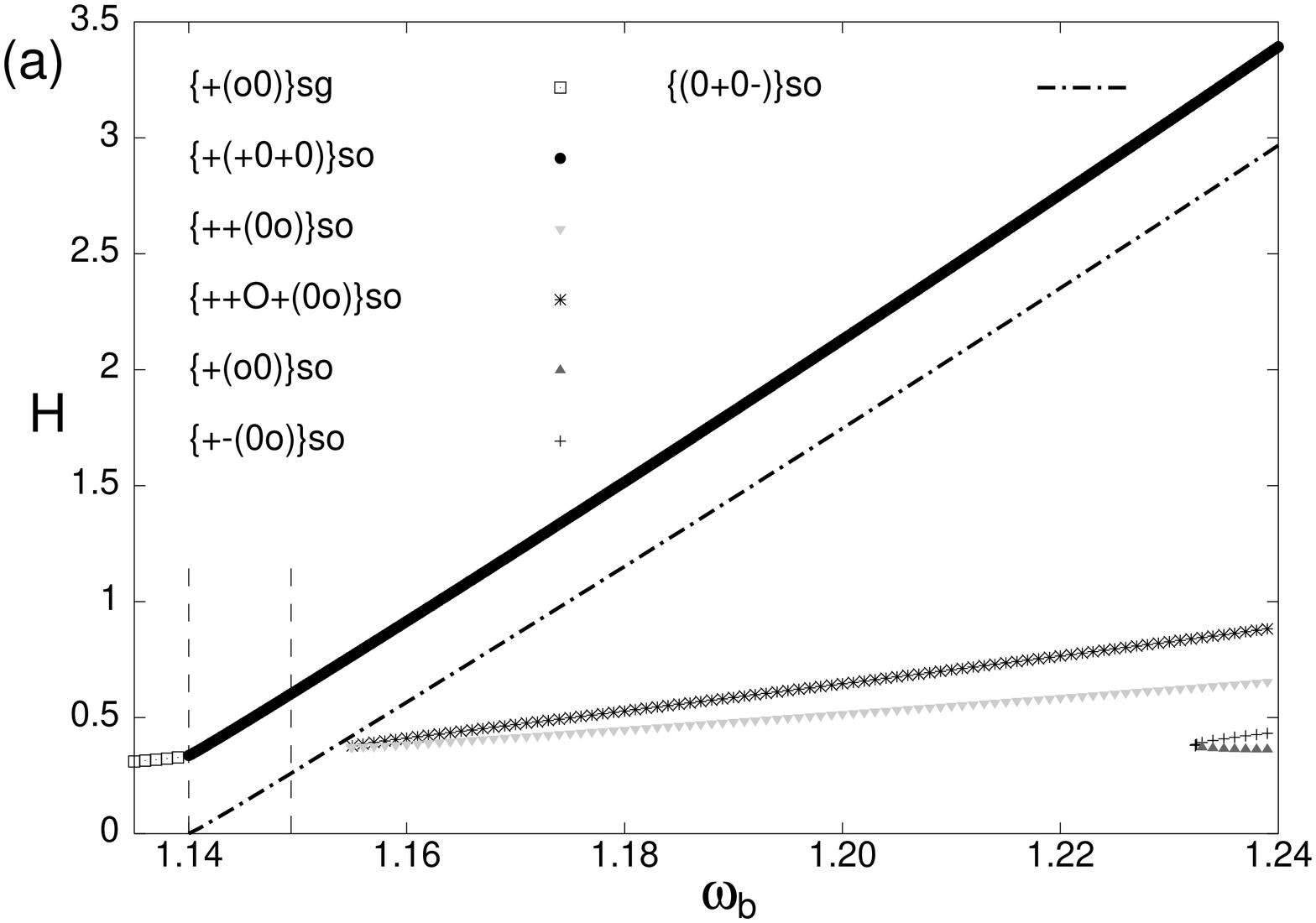}
  \includegraphics[scale=0.34,angle=270]{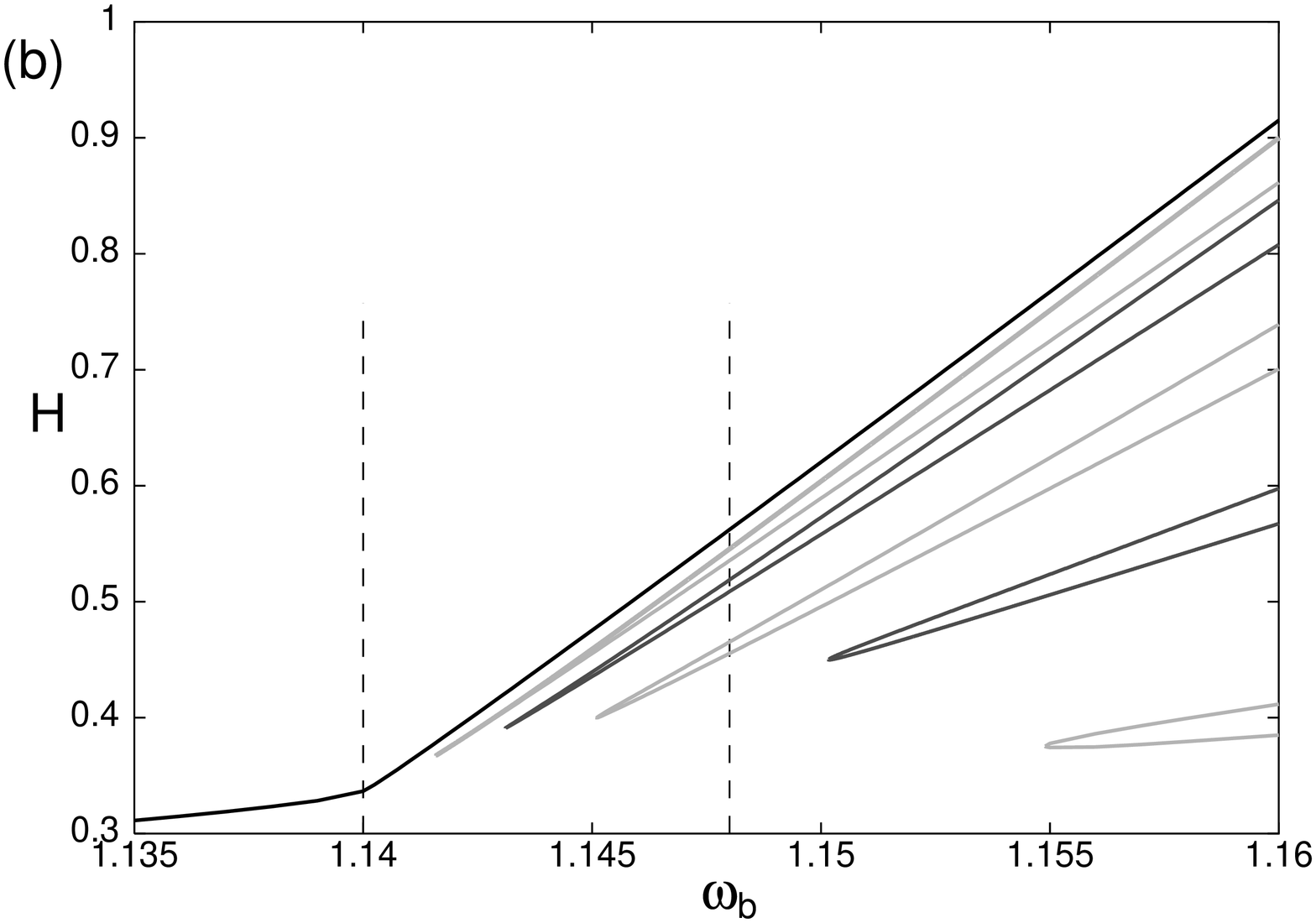} 
\caption{Hamiltonian 
 $H$ (\ref{energy}) vs.  
frequency $\omega_b$ 
for different solutions. 
Vertical lines show the boundaries of the upper branch of
the linear spectrum  $\omega_2=1.14$, 
$\omega_u \approx 1.148$.
Curves in (b) show, from top to bottom: 
continuation DGB $\{\Uparrow (0\mathbb{O})\}_S^G$  
$\rightarrow$ DOGB 
$\{\Uparrow(\uparrow\mathbb{O}\downarrow\mathbb{O})\}_S^O$, 
bifurcation loops for DOGBs with 1,2,4,8 and 12 'defects' 
(see text for explanation). $\delta^2=0.1, C=0.02, N=50$.
}
\label{fig:bifur}
\end{figure*}

As  described above, gap breathers
$\{\Uparrow (0\mathbb{O})\}_S^G$ gain non-decaying tails with wave number 
$q=\pi/2$
when continued into 'normal' DOGBs above the upper gap boundary,
and their energies
rapidly grow. 
However, there exist also 
localized brea\-thers with frequencies above the linear
spectrum -- 'on-top' brea\-thers (DOTB), 
with exponentially decaying tails of wave number $q=\pi$ (for
stationary solutions). 
The simplest example is the single-site DOTB with coding sequence
$\{\Uparrow (0\mathbb{O})\}_S^O$.
Attempting a continuation of this solution in frequency down to
the linear spectrum, it bifurcates with another on-top
breather $\{\Uparrow \downarrow (\mathbb{O}0)\}_S^O$
rather far from the 'top' frequency
$\omega_u$ of linear waves (Fig.~\ref{fig:bifur}(a), lowest curves).
Another DOTB
$\{\Uparrow \uparrow (\mathbb{O}0)\}_S^O$
(three-site breather) can be continued much
further down in frequency, bifurcating with the DOTB
$\{\Uparrow \uparrow \mathbb{O} \uparrow (\mathbb{O} 0)\}_S^O$
close to the point $\omega=\omega_u$ 
(Fig.~\ref{fig:bifur}(a), middle curves).
The slope of the curve $H(\omega_b)$ for the DOTB
$\{\Uparrow \uparrow (\mathbb{O}0)\}_S^O$ is approximately
the same as for the DGB $\{\Uparrow (0\mathbb{O})\}_S^G$.
In that sense this on-top breather,
rather than the DOTB $\{\Uparrow (0\mathbb{O})\}_S^O$,
can be considered as the 'on-top counterpart' of the gap breather 
(however, this is true only for small coupling, see below).

Besides the 'normal' DOGBs, another type of out-gap breather
with tail wave number  $q=\pi/2$ 
exists above the gap -- 
the DOGB $\{(\mathbb{O}\uparrow\mathbb{O}\downarrow)\}_S^O$ (and its 
antisymmetric counterpart 
$\{0(\mathbb{O}\uparrow\mathbb{O}\downarrow)\}_A^O$). The
dependence 
$H(\omega_b)$ for the continuation in frequency of this DOGB at fixed 
coupling $C=0.02$ is shown in Fig.~\ref{fig:bifur}(a) (dash-dotted line). 
Originating from the linear wave with $q=\pi/2$
at the upper gap boundary,  it 
represents a kink-like 
excitation of 'light' sites ('phase domain wall') accompanied by 
localized excitation of 'heavy' sites at non-zero
coupling,  and
corresponds to the kink-like out-gap soliton in the 
continuous limit (e.g. \cite{Chub,mePRE}). However, above
a critical coupling $C_{cr}\approx 0.188$ 
the DOGB $\{(\mathbb{O}\uparrow\mathbb{O}\downarrow)\}_S^O$ does 
not exist at $\Delta\omega\rightarrow 0$, since being continued in frequency 
down towards the upper gap boundary at fixed 
coupling $C>C_{cr}$, it bifurcates with the solution
$\{\Downarrow \downarrow (\mathbb{O} \downarrow \mathbb{O} \uparrow)\}_S^O$ 
at $\Delta\omega>0$. Instead, at $C>C_{cr}$ the solution 
$\{\Downarrow 0 (\mathbb{O} \downarrow \mathbb{O} 
\uparrow)\}_S^O$ can be continued  down to 
the upper gap boundary, corresponding
to the kink-like soliton in this regime. This is similar to
the switching of the coding sequence for the continuous-like 
'normal' DOGB from 
$\{\Uparrow (\uparrow \mathbb{O} \downarrow \mathbb{O}) \}_S^O$ to
$\{\Uparrow 0 \Downarrow  (\downarrow \mathbb{O} \uparrow \mathbb{O}) 
\}_S^O$ discussed in Sec.~\ref{subsec_freq}.

Apart from DOGBs with tail wave numbers $q=\pi/2$ and DOTBs with $q=\pi$, 
one can
also construct DOGBs with any other tail structure, if
the chosen wave number is allowed for the particular system size.
As the coding sequences become 
complicated,  we introduce  a
simplified notation. Restricting to symmetric breather configurations we 
omit the
subscripts $S$, and likewise the superscripts $O$ will be omitted 
as all DOGBs are constructed at frequencies above the gap.
For DOGBs with frequencies inside the linear spectrum
the upper band sub-field ('light' amplitudes) should have non-zero tails
because of resonances with linear waves. Therefore, the codes
for 'light' amplitudes should be, generally, non-zero.
Restricting to DOGBs which are 'single-site' in the lower band sub-field, 
all codes for 'heavy' sites are zero except for the central excited site, 
and the codes for  'heavy' amplitudes can be omitted.
For example, the simplified coding sequence for a
DOGB $\{\Uparrow(0\mathbb{O})\}_S^O$ is $[(0)]$, and for
DOGB $\{\Uparrow(\uparrow\mathbb{O}\downarrow\mathbb{O})\}_S^O$ is
$[(\uparrow\downarrow)]$ (we use different types of brackets to
distinguish normal and simplified coding sequences).

To
understand the total bifurcation picture of DOGBs with
different tails, we construct new solutions by inserting
'defects' into the tail structure of a DOGB $[(\uparrow\downarrow)]$.
Considering this sequence as the 'right' one, we  replace the
'right' code at some particular place $k$ with a 'wrong' one together with
introducing an additional phase shift, i.e. a change of the signs of
all the subsequent codes (i.e. we \emph{insert} a defect into the 'right'
coding sequence to obtain a standing wave with wave number smaller
than $q=\pi/2$).
For instance, denoting the number of codes in the simplified coding
sequence as $N_L$ (i.e. the number of 'light' sites to the right of the 
breather center, e.g.\ $N_L=13$ if $N=50$),
 the solution with one 'defect' at the position $k$
 (in the {\em simplified} coding
sequence, corresponding to site $n=N_c+2k-1$)
has
the simplified coding sequence:
$[(r_i)_{i=1,2,...,k-1},w_k,(-r_i)_{i=k+1,k+2,...,N_L}]$. Here $r_i$
is the
'right' code at the $i$-th position, and $w_k$ is a 'wrong' code
($w_k$ is '$0$' or '$\downarrow$' if $r_k$ is '$\uparrow$', $w_k$ is
'$0$' or '$\uparrow$' if $r_k$ is
'$\downarrow$'). A DOGB $[0(\uparrow\downarrow)]$ is an example of a solution
with a defect at the first position, while a DOGB
$[\uparrow\downarrow\uparrow\uparrow(\downarrow\uparrow)]$
has a defect at the 4th postition. Similarly DOGBs
with more defects can be constructed. A
single-site DOTB $[(0)]$ can be considered as a solution
with $N_L$ 'defects'.

We find, that being continued down in frequency towards the
linear spectrum, each  solution with 'defects'
bifurcates with a solution having the same number of 'defects', 
with coding sequences differing only
at the 'defects' nearest to the center.
For example, the DOGB $[\uparrow\uparrow\downarrow\uparrow 0
\downarrow\downarrow(\uparrow\downarrow)]$ with three 'defects' at
2nd, 5th and 7th positions bifurcates with the DOGB
$[\uparrow 0 \downarrow\uparrow 0
\downarrow\downarrow(\uparrow\downarrow)]$,
with a different type of 'defect' at the 2nd
position. The resulting bifurcation picture is shown
in Fig.~\ref{fig:bifur}(b), where bifurcation loops for solutions with
1,2,4,8 and 12 'defects' are plotted in different greyscales, while the
solid black curve corresponds to the transition DGB
$\{\Uparrow(0\mathbb{O})\}_S^G$
$\rightarrow$ DOGB $[(\uparrow\downarrow)]$.
Note that in Fig.~\ref{fig:bifur} $N_L=13$,
and the bifurcation loop for solutions with 12
'defects' in Fig.~\ref{fig:bifur}(b) is the same as 
that for the DOTBs $[\uparrow(0)]$ and $[\uparrow\uparrow(0)]$ 
in  Fig.~\ref{fig:bifur}(a).

\begin{figure}
\rotatebox{270}{
\resizebox{0.72\columnwidth}{!}{%
  \includegraphics{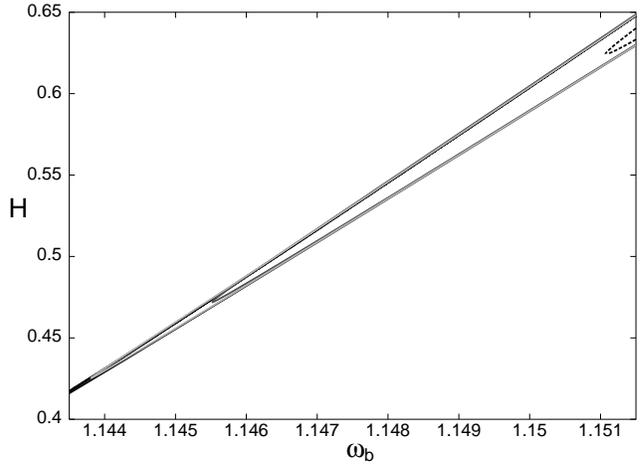}
}
}
\caption{The intrinsic structure of the bifurcation loop for the DOGB with 
one
'defect' in
Fig.~\ref{fig:bifur}(b). Dashed black, solid dark-grey, solid light-grey and solid
black lines represent solutions with
one 'defect' at 1st, 2nd, 3rd and 9th positions respectively.
}
\label{fig:bifur_intr}
\end{figure}

Each bifurcation loop corresponding to a particular number of 'defects'
has a complex intrinsic structure, as these 'defects' can be at different
positions. For each number of defects one can find some 'optimal'
configuration, so that the solution with 'defects' placed on these
'optimal' positions can be continued further down in frequency than all
other solutions with the same number of 'defects'. 
The loops shown in Fig.~\ref{fig:bifur}(b) represent such optimal 
configurations.
In Fig.~\ref{fig:bifur_intr} the intrinsic structure of the bifurcation loop
for solutions with one 'defect' at 1st, 2nd, 3rd and 9th ('optimal')
positions, respectively, is shown.  

Increasing the coupling, the bifurcation picture remains qualitatively 
the same. However, the slope of the curve $H(\omega_b)$ for the DGB 
$\{\Uparrow (0\mathbb{O})\}_S^G$ changes, so that 
for larger coupling, when the DGB has 'continuous-like' structure with 
upper-band sub-field  peaks at $n=N_c\pm 3$, its
'on-top' counterpart (with 
approximately the same slope for $H(\omega_b)$) is the DOTB
$\{\Uparrow \uparrow \mathbb{O} \downarrow (\mathbb{O} 0)\}_S^O$. 
Similar changes of the DGB's 'on-top' counterpart 
are expected for a further increase of coupling, 
as the peaks of the upper-band sub-field will 
move further away from the breather center.

The bifurcation scheme for out-gap breathers in
Fig.\ \ref{fig:bifur}(b) is qualitatively 
similar to that for phonobreathers with
different tails resulting from second-harmonic resonances in 
mono\-atomic KG chains\cite{phantom}. However,
in the modulated DNLS model out-gap breather tails
apparently always have non-zero
asymptotes, and thus we found no analogue to 'phantom' breathers
\cite{phantom} with
vanishing tails.

\subsection{Stability of out-gap breathers}
\label{subsec_DOGB}

We analyze the stability of 
DOGBs by continuation in coupling 
at fixed frequency detuning (\ref{dw}), focusing on 
 the three pairs of most important solutions: 'normal' DOGBs
$\{\Uparrow(\uparrow\mathbb{O}\downarrow\mathbb{O})\}_S^O$,
$\{0\Uparrow(\uparrow\mathbb{O}\downarrow\mathbb{O})\}_A^O$,
being the smooth continuation of symmetric and antisymmetric DGBs in
frequency above the gap for small coupling; DOTBs
$\{\Uparrow(0\mathbb{O})\}_S^O$, $\{0\Uparrow(0\mathbb{O})\}_A^O$,
being the fundamental 'on-top' breathers; and DOTBs
$\{\Uparrow\uparrow(\mathbb{O}0)\}_S^O$,
$\{0\Uparrow\uparrow(\mathbb{O}0)\}_A^O$,
being the 'on-top counterparts' of DGBs for small coupling.
We fix the 
system size $N=242$ and gap parameter $\delta^2=0.1$.
\begin{figure*}
\rotatebox{270}{
\resizebox{0.7\columnwidth}{!}{%
  \includegraphics{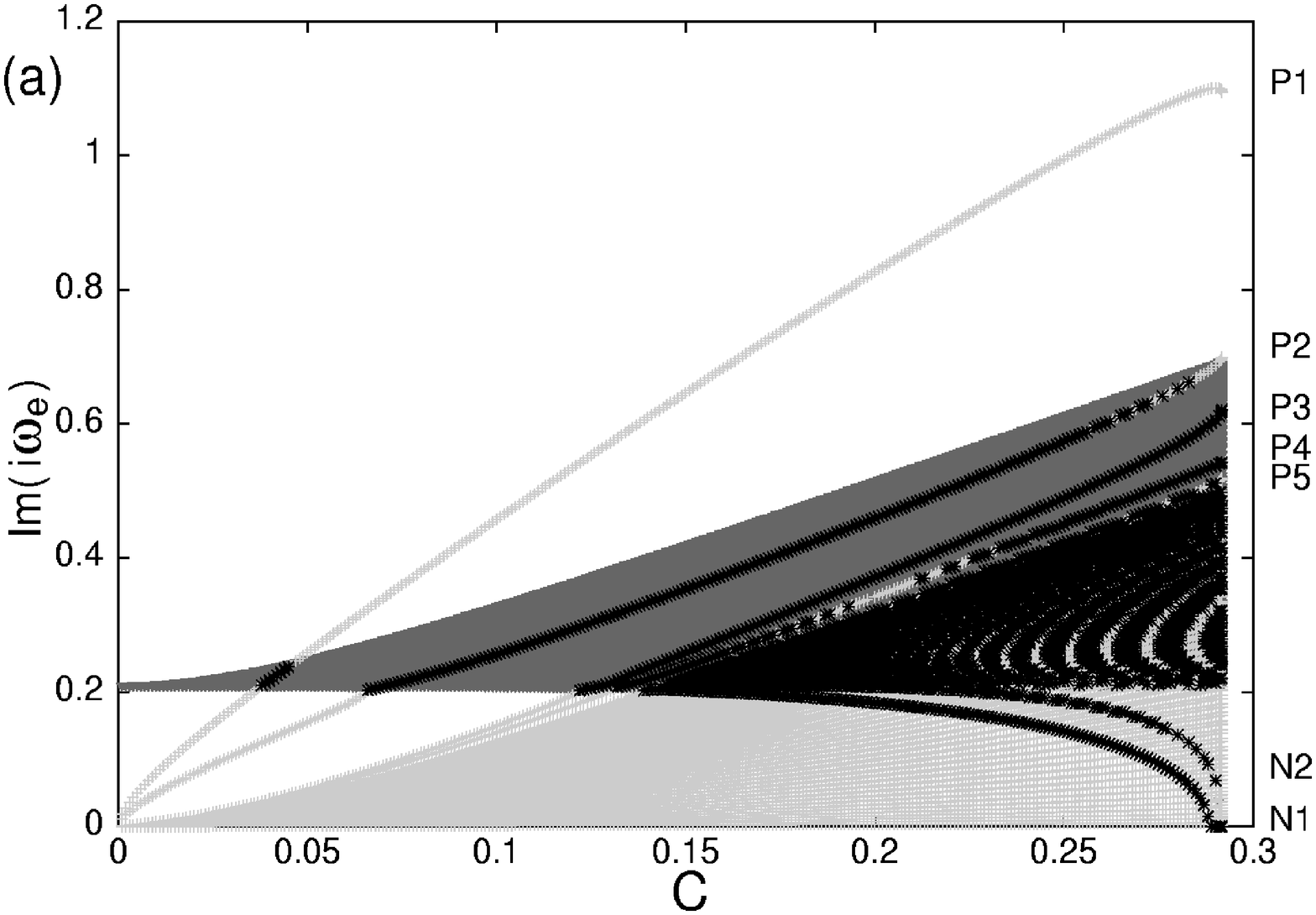}
}
\resizebox{0.7\columnwidth}{!}{%
  \includegraphics{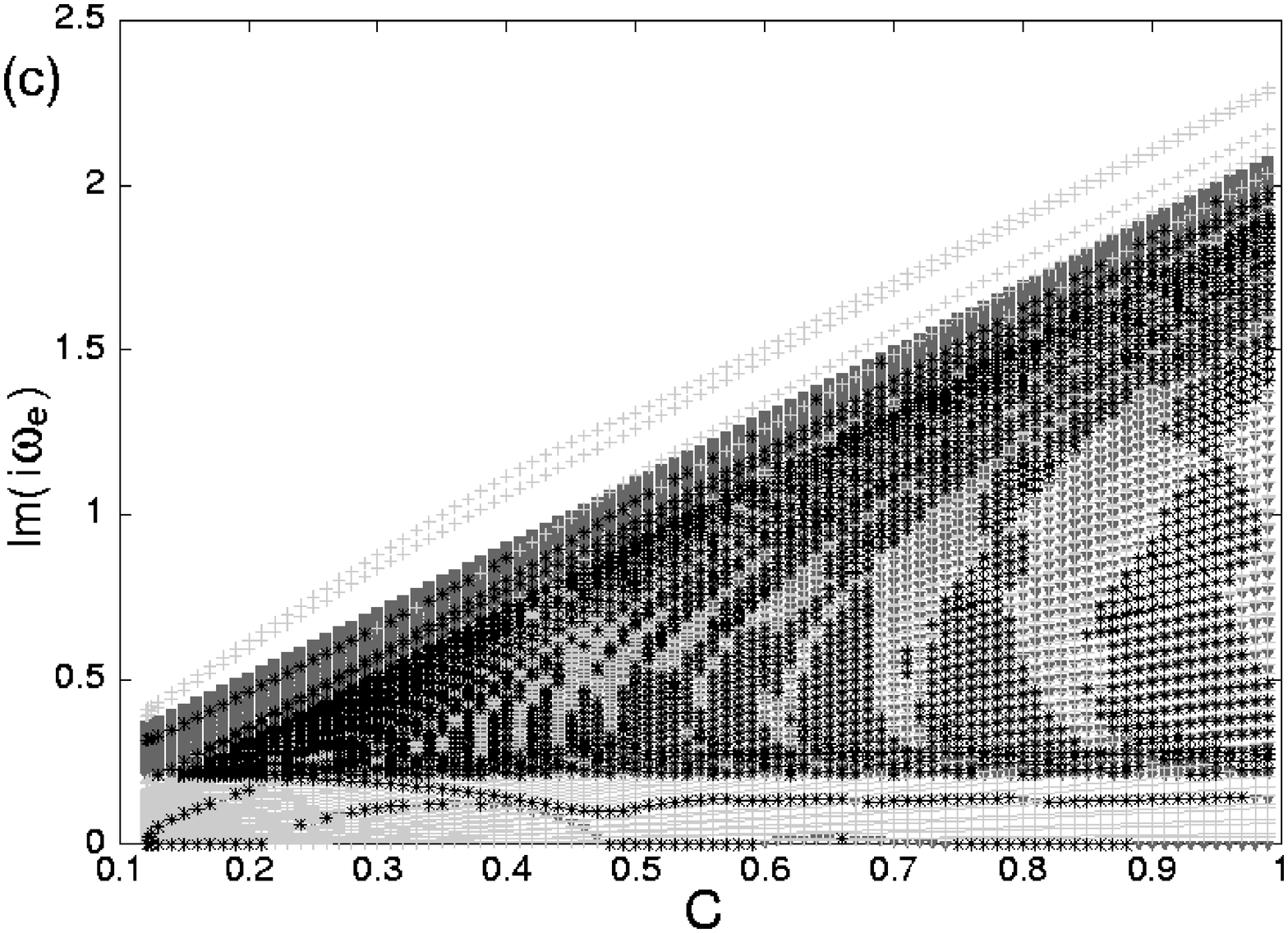}
}
}
\rotatebox{270}{
\resizebox{0.7\columnwidth}{!}{%
  \includegraphics{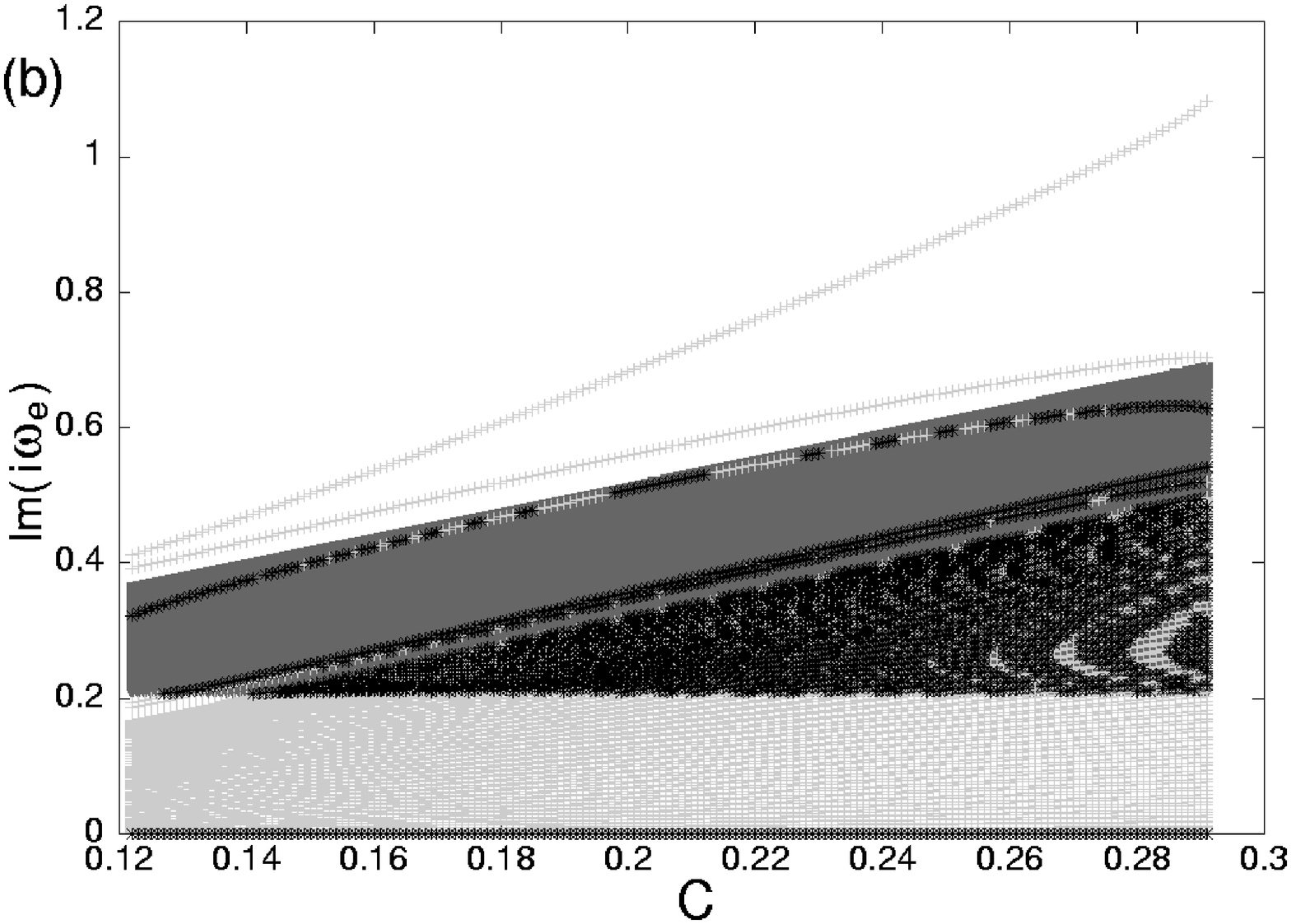}
}
\resizebox{0.7\columnwidth}{!}{%
  \includegraphics{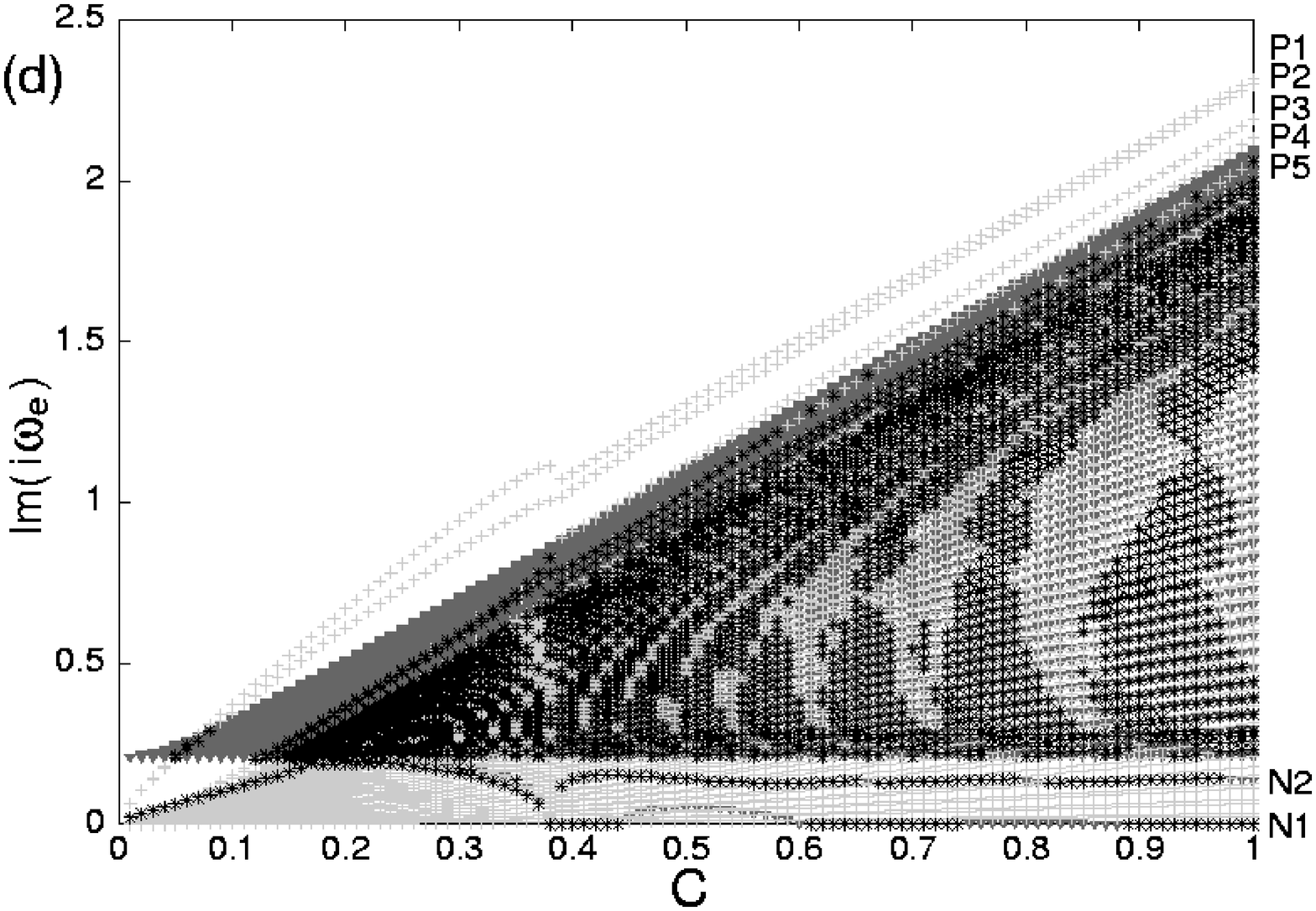}
}
}
\caption{Imaginary parts of eigenvalues $i\omega_e$ of (\ref{eq_per_matr})
for symmetric discrete-like (a), intermediate (b), continuous-like (c) and
antisymmetric (d) 'normal' DOGBs
at frequency deviation (\ref{dw})
$\Delta\omega=0.1\cdot \delta^2$.
Light-grey crosses (dark-grey triangles) correspond to eigenvalues with 
$\kappa=+1$ ($\kappa=-1$), 
and black stars to unstable eigenvalues.
(a)-(c) correspond to  upper right loop in Fig.~\ref{fig:en_Dis-Con2}(a).}
\label{fig:normal_DOGB_eigen}
\end{figure*}

\subsubsection{'Normal' DOGBs}
\label{subsubsec_normalDOGB}

In Fig.~\ref{fig:normal_DOGB_eigen} we show 
stability results
for symmetric discrete-like,
intermediate, continuous-like, and anti\-symmetric out-gap breathers 
at frequency detuning
$\Delta\omega=0.1\cdot\delta^2$. (Note, that in 'normal' DOGBs the 
tail wave-number
is $q=\pi/2$, so according to
(\ref{dw}) 
$\Delta\omega=\omega_b-\omega_o(\pi/2)\equiv\omega_b-\omega_2$.)
As for DGBs  (Sec.~\ref{subsec_DGB}), there are two bands
of extended eigenvalues with opposite Krein signatures:
\begin{eqnarray}
\label{bands_DOGB}
&&i\omega_{\mp}=\left\{\frac{\left(\Delta\omega+2\delta^2\right)^2+
8C^2\cos^2 (q)}{2}\pm \right.\\
\nonumber
&&\left.\qquad\pm\frac12\sqrt{\left(\Delta\omega+2\delta^2\right)^4+
16C^2\cos^2
(q)\left(4\delta^4-\Delta\omega^2\right)}\right\}^{\frac12},
\end{eqnarray}
with all wavenumbers $q$ allowed for the particular system
size.
But now the band $\{i\omega_+\}$ with positive
Krein signature always has its lower boundary at zero.
Moreover, as for non-localized solutions Krein instabilities corresponding 
to extended
eigenmodes survive also for infinite systems (e.g.\cite{Morgante}),
overlapping of the extended 
bands  is more essential for extended DOGBs than for localized DGBs.

Increasing the coupling,  several localized modes bifurcate
from the top of the band with $\kappa=+1$
($P1$-$P5$
in Fig.~\ref{fig:normal_DOGB_eigen} (a),(d)) and
from the bottom of the band with $\kappa=-1$
 ($N1$ and $N2$
in Fig.~\ref{fig:normal_DOGB_eigen} (a),(d)). Penetrating the bands
with
opposite Krein signature, these  modes produce oscillatory
instabilities. The $N1$ mode also causes a real instability with
an eigenvector of opposite symmetry to the DOGB,
colliding with its complex conjugate at
zero (PN instability yielding
'exchange of stability\footnote{Krein instabilities generally also exist 
here, see 
Fig.~\ref{fig:normal_DOGB_eigen}.}' 
between symmetric and antisymmetric DOGBs, cf. Sec.\ \ref{subsec_DGB}).
Similarly, the $N2$ mode causes a real instability with
an eigenvector of the same symmetry as the DOGB,
in the regime of transition from discrete-like to continuous-like
solutions.

However, the behaviour of the localized modes at small coupling
is different for gap and out-gap breathers.
For symmetric DGBs and DOGBs, in the limit of small $C$ 
only two modes can stay localized: $P1$ and $P2$, 
associated with symmetric and antisymmetric oscillations on the sites
closest to the breather center.
In the case of DOGBs, in contrast to DGBs, there is no threshold 
$\Delta\omega_{cr}$, and 
these two eigenvalues
bifurcate from the extended band immediately
when $C$ becomes non-zero, for all
DOGB frequencies (see Appendix~\ref{append}, Table~\ref{tab:1}).
The
differences between each of the $P1$ and $P2$ eigenvalues and the top
boundary of the extended band are in this case
proportional to $\sqrt{C}$ at small values of $C$
(see Fig.~\ref{fig:normal_DOGB_eigen} and Table~\ref{tab:1}; 
compare with $C^2$-dependence for DGBs as discussed in Sec.\ \ref{subsec_DGB})
\begin{figure}
\rotatebox{270}{
\resizebox{0.7\columnwidth}{!}{%
  \includegraphics{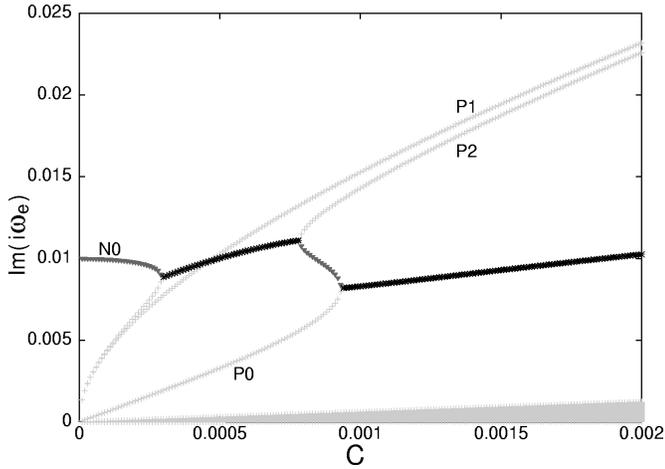}
}
}
\caption{Blow-up of a regime close to the origin of 
Fig.\ \ref{fig:normal_DOGB_eigen}(d).}
\label{fig:normal_DOGB_asym}
\end{figure}

For antisymmetric 'normal' DOGB 
$\{0\Uparrow(\uparrow\mathbb{O}\downarrow\mathbb{O})\}_A^O$ 
an additional mode $N0$
appears (see Fig.~\ref{fig:normal_DOGB_asym}), localized on the 'central' 
site $N_c$ which is
the only 'light' site with code '0'.
 Note,
that in an antisymmetric DGB $\{0\Uparrow(0\mathbb{O})\}_A^G$ all
'light'
sites have zero codes, therefore there is no such localized
mode for breather frequencies inside the gap.

Another
important difference between  antisymmetric 'normal' DOGBs
and DGBs is, that while the latter always have a real instability 
at small coupling (see Sec.\ \ref{subsec_DGB}), 
the former are always stable 
in a small interval of $C$: $[0,C_0]$ 
(see Fig.~\ref{fig:normal_DOGB_asym}). A localized mode $P0$,
with eigenvector of opposite symmetry to the DOGB, now
bifurcates from the upper boundary of the positive Krein signature band, 
similarly to the $P1$ and $P2$
modes. The eigenfrequency is proportional to $C$ for small $C$
(see Table~\ref{tab:1} in Appendix~\ref{append}, eigenmode $(P0,A,O)$), 
and the
eigenvector for $C\rightarrow 0$
involves oscillations not only in the 'excited'
'heavy' sites, but also in the subsequent 'light' sites.
At a certain non-zero value of $C$ it collides with the $N0$ mode
producing an oscillatory instability, associated with
a resonance between oscillations in the central 'light' site and
symmetric coupled oscillations in the two neighboring 'heavy' sites.

The $N1$ localized mode, which is responsible for the 'exchange of
stability' between antisymmetric and symmetric 'normal' DOGBs, now
appears only at large coupling, in the continuous-like regime. 

\subsubsection{'On-top' DOGBs}
\label{subsubsec_ontopDOGB}

\begin{figure*}
\rotatebox{270}{
\resizebox{0.7\columnwidth}{!}{%
  \includegraphics{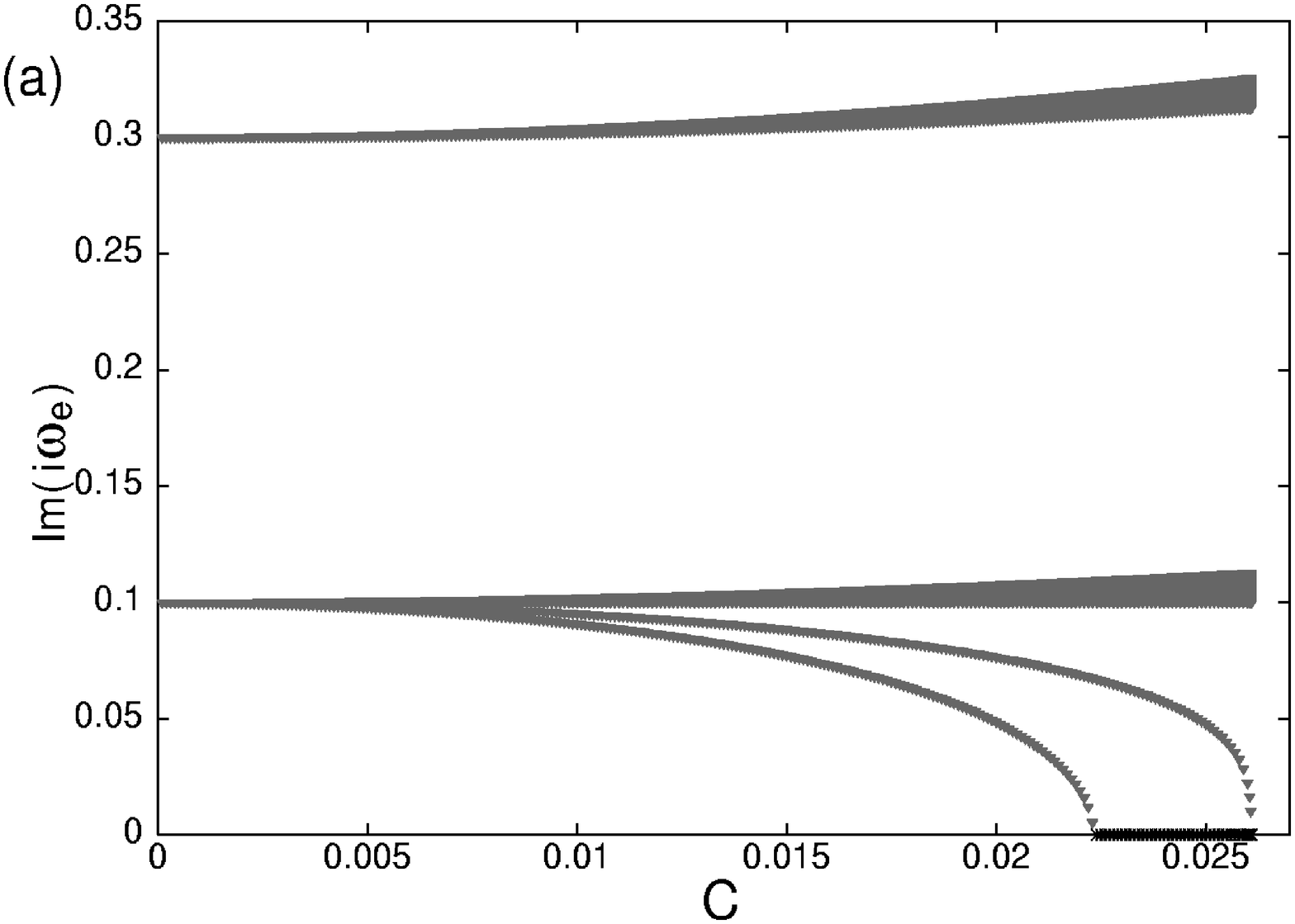}
}
\resizebox{0.7\columnwidth}{!}{%
  \includegraphics{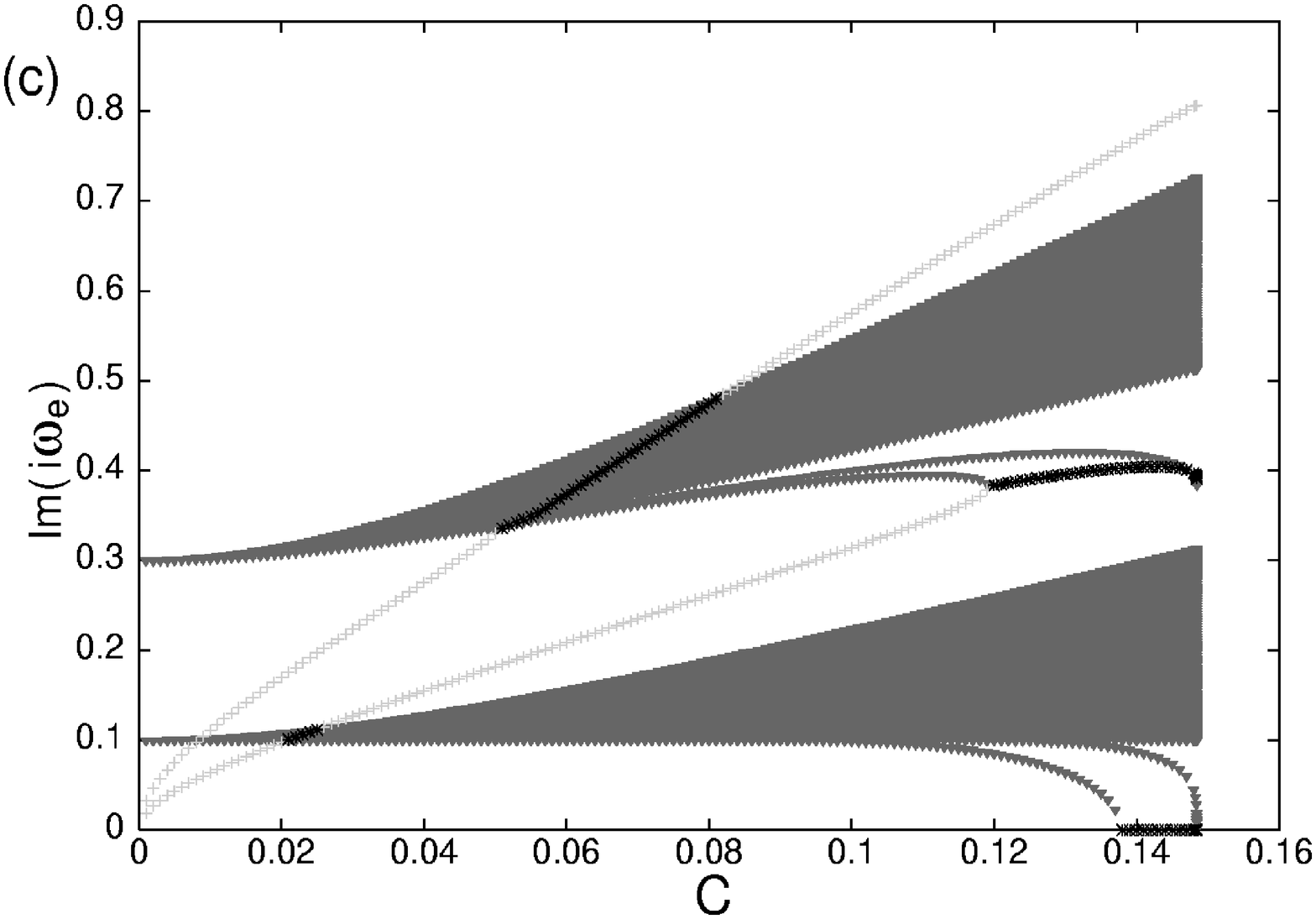}
}
}
\rotatebox{270}{
\resizebox{0.7\columnwidth}{!}{%
  \includegraphics{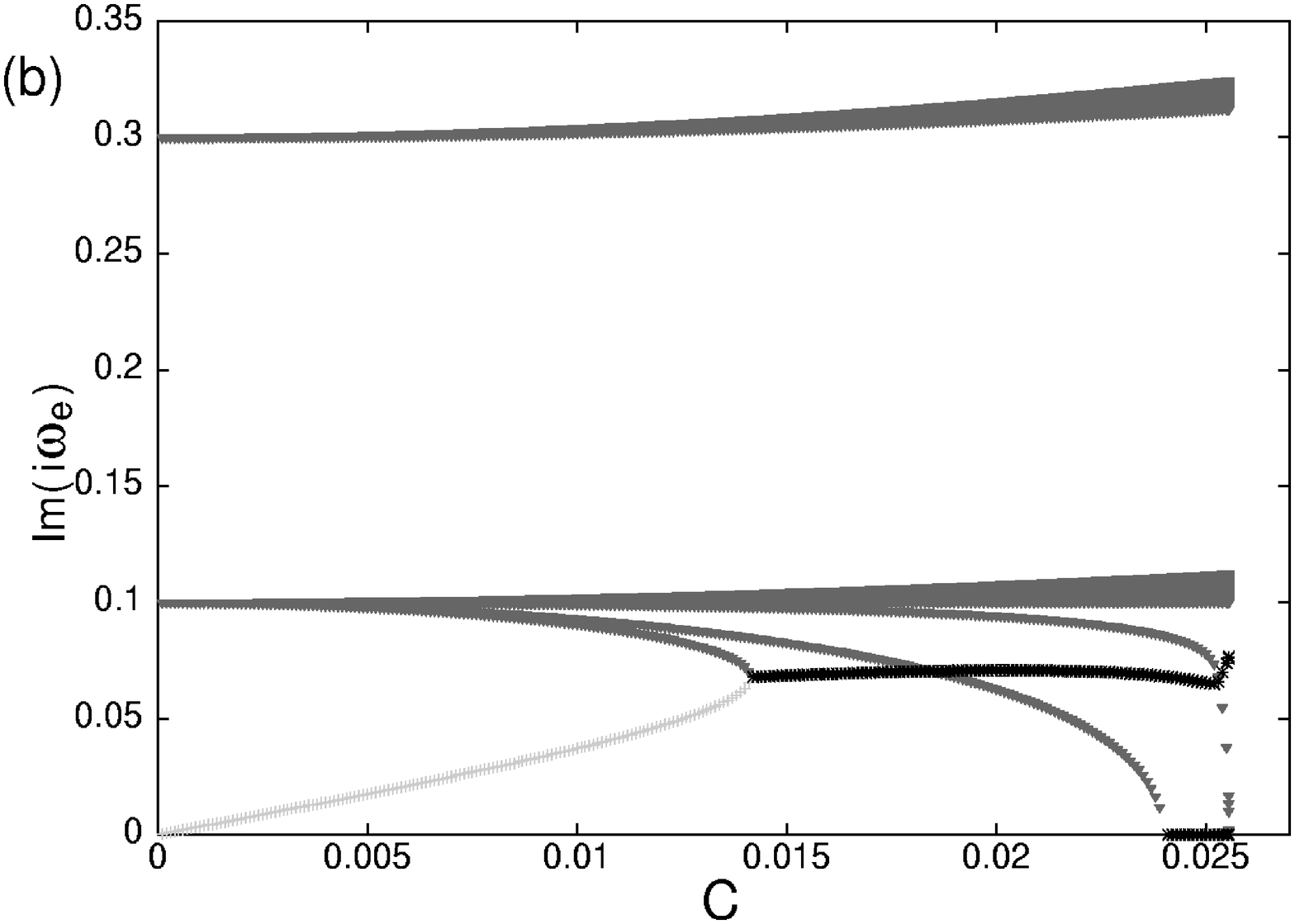}
}
\resizebox{0.7\columnwidth}{!}{%
  \includegraphics{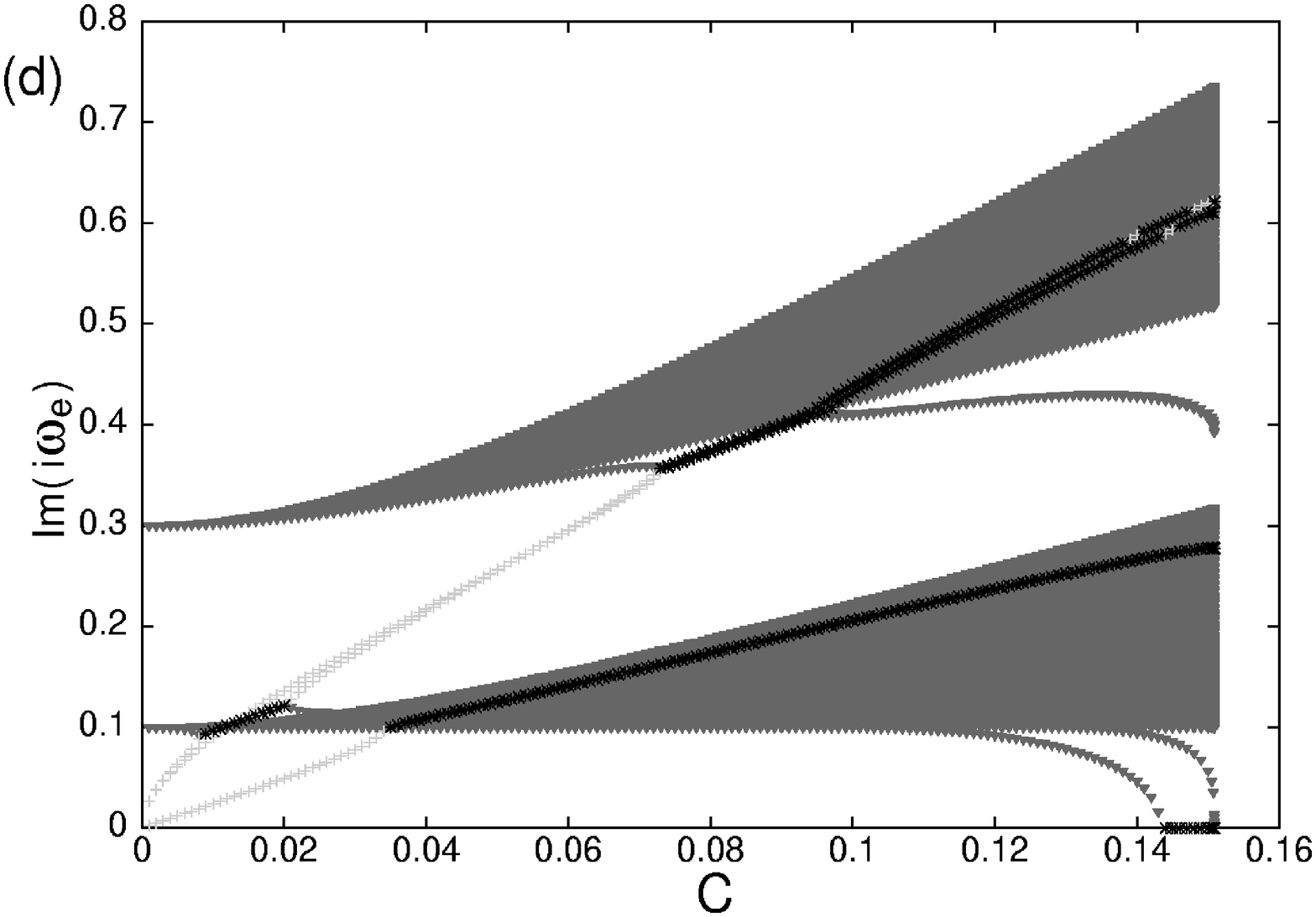}
}
}
\caption{Imaginary parts of eigenvalues $i\omega_e$ of (\ref{eq_per_matr})
for continuation in coupling of DOTBs $\{\Uparrow(0\mathbb{O})\}_S^O$
(a), $\{0\Uparrow(0\mathbb{O})\}_A^O$ (b),
$\{\Uparrow\uparrow(\mathbb{O}0)\}_S^O$ (c),
$\{0\Uparrow\uparrow(\mathbb{O}0)\}_A^O$ (d)
at frequency deviation (\ref{dw}) 
$\Delta\omega=\delta^2$ (note, that $q=\pi$ for DOTBs).
Light-grey crosses (dark-grey triangles) correspond to eigenvalues with 
$\kappa=+1$ ($\kappa=-1$), 
and black stars correspond to unstable eigenvalues.}
\label{fig:ontop_DOGB_eigen}
\end{figure*}

As indicated in Sec.~\ref{subsec_freq2}, 'on-top' breathers cannot
be continued in frequency down to the linear spectrum\footnote{We
here only discuss DOTBs with nonzero oscillations at 'heavy' sites as 
$C\rightarrow 0$. There are also solutions with nonzero codes only at 
'light' sites,
e.g. $\{\uparrow(\mathbb{O}0)\}_S^O$, which can be continued down to 
the linear spectrum.}. 
Analogously they cannot be continued in coupling up to arbitrary
values at fixed frequency detuning $\Delta\omega$,
since they bifurcate with other DOTBs in the same way as 
when continued in frequency at fixed $C$
(see Fig.~\ref{fig:bifur}(a)). By contrast, the DOTBs can be
smoothly continued by simultaneously increasing the frequency detuning 
and the coupling at fixed value of the ratio
$\Delta\omega/C$. This corresponds to the continuation towards the
'non-modulated' limit $\delta^2/C\rightarrow 0$ (see Eq.
(\ref{DNLS-algebra})), i.e. towards usual 'on-top' breathers in
homogeneous media. Only for $\delta^2/C\ = 0$ the smooth
continuation in coupling at fixed frequency detuning can
be performed; at any non-zero $\delta^2/C$ the exact
limit $\Delta\omega/C\rightarrow 0$ cannot be reached because of the
bifurcations.

In Fig.~\ref{fig:ontop_DOGB_eigen} results of continuation in
coupling of DOTBs $\{\Uparrow(0\mathbb{O})\}_S^O$,
$\{0\Uparrow(0\mathbb{O})\}_A^O$,
$\{\Uparrow\uparrow(\mathbb{O}0)\}_S^O$ and
$\{0\Uparrow\uparrow(\mathbb{O}0)\}_A^O$ at fixed
frequency detuning  are shown. Each DOTB is
continued
from $C=0$ up to the bifurcation point, where it bifurcates with
DOTBs $\{\Uparrow\downarrow(\mathbb{O}0)\}_S^O$,
$\{0\Uparrow\downarrow(\mathbb{O}0)\}_A^O$,
$\{\Uparrow\uparrow\mathbb{O}\uparrow(\mathbb{O}0)\}_S^O$ and
$\{0\Uparrow\uparrow\mathbb{O}\uparrow(\mathbb{O}0)\}_A^O$,
respectively.
For all DOTBs the extended eigenvalues
have the same (negative) Krein signature. Still they can be
divided into two bands ('upper' and 'lower'):
\begin{eqnarray}
\label{bands_ontop}
i\omega_{up}&=&\omega_b-\omega_o(q),\\
\nonumber
i\omega_{lo}&=&\omega_b-\omega_a(q),
\end{eqnarray}
where $\omega_{o,a}(q)$ are frequencies from the upper and
lower bands of the linear spectrum (\ref{dispersion}).

Close to the bifurcation point, all the described DOTBs possess
two real instabilities with symmetric and antisymmetric eigenvectors,
respectively.
These instabilities are similar to those produced by the $N1$ and $N2$
modes in the case of DGBs and 'normal' DOGBs in the region
of transition from discrete- to continuous-like solutions.

Like 'normal' antisymmetric DOGBs, antisymmetric DOTBs have
no instabilities at small coupling, so
for each pair of symmetric and antisymmetric DOTBs there is
an interval $[0,C_0]$ of simultaneous stability. 
The localized mode associated with symmetric coupled
oscillations in the two 'excited' 'heavy' sites (similar to the $P0$ mode
for a 'normal' antisymmetric DOGB)
bifurcates from zero and, having $\kappa>0$,
collides with another localized mode 
associated with oscillations in the 'central' 'light' site
(similar to the $N0$ mode). This results in an oscillatory
instability of antisymmetric DOTBs
(Fig.~\ref{fig:ontop_DOGB_eigen}(b), (d)).
The DOTBs $\{\Uparrow\uparrow(\mathbb{O}0)\}_S^O$ and
$\{0\Uparrow\uparrow(\mathbb{O}0)\}_A^O$ have two additional
localized modes with $\kappa>0$, 
similar to the $P1$ and $P2$ modes described in
Sec.~\ref{subsubsec_normalDOGB}. Penetrating the extended bands or
colliding with other localized modes (with $\kappa <0$) 
they also produce oscillatory instabilities
(Fig.~\ref{fig:ontop_DOGB_eigen}(c), (d)).

%
%

\section{Conclusions}
\label{sec_conclude}
We have investigated the linear stability properties of
discrete gap and out-gap breathers in the modulated DNLS model, which
in particular describes electromagnetic wave
propagation in an array of weakly coupled optical waveguides of
different widths. The coupling constant was varied from
$C=0$ (anti-continuous limit) up to values, at which the
localization length of the breather becomes much larger than
the array constant (continuous limit).

Different types of oscillatory and real instabilities of DGBs were
revealed, similar to those described for DGBs in the diatomic
Klein-Gordon model \cite{we}. However, in contrast to the KG case, 
DGBs in infinite DNLS chains do not possess any
oscillatory instability 
when the breather frequency is in the lower
half of the gap.

The transition from gap to out-gap breathers was
studied by varying the breather frequency at different fixed values of
the coupling. Such transitions were found to occur in
different ways, depending on the value of the coupling, and a relation
to the transition from discrete- to continuous-like breathers 
(described earlier for DGBs in the diatomic KG model \cite{we}) was 
revealed.

Discrete out-gap breathers with  tails of different wave numbers were
investigated, and a general bifurcation picture described.
DOGBs were shown to, generally,
possess the same types of oscillatory and real instabilities as
DGBs. Still, there are important
differences between stability properties of DGBs and DOGBs.
Unlike antisymmetric DGBs, antisymmetric
DOGBs are linearly stable for small coupling, 
so there is an interval in coupling $[0,C_0]$ 
where symmetric and antisymmetric DOGBs are stable
simultaneously. Also, since DOGBs  generally are 
non-localized, they possess strong oscillatory instabilities 
associated
with resonances between extended lattice modes of the upper- and
lower-band
sub-fields. For DGBs such instabilities do not survive in the
limit of  infinite chain.

The process of breather frequency penetration into the linear 
spectrum and
bifurcation of a DGB into a DOGB was found to have 
many qualitative similarities
with the breather second harmonic penetration into the linear spectrum 
and formation of  'phonobreathers' with non-zero tails, 
discussed for a mono\-atomic KG chain in \cite{phantom}.
However, we found no
analogue to 'phantom' breathers with vanishing tails, since 
all DOGBs with tails of different 
wave numbers  apparently are non-localized at frequencies inside the linear 
spectrum. The question about the existence
of such localized excitations inside the linear wave spectrum in
modulated models is thus still unresolved.

Another interesting and important question, which we address for future
investigations, concerns the possibility of finding exact moving gap and 
out-gap breathers. As we have shown here, DGBs and DOGBs
in the modulated DNLS
model exhibit the 'exchange of stability' between symmetric and
antisymmetric modes, and thus a considerably increased mobility of these 
excitations is 
expected for parameter values
close to the stability exchange point \cite{Cretegny_PhD}. 
We hope that our results will stimulate experimental activity to directly 
observe this phenomenon, e.g. in coupled waveguide arrays.

We recently demonstrated such DGB 
mobility in the diatomic
KG chain \cite{we}, and connected
it with a vanishing of the PN barrier. 
However, the determination of the PN potential, 
and the PN barrier as the energy difference between symmetric and 
antisymmetric breathers, is not unambiguous in the KG model,  
since for fixed values of the
system parameters there exist continuous families of 
symmetric and antisymmetric 
DBs with different frequencies. 
In the modulated DNLS 
model, the existence of an additional conserved quantity, 
the excitation norm, 
makes it natural to define the PN barrier as the difference between energies 
of
symmetric and antisymmetric DBs with the same norm\cite{KC93}.
Thus it is of interest to study more in details the 'exchange of stability' 
in
the modulated DNLS model, and to look for 
exact moving breathers in this regime.

Another interesting effect, which also should be experimentally 
observable, arises as a consequence of the existence of DGB internal 
modes  with frequencies above the continuous 
spectrum (e.g.\ $P1$-$P5$ in Fig.~\ref{fig:DGB_eigen}). Since then also 
all multiples of the mode frequency will lie outside the continuous spectrum, 
similar arguments as in \cite{KW03} prove the existence of {\em 
exact localized quasiperiodic gap breathers}, with two, generally 
incommensurate, frequencies. These solutions are characterized by persistent 
large-amplitude internal intensity oscillations. It is particularly 
interesting to note, that since e.g. the $P1$ mode of the symmetric DGB 
is always above the continuous spectrum in the lower half of the gap 
when
$-1.56 \lesssim \Delta\omega/\delta^2 < -1$, quasiperiodic solutions 
bifurcating from this mode should exist 
in continuous families, arbitrarily close as well to the anticontinuous 
as to the continuous limit. The properties of such solutions, which cannot 
exist in KG models, will be discussed elsewhere.

\begin{acknowledgement}
We acknowledge support from the Royal Swedish Academy of Sciences 
and ``Knut och Alice Wallenbergs jubileumsfond''.
A.V.G. acknowledges support from the Swedish Institute. M.J.
acknowledges support from the Swedish Research Council.
\end{acknowledgement}

\appendix
\section{Calculation of localized
modes at small $C$}
\label{append}
For localized eigenmodes at small $C$, the eigenvalue
problem (\ref{eq_per_real2}) can
be simplified. Indeed, at small coupling these eigenmodes
are mainly localized on a few central sites (if the
corresponding eigenvalues lie inside the extended band, these
eigenmodes are not purely localized, but the central
site amplitudes are much larger than the tails).
In this case one can put $(a,b)_n\equiv0$ for
$\left| n-N_c \right|\ge N_\epsilon$, where $N_\epsilon$ characterizes the
localization length of an eigenmode. Moreover, using $C$ as a
small parameter, one can obtain approximate expressions for the central site
amplitudes of the unperturbed solution
$\phi_n^{(0)}$, for DGB $\{\Uparrow(0\mathbb{O})\}_S^G$:
\begin{eqnarray}
\label{a1}
&&\phi_{N_c}^2=2\delta^2+\Delta\omega-\frac{2}{\Delta\omega}C^2+\mathcal{O}
\left(C^3\right),
\\
\nonumber
&&\phi_{N_c\pm1}^2=\frac{2\delta^2+\Delta\omega}{\Delta\omega^2}C^2+\mathcal{O}
\left(C^3\right),
\end{eqnarray}
DGB $\{0\Uparrow(0\mathbb{O})\}_A^G$:
\begin{eqnarray}
\nonumber
&&\phi_{N_c}^2\equiv 0, \\
\label{a2}
&&\phi_{N_c\pm 1}^2=
2\delta^2+\Delta\omega-\frac{1}{\Delta\omega}C^2+\mathcal{O}
\left(C^3\right),
\\
\nonumber
&&\phi_{N_c\pm 2}^2=
\frac{2\delta^2+\Delta\omega}{\Delta\omega^2}C^2+\mathcal{O}
\left(C^3\right),
\end{eqnarray}
DOGB $\{\Uparrow(\uparrow\mathbb{O}\downarrow\mathbb{O})\}_S^O$:
\begin{eqnarray}
\label{a3}
&&\phi_{N_c}^2=2\delta^2+\Delta\omega+2\sqrt{\frac{\Delta\omega}
{2\delta^2+\Delta\omega}}C+
\mathcal{O}\left(C^2\right),
\\
\nonumber
&&\phi_{N_c\pm1}^2=\Delta\omega+2\sqrt{
\frac{2\delta^2+\Delta\omega}{\Delta\omega}}C
+\mathcal{O}\left(C^2\right),
\end{eqnarray}
and DOGB $\{0\Uparrow(\uparrow\mathbb{O}\downarrow\mathbb{O})\}_A^O$:
\begin{eqnarray}
\nonumber
&&\phi_{N_c}^2\equiv 0, \\
\label{a4}
&&\phi_{N_c\pm 1}^2=2\delta^2+\Delta\omega+\sqrt{\frac{\Delta\omega}
{2\delta^2+\Delta\omega}}C+
\mathcal{O}\left(C^2\right),
\\
\nonumber
&&\phi_{N_c\pm 2}^2=
\Delta\omega+\sqrt{
\frac{2\delta^2+\Delta\omega}{\Delta\omega}}C
+\mathcal{O}\left(C^2\right).
\end{eqnarray}
Note that $\Delta\omega<0$ ($\Delta\omega>0$) for gap (out-gap) breathers.

Thus, in the case of a sym\-metric DGB $\{\Uparrow(0\mathbb{O})\}_S^G$,
the $P1$ mode is mainly
localized on the central
'heavy' site with $n=N_c$, the two neighboring 'light' sites with
$n=N_c\pm 1$, and the subsequent pair of 'heavy' sites with
$n=N_c\pm 2$. Therefore, for small $C$
one can put $(a,b)_n\equiv 0$ when $\left| n-N_c
\right|\ge3$ in equations (\ref{eq_per_real2}). Substituting
the approximate breather solution (\ref{a1})
into (\ref{eq_per_real2}) and using the sym\-me\-try of the $P1$ mode:
$(a,b)_{N_c-i}=(a,b)_{N_c+i}$, we obtain the corresponding eigenfrequency
as 
\begin{equation}
\label{p1_S_G}
\omega_e^{\{P1,S,G\}}=-\Delta\omega+\frac{2C^2}{\Delta\omega^2}\left(
4\delta^2+\Delta\omega \right) + \mathcal{O}\left(C^3\right).
\end{equation}
Here the superscript marks the type of  mode ($P1$ or $P2$), the
symmetry of the breather
$\phi_n^{(0)}$ ($S$ [$A$] for symmetric [antisymmetric]
breathers), and the type of breather
($G$ and $O$ for gap and out-gap
breathers, respectively).
\begin{table*}
\caption{Localized eigenmodes at small coupling. 
Eigenfrequencies are given at 
lowest significant order of $C$,
with the notations: $\Delta_1\equiv 2\delta^2+\Delta\omega$ (i.e. 
frequency detuning measured from the lower gap edge),
$\mu_{(\pm)}\equiv\left[16\delta^4+\Delta_1^2 \pm
\sqrt{128\delta^8+\Delta_1^4}
\right]/\left[4\Delta\omega^2\delta^2\right]$. Schematic eigenvector 
structures are shown for the central part; higher arrows denote sites 
with nonzero oscillations at $C\rightarrow 0$. 
}
\label{tab:1}       

\begin{tabular}{p{0.15\textwidth}|p{0.28\textwidth}|p{0.39\textwidth}|
p{0.08\textwidth}}
\hline
\hline
Eigen\-mode & Schematic eigenvector structure& 
Eigenfrequency $\omega_e$ &
$\Delta\omega_{cr}$
\\
\hline
\hline
$(P1,S,G)$ 
& 
$...\mathbb{O} 0 \, {\scriptstyle \Uparrow} \uparrow 
{\scriptstyle \Downarrow}  \uparrow {\scriptstyle \Uparrow}\, 0 \mathbb{O}...$
&
$-\Delta\omega+2C^2\left(
4\delta^2+\Delta\omega \right)/\Delta\omega^2$
&
$-1.56\cdot\delta^2$
\\
\hline
$(P2,S,G)$ 
& 
$...\mathbb{O} 0 \, {\scriptstyle \Uparrow} \downarrow \mathbb{O} \uparrow 
{\scriptstyle \Downarrow}\, 0 \mathbb{O} ...$
&
$-\Delta\omega+C^2 \left(4\delta^4+
\Delta_1^2 \right)/\left[
2\delta^2\Delta\omega^2\right]$
& 
$-1.1\cdot\delta^2$
\\
\hline
$(P1,A,G)$ 
& 
$...\mathbb{O} 0 \, {\scriptstyle \Downarrow} \uparrow 
{\scriptstyle \Downarrow} \uparrow {\scriptstyle \Downarrow} \uparrow 
{\scriptstyle \Downarrow}\, 0 \mathbb{O}...$
&
$-\Delta\omega+C^2 \mu_{(+)}$
&                                                                       
$-1.85\cdot\delta^2$
\\
\hline
$(P2,A,G)$ 
& 
$...\mathbb{O} 0 \, 
{\scriptstyle \Downarrow} \uparrow {\scriptstyle \Downarrow} 
\, 0 \, {\scriptstyle \Uparrow} \downarrow 
{\scriptstyle \Uparrow} \, 0 \mathbb{O}...$
&
$-\Delta\omega+
C^2\left(
12\delta^4+4\Delta\omega\delta^2+\Delta\omega^2\right)/
\left[2\Delta\omega^2\delta^2\right]$
& 
$-1.44\cdot\delta^2$                                                                 
\\
\hline
$(P3,A,G)$ 
& 
$...\mathbb{O} 0 \, 
{\scriptstyle \Downarrow} \uparrow {\scriptstyle \Downarrow} 
\uparrow {\scriptstyle \Downarrow} \uparrow 
{\scriptstyle \Downarrow} \, 0 \mathbb{O}...$
&
$-\Delta\omega+C^2 \mu_{(-)}$
& 
$-0.86\cdot\delta^2$                                                                 
\\
\hline
$(N1,A,G)$ 
& 
$...0 \mathbb{O} \, {\scriptstyle \uparrow} \Uparrow {\scriptstyle \uparrow} 
\Uparrow {\scriptstyle \uparrow} \, \mathbb{O} 0 
...$
&
$i2C\sqrt{\Delta_1}/\sqrt{-\Delta\omega}$
& 
$-2\cdot\delta^2$
\\
\hline
$(P1,S,O)$ 
& 
$...\mathbb{O} 0 \, {\scriptstyle \Downarrow} \downarrow \Uparrow 
\downarrow {\scriptstyle \Downarrow} \,  0 \mathbb{O}...$
&
$\left\{6\sqrt{\Delta\omega\Delta_1}C\right\}^{0.5}$
& 
$0$
\\
\hline
$(P2,S,O)$ 
& 
$...\mathbb{O} 0 \, {\scriptstyle \Uparrow} \downarrow \mathbb{O} \uparrow 
{\scriptstyle \Downarrow} \, 0 \mathbb{O}...$
&
$\left\{2\sqrt{\Delta\omega\Delta_1}C\right\}^{0.5}$
& 
$0$
\\
\hline
$(P1,A,O)$ 
& 
$...0 \mathbb{O} \uparrow\Downarrow 0 \Uparrow \downarrow \mathbb{O} 0...$
&
$\left\{4\sqrt{\Delta\omega\Delta_1}C\right\}^{0.5}$
& 
$0$
\\
\hline
$(P2,A,O)$ 
& 
$...\mathbb{O} 0 \, {\scriptstyle \Uparrow} \downarrow \Uparrow 
{\scriptstyle \uparrow} \Uparrow \downarrow 
{\scriptstyle \Uparrow}\, 0 \mathbb{O}...$
&
$\left\{4\sqrt{\Delta\omega\Delta_1}C\right\}^{0.5}$
& 
$0$
\\
\hline
$(P0,A,O)$ 
& 
$...\mathbb{O} 0 \, {\scriptstyle \Downarrow} \uparrow\Uparrow 
{\scriptstyle \uparrow} \Uparrow \uparrow 
{\scriptstyle \Downarrow} \, 0 \mathbb{O}...$
&
$C\sqrt{2\Delta_1^2-\Delta\omega^2}/
\sqrt{\Delta\omega\Delta_1}$
&
$0$
\\
\hline
$(N0,A,O)$ 
& 
$...0 \mathbb{O} \, {\scriptstyle \downarrow}\, {\scriptstyle \Uparrow} 
\uparrow {\scriptstyle \Uparrow}\, {\scriptstyle \downarrow} \, 
 \mathbb{O} 0...$
&
$\Delta\omega+\mathcal{O}(C^4)$
&
$0$
\\
\hline
\hline
\end{tabular}
\end{table*}

Analogously, for the antisymmetric $P2$ mode we have:
\begin{eqnarray}
\label{p2_S_G}
&&\omega_e^{\{P2,S,G\}}=\\
\nonumber
&&\qquad =-\Delta\omega+\frac{C^2 \left(
8\delta^4+4\delta^2\Delta\omega+\Delta\omega^2 \right)}
{2\delta^2\Delta\omega^2} +
\mathcal{O}\left(C^3\right).
\end{eqnarray}

To see if a particular eigenmode is localized or not at small
 $C$, i.e. whether
the corresponding eigenvalue lies inside or outside the extended band,
we check the difference between the  eigenvalue
and the band edge  $\omega_+^{\{max,G\}}$ (\ref{ev_ext}):
\begin{equation}
\label{bandedge}
\omega_+^{\{max,G\}}=\omega_o(\pi)-\omega_b
=-\Delta\omega-\delta^2+\sqrt{\delta^4+4C^2},
\end{equation}
from which it bifurcates (see Fig.~\ref{fig:DGB_eigen}). This yields 
for the $P1$
mode
\begin{eqnarray}
\label{dw_p1_s_g}
&&\omega_e^{\{P1,S,G\}}-\omega_+^{\{max,G\}}=
\\
\nonumber
&&\qquad=\frac{2C^2}{\Delta\omega^2\delta^2}\left(4\delta^4+\delta^2\Delta
\omega - \Delta\omega^2 \right) + \mathcal{O}\left(C^3\right).
\end{eqnarray}
Consequently, when the frequency detuning of a  symmetric DGB
is above the critical
value $\Delta\omega>\Delta\omega_{cr}^{\{P1,S,G\}}\approx
-1.56\cdot\delta^2$,
the $P1$ eigenmode is localized at any non-zero
value of $C$. By contrast, when
$\Delta\omega\le\Delta\omega_{cr}^{\{P1,G\}}$, the $P1$ eigenmode is not
localized at small values of $C$.

Similar results are obtained for the $P2$ mode:
\begin{eqnarray}
\label{dw_p2_s_g}
&&\omega_e^{\{P2,S,G\}}-\omega_+^{\{max,G\}}=
\\
\nonumber
&&\qquad=\frac{C^2}{2\Delta\omega^2\delta^2}\left(8\delta^4+4\delta^2\Delta
\omega - 3\Delta\omega^2 \right) + \mathcal{O}\left(C^3\right).
\end{eqnarray}
The critical value of the frequency detuning for this mode is equal to
$\Delta\omega_{cr}^{\{P2,S,G\}}\approx
-1.1\cdot\delta^2$.

Using the same technique, expressions 
for the eigenvalues and eigenvectors corresponding to localized
modes at small coupling
can be obtained for antisymmetric DGB
$\{0\Uparrow(0\mathbb{O})\}_A^G$, and for out-gap breathers 
$\{\Uparrow(\uparrow\mathbb{O}\downarrow\mathbb{O})\}_S^O$,
$\{0\Uparrow(\uparrow\mathbb{O}\downarrow\mathbb{O})\}_A^O$. The 
results are presented in Table~\ref{tab:1}, where for each  
eigenmode the schematic eigenvector and eigenvalue 
are given at the lowest significant order of $C$, 
together with the approximate value of the critical frequency detuning 
$\Delta\omega_{cr}$, above which the mode is localized 
at small $C$. Note in particular that the antisymmetric
DGB $\{0\Uparrow (0\mathbb{O})\}_A^G$ has an imaginary eigenfrequency 
corresponding to the $N1$ mode, and thus it is unstable for small $C$. 
All other eigenfrequencies are real, proving the linear stability of 
the other three solutions for small $C$.


\begin{thebibliography}{99}

\bibitem{Aubry} S.\ Aubry, Physica D \textbf{103}, 201 (1997).
\bibitem{Flach} S.\ Flach, C.R.\ Willis, Phys.\ Rep.\ \textbf{295}, 181
(1998).
\bibitem{Entrap} G.P.\ Tsironis, S.\ Aubry, Phys.\ Rev.\ Lett.\ \textbf{77},
5225 (1996).
\bibitem{Chen} D.\ Chen, S.\ Aubry, G.P.\ Tsironis, 
Phys.\ Rev.\ Lett.\ \textbf{77}, 4776 (1996).
\bibitem{Cretegny_PhD} T.\ Cretegny, Ph.D.\ thesis, {\'E}cole Normale
Sup{\'e}rieure de Lyon, France, 1998 (in French).
\bibitem{Chen&Mills} W.\ Chen, D.L.\ Mills, Phys.\ Rev.\ Lett.\ \textbf{58},
 160 (1987).
\bibitem{Chub} O.A.\ Chubykalo, Yu.S.\ Kivshar, Phys.\ Rev.\ E \textbf{48},
4128 (1993); \textbf{49}, 5906(E) (1994).
\bibitem{Kiselev1} S.A.\ Kiselev, S.R.\ Bickham, A.J.\ Sievers, 
Phys.\ Rev.\ B \textbf{48}, 13508 (1993).
\bibitem{Aoki} M.\ Aoki, S.\ Takeno, A.J.\ Sievers, J.\ Phys.\ Soc.\ Jpn
\textbf{62}, 4295 (1993).
\bibitem{Kiselev2} S.A.\ Kiselev, S.R.\ Bickham, A.J.\ Sievers, 
Phys.\ Rev.\ B \textbf{50}, 9135 (1994).
\bibitem{Franchini1} A.\ Franchini, V.\ Bortolani, R.F.\ Wallis, Phys.
Rev.\ B \textbf{53}, 5420 (1996).
\bibitem{Livi} R.\ Livi, M.\ Spicci, R.S.\ MacKay, Nonlinearity
\textbf{10}, 1421 (1997).
\bibitem{James} G.\ James, P.\ Noble (submitted to Physica D).
\bibitem{Cretegny} T.\ Cretegny, R.\ Livi, M.\ Spicci, Physica D
\textbf{119}, 88 (1998).
\bibitem{Zolotaryuk} A.V.\ Zolotaryuk, P.\ Maniadis, G.P.\ Tsironis,
Physica B \textbf{296}, 251 (2001).
\bibitem{Maniadis} P.\ Maniadis, A.V.\ Zolotaryuk, G.P.\ Tsironis,
Phys.\ Rev.\ E\ \textbf{67}, 046612 (2003).
\bibitem{we} A.V.\ Gorbach, M.\ Johansson, Phys.\ Rev.\ E \textbf{67}, 
066608 (2003).
\bibitem{DNLSopt} A.A.\ Sukhorukov, Yu.S.\ Kivshar, H.S.\ Eisenberg, 
Y.\ Silberberg, IEEE J.\ Quantum Electron.\ \textbf{39}, 31 
(2003), and references therein.
\bibitem{DGBopt} A.A.\ Sukhorukov, Yu.S.\ Kivshar, Opt.\ Lett.\ \textbf{27}, 
2112 (2002).
\bibitem{newKivshar} A.A.\ Sukhorukov, Yu.S.\ Kivshar,
arXiv:nlin.PS/0303054 (2003).
\bibitem{SKPRL} A.A.\ Sukhorukov, Yu.S.\ Kivshar, 
Phys.\ Rev.\ Lett.\ \textbf{91}, 113902 (2003).
\bibitem{Kevrekidis} P.G.\ Kevrekidis, B.A.\ Malomed, Z.\ Musslimani, 
Eur.\ Phys.\ J.\ D \textbf{23}, 421 (2003).
\bibitem{Trombeltoni} A.\ Trombettoni, A.\ Smerzi, 
Phys.\ Rev.\ Lett.\ \textbf{86}, 2353 (2001); A.\ Smerzi, A.\ Trombettoni, Chaos \textbf{13}, 
766 (2003).
\bibitem{Cataliotti} F.S.\ Cataliotti et.\ al., Science \textbf{293}, 843
(2001). 
\bibitem{EJ} J.C.\ Eilbeck, M.\ Johansson, in "Localization and Energy 
Transfer in Nonlinear Systems", edited by L.\ V\'azquez, M.P.\ Zorzano, 
R.\ MacKay (World Scientific, Singapore, in press); arXiv: 
nlin.PS/0211049 (2002).
\bibitem{Peyrard} Yu.S.\ Kivshar, M.\ Peyrard, Phys.\ Rev.\ A \textbf{46},
3198 (1992);
Yu.S.\ Kivshar, Phys.\ Lett.\ A \textbf{173}, 172 (1993);
I.\ Daumont, T.\ Dauxois, M.\ Peyrard, Nonlinearity
\textbf{10}, 617 (1997).
\bibitem{Morgante} A.M.\ Morgante, M.\ Johansson, G.\ Kopidakis, S.\ Aubry, 
Physica D \textbf{162}, 53 (2002).
\bibitem{KF92} Yu.S.\ Kivshar and N.\ Flytzanis, 
Phys.\ Rev.\ A \textbf{46}, 7972 (1992).
\bibitem{mePRE} A.S.\ Kovalev, O.V.\ Usatenko, A.V.\ Gorbatch, 
Phys.\ Rev.\ E \textbf{60}, 2309 (1999).
\bibitem{Peyraurd} J.\ Coste, J.\ Peyraud, Phys.\ Rev.\ B \textbf{39},  13086
(1989); \textbf{39}, 13096 (1989).
\bibitem{Bose_outgap} A.V.\ Yulin, D.V.\ Skryabin, Phys.\ Rev.\ A
\textbf{67}, 023611 (2003).
\bibitem{embed} J.\ Yang, B.A.\ Malomed, D.J.\ Kaup, A.R.\ Champneys, 
Math.\ Comput.\ Simul.\ \textbf{56}, 585 (2001). 
\bibitem{Franchini2} A.\ Franchini, V.\ Bortolani, R.F.\ Wallis, J.\ Phys.: 
Condens.\ Matter \textbf{14}, 145 (2002). 
\bibitem{CE85} J.\ Carr, J.C.\ Eilbeck, Phys.\ Lett.\ A \textbf{109}, 201 
(1985).
\bibitem{Bri} T.J.\ Bridges, Proc.\ R.\ Soc.\ Lond.\ A \textbf{453}, 1365
(1997).
\bibitem{Skryabin} D.V.\ Skryabin, J.\ Opt.\ Soc.\ Am.\ B \textbf{19}, 529
(2002).
\bibitem{Marin&Aubry} J.L.\ Mar\'{\i}n, S.\ Aubry, Nonlinearity
\textbf{9}, 1501 (1996).
\bibitem{Eilbeck} J.C.\ Eilbeck, P.S.\ Lomdahl, A.C.\ Scott, Phys.\ Rev.\ B
\textbf{30}, 4703 (1984); Physica D \textbf{16}, 318 (1985).
\bibitem{phantom} A.M.\ Morgante, M.\ Johansson, S.\ Aubry, G.\ Kopidakis,
J.\ Phys.\ A: Math.\ Gen.\ \textbf{35}, 4999 (2002).
\bibitem{size_effect} J.L \ Mar\'{\i}n, S \ Aubry, Physica D
\textbf{119}, 163 (1998).
\bibitem{Vakh} N.G \ Vakhitov, A.A \ Kolokolov \ Radiophys \ Quantum
Elec\-tron \ \textbf{16}, 783 (1975). [Translated from Izvestiya
Vys\-shikh Ucheb\-nykh Zave\-denii, Ra\-dio\-fi\-zi\-ka 
\textbf{16}, 1020 (1973).]
\bibitem{KC93} Yu.S.\ Kivshar and D.K.\ Campbell, Phys.\ Rev.\ E 
\textbf{48}, 3077 (1993).
\bibitem{KW03} P.G.\ Kevrekidis and M.I.\ Weinstein, Math.\ Comput.\ 
Simul.\ {\bf 62}, 65 (2003).
\end{thebibliography}
\end{document}